## Research Article

# SALT spectroscopic follow-up of the G4Jy sample

Sarah V. White[1,2], Kshitij Thorat[3], Moses Mogotsi[1], Rosalind Skelton[1], Solohery Randriamampandry[1,4], Encarni Romero-Colmenero[1], Precious K. Sejake[3], Francesco Massaro[5,6,7], Abigail García-Pérez[5,8,9], Ana Jiménez-Gallardo[9], Harold Peña-Herazo[10] and Edward N. Taylor[11]

[1]South African Astronomical Observatory (SAAO), Observatory, Cape Town, South Africa, [2]Department of Physics and Electronics, Rhodes University, Makhanda, South Africa, [3]Department of Physics, University of Pretoria, Hatfield, Pretoria, South Africa, [4]A& A, Department of Physics, Faculty of Sciences, University of Antananarivo, Antananarivo, Madagascar, [5]Dipartimento di Fisica, Università degli Studi di Torino, Torino, Italy, [6]Istituto Nazionale di Astrofisica (INAF) – Osservatorio Astrofisico di Torino, Pino Torinese, Italy, [7]Istituto Nazionale di Fisica Nucleare (INFN) – Sezione di Torino, Torino, Italy, [8]Instituto Nacional de Astrofísica, Óptica y Electrónica, Tonantzintla, Puebla, México, [9]European Southern Observatory (ESO), Vitacura, Región Metropolitana, Chile, [10]East Asian Observatory (EAO), Hilo, HI, USA and [11]Centre for Astrophysics and Supercomputing, Swinburne University of Technology, Hawthorn, Melbourne, Australia

### Abstract

The GLEAM 4-Jy (G4Jy) Sample is a thorough compilation of the 'brightest' radio sources in the southern sky (Declination $< 30°$), as measured at 151 MHz ($S_{151\,\mathrm{MHz}} > 4.0$ Jy) with the Murchison Widefield Array (MWA), through the GaLactic and Extragalactic All-sky MWA (GLEAM) Survey. In addition to flux-density measurements, the G4Jy catalogue (https://github.com/svw26/G4Jy.) provides host-galaxy identifications (through careful visual-inspection) and four sets of spectral indices. Despite their brightness in the radio, many of these sources are poorly studied, with the vast majority lacking a spectroscopic redshift in published work. This is crucial for studying the intrinsic properties of the sources, and so we conduct a multi-semester observing campaign on the Southern African Large Telescope (SALT), with optical spectroscopy enabling us to provide new redshifts to the astronomical community. Initial results show that not all of the host galaxies exhibit emission-line spectra in the optical (∼4 500–7 500Å), which illustrates the importance of radio-frequency selection (rather than optical selection) for creating an *unbiased* sample of active galactic nuclei. By combining SALT redshifts with those from the 6-degree Field Galaxy Survey (6dFGS) and the Sloan Digital Sky Survey (SDSS), we calculate radio luminosities and linear sizes for 299 G4Jy sources (which includes one newly-discovered giant radio-galaxy, G4Jy 604). Furthermore, with the highest redshift acquired (so far) being $z \sim 2.2$ from SDSS, we look forward to evolution studies of this complete sample, as well as breaking degeneracies in radio properties with respect to, for example, the galaxy environment.

**Keywords:** galaxies: active; radio continuum: galaxies; techniques: spectroscopic; catalogues

## 1. Introduction

Powerful active-galactic-nuclei (AGN) feature heavily in our understanding of galaxy evolution, with AGN activity thought to promote (e.g. Ishibashi & Fabian 2012; Silk 2013) or suppress (e.g. Croton et al. 2006; Davies et al. 2020; Lammers et al. 2023) star formation in the host galaxy. Meanwhile, radio observations are unaffected by dust obscuration, and so allow such activity to be detected out to higher redshift than is possible at other wavelengths (e.g. Collier et al. 2014; Singh et al. 2014). This includes finding high-redshift (proto-)clusters, by exploiting the tendency of 'radio-loud' AGN to reside in dense environments (Venemans et al. 2007; Wylezalek et al. 2013; Hlavacek-Larrondo et al. 2015). However, when studying the properties of powerful AGN as a function of redshift and/or environment, detailed research is hindered by small-number statistics.

**Corresponding author:** Sarah V. White; Email: sarahwhite.astro@gmail.com

Currently, the revised Third Cambridge Catalogue of Radio Sources (3CRR; Laing, Riley, & Longair 1983) is the most-prominent, low-frequency radio-source sample ($S_{178\,\mathrm{MHz}} > 10.9$ Jy) that is optically complete, but this consists of only 173 sources (all in the northern hemisphere). White et al. (2018, 2020a,b) have created a sample that is over 10 times larger – the GLEAM 4-Jy (G4Jy) Sample – using observations from the Murchison Widefield Array (MWA; Tingay et al. 2013) over the entire southern sky (Dec. $< 30°$). Thanks to these measurements at low radio-frequencies, we can select radio-loud AGN in an orientation-independent way (Barthel 1989). This is because the low-frequency emission of powerful AGN is dominated by the radio lobes, which are not subject to relativistic beaming (Rees 1966). The same cannot be said for the radio core, 'hotspots', and jets that dominate the emission of sources at high radio-frequencies. As a result of this beaming effect, radio sources selected at high frequencies tend to be biased towards AGN that have their jet axis close to the line-of-sight (e.g. Lister 2003). Therefore, with 1 863 of the brightest radio-sources at low frequencies making up the G4Jy Sample, we can test models of powerful AGN more robustly than previously (e.g. Mullin, et al., 2008; Wang & Kaiser 2008; Best & Heckman 2012; Shabala et al. 2020). However, the paucity of existing optical spectroscopy over the southern sky [e.g. through the 6-degree Field Galaxy Survey, 6dFGS; Jones et al. 2009] significantly limits the breadth of science that can be undertaken, hence our follow-up of a large fraction of G4Jy sources (Sejake et al. in preparation) using the Southern African Large Telescope (SALT).

In this paper, our results include some of the very 'brightest' of the G4Jy Sample, as determined through their integrated flux-density at ∼178 MHz. Applying a threshold of 9.0 Jy[a] here allows us to define a subsample that is equivalent to the first revised version of the Third Cambridge catalogue of radio sources (3CR survey; Bennett 1962; Spinrad et al. 1985) of the northern hemisphere. Hence, they are referred to as *the G4Jy-3CRE subset* (Massaro et al. 2023a). To ensure that there is not overlap with the 298 extragalactic sources that comprise the 3CR sample, a criterion is applied with respect to the sky coverage (Dec. $< -5°$), resulting in a list of 264 G4Jy sources. These sources are being studied in the X-ray (e.g. Massaro et al. 2023b) in order to better-understand how radio jets interact with their surroundings, with additional optical spectroscopy being provided via numerous telescopes (e.g. García-Pérez et al. 2024). The intrinsic radio-properties of these G4Jy-3CRE sources will be presented (in more-complete form) in future work.

We also plan to extend X-ray follow-up to the wider G4Jy Sample as a whole. However, to ensure good signal-to-noise ratios we need to have high spectroscopic completeness so that sources are not biased with respect to dust obscuration. This is also necessary for creating appropriate subsamples for observations with the Atacama Large Millimeter/submillimeter Array (ALMA), which would allow us to address questions about the amount of gas that is available for black-hole accretion and star formation. The wide range in optical magnitudes for our sources makes the G4Jy Sample particularly suitable for large-scale follow-up with SALT, offering us a firm foundation for combining with additional spectroscopy in the future.

### 1.1. Paper outline

In the next section (Section 2), we describe the criteria that were applied to the G4Jy Sample in order to generate the target list for our spectroscopic campaign with SALT. The resulting spectra are presented in Appendix B, and the acquired redshifts are summarised in Section 3. Section 3 also discusses the radio luminosities and linear sizes for G4Jy sources, enabled by the collation of 299 redshifts. J2000 co-ordinates and AB magnitudes are used throughout this work, and we use a ΛCDM cosmology, with $H_0 = 70$ km s$^{-1}$ Mpc$^{-1}$, $\Omega_m = 0.3$, $\Omega_\Lambda = 0.7$.

## 2. Target selection and data acquisition

In creating the G4Jy Sample, White et al. (2020a,b) used the GaLactic and Extragalactic All-sky MWA (GLEAM) catalogue (Hurley-Walker et al. 2017) to select 1 863 radio-sources brighter than 4.0 Jy at 151 MHz (with a spatial resolution of ∼ arcmin). They then used higher-resolution images from:

[a]Noting the calibration differences found by Roger, Costain, & Bridle (1973) with respect to the original 3C observations, application of a 9.8-Jy threshold results in 181 G4Jy sources at Dec. $< -5°$ (Massaro et al. 2023a).

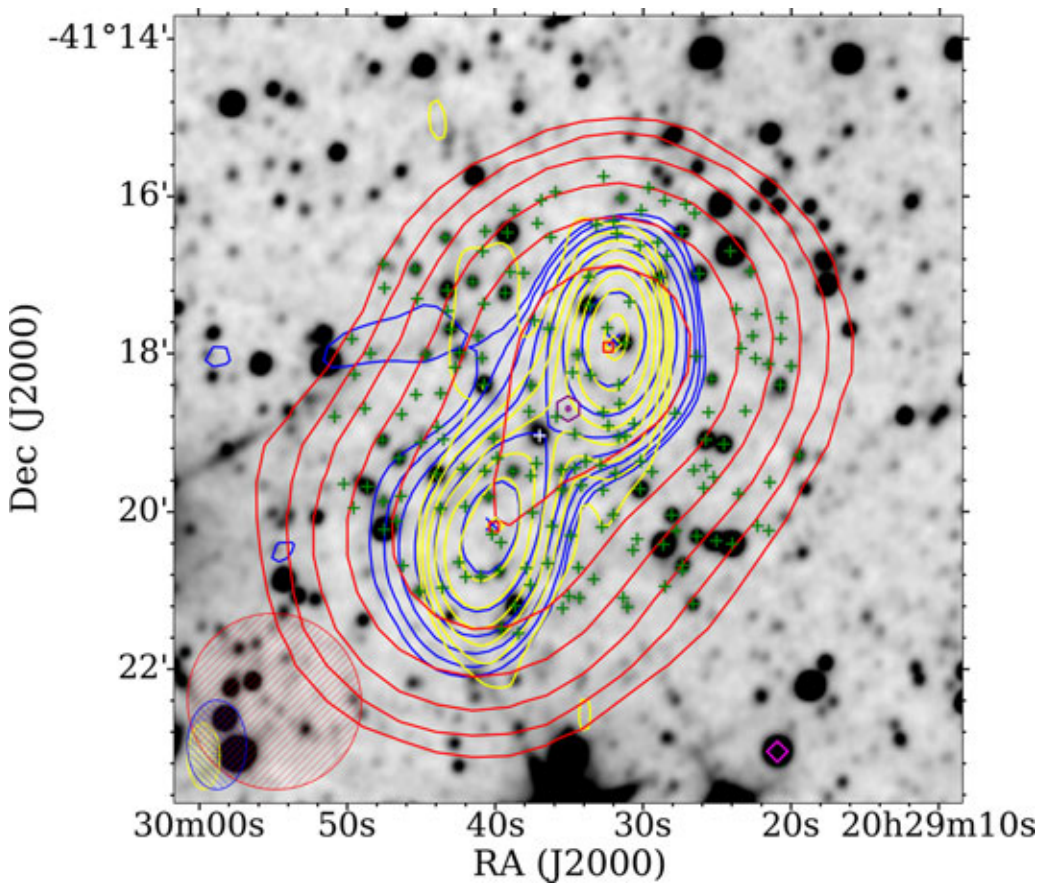

**Figure 1.** An example overlay showing how different sets of radio contours (GLEAM [200 MHz] in red, SUMSS [843 MHz] in blue, and TGSS [150 MHz] in yellow) were used to assess the morphology of a G4Jy source (G4Jy 1628), with the respective beam-sizes of the different radio surveys shown in the bottom left-hand corner. The underlying, inverted-greyscale image is from the W1 band of AllWISE, with green plusses ('+') marking AllWISE catalogue positions within 3 arcmin of the radio-centroid position (purple hexagon). This enabled White et al. (2020a,b) to identify the appropriate host galaxy of the radio emission (white '+'), which was followed by thorough checks against published studies before being included the G4Jy catalogue.

1. the TIFR GMRT Sky Survey (TGSS) first alternative data release (ADR1; Intema et al. 2017), at ∼25 arcsec resolution,
2. the Sydney University Molonglo Sky Survey (SUMSS) catalogue (Mauch et al. 2003; Murphy et al. 2007), at ∼45 arcsec, and
3. the NRAO (National Radio Astronomy Observatory) VLA Sky Survey (NVSS; Condon et al. 1998) at ∼45 arcsec,

to determine the radio morphology of the sources, and help to identify the host galaxies (Fig. 1). The latter was done through careful visual inspection, using W1-band images from AllWISE (Cutri et al. 2013) to *avoid being biased against the most dust-obscured sources*. The result is that 1 606 of the 1 863 sources have identifications in the (original) G4Jy catalogue (White et al. 2020a,b), 1 253 of which are at Declinations accessible by SALT ($-76° <$ Dec. $< 11°$). Considering the 1 606 sources, we remove 136 with spectra from the 6-degree Field Galaxy Survey (6dFGS; Jones et al. 2009), and 104 with Sloan Digital Sky Survey (SDSS) spectra (DR12; Alam et al. 2015). This illustrates that we still require optical spectroscopy for the vast majority of the sample, primarily in order to obtain robust redshifts and (where possible) determine the accretion modes of these radio galaxies.

When compiling *R*-band magnitudes for the sample (Fig. 2) in January 2020 (SALT proposal: 2020-1-MLT-008, PI: White), we cross-matched the AllWISE host-galaxy positions with the following datasets, using a radius of 1 arcsec: SuperCOSMOS (Hambly et al. 2001), the National Optical Astronomy Observatory (NOAO) Source Catalog (NSC) DR1 (Nidever et al. 2018), the Dark Energy Spectroscopic Instrument (DESI) Legacy Imaging Surveys (LS) DR8 (Dey et al. 2019), SDSS DR12 (Alam et al. 2015), SkyMapper DR1.1 (Wolf et al. 2018) and PanSTARRS (Flewelling et al. 2020) photometry newly-extracted using ProFound (Robotham et al. 2018). (Further details are provided in Appendix A.) The scatter in magnitudes from the different datasets was then assessed and used to convert everything onto the SuperCOSMOS magnitude scale. The aim of our SALT follow-up is to get good completeness for the G4Jy Sample down to $R = 20.0$ mag (the optical limit of the SuperCOSMOS data), and doing so gives us a subsample of 706 sources (at $-76^{\circ} <$ Dec. $< 11^{\circ}$, and including the 158 sources with 6dFGS/SDSS spectra in this range). An additional 38 G4Jy sources were added to the SALT target-list, thanks to new host-galaxy identifications provided by Sejake et al. (2023) via 1300-MHz MeerKAT imaging (of $\sim$7-arcsec spatial resolution). For a summary of the number of sources considered, please see Table 1.

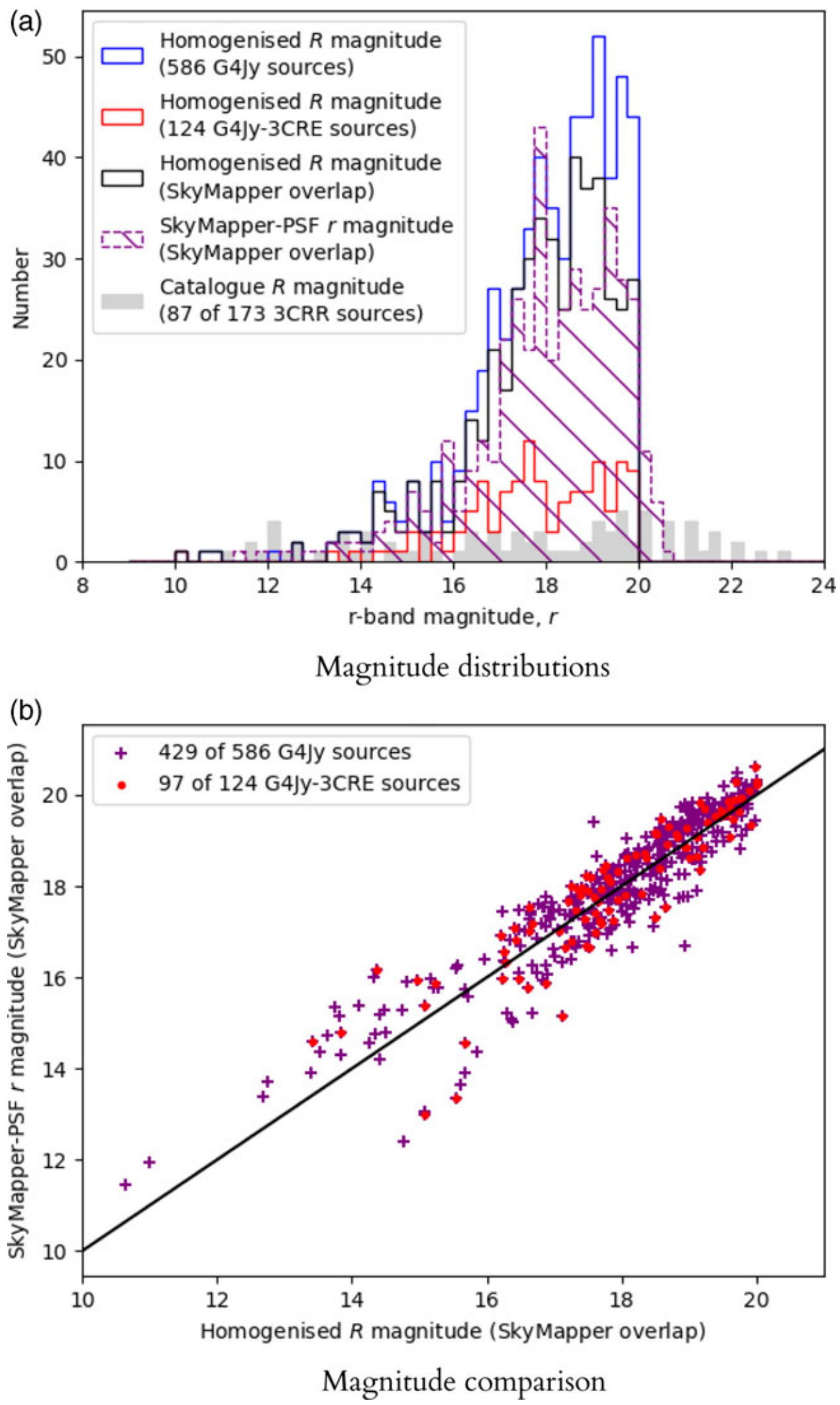

**Figure 2.** (a) Distributions of the $R$-band magnitudes (Section 2) for different subsets of the G4Jy Sample ($-76^{\circ} <$ Dec. $< 11^{\circ}$, restricted to $R \sim 20$), with the magnitude distribution for 3CRR sources (without restriction) added for comparison (grey histogram). 'Homogenised' refers to magnitudes from different surveys being put on the SuperCOSMOS scale (Hambly et al. 2001), and 'SkyMapper overlap' refers to the 429 G4Jy-on-SALT targets that appear in the SkyMapper survey. (b) A comparison of the SkyMapper-PSF $r$-magnitude with the homogenised $R$-band magnitude for the G4Jy targets that appear in the SkyMapper survey (purple '+'). Sources that also belong to the G4Jy-3CRE subset (Massaro et al. 2023a) are indicated by red dots.

**Table 1.** A summary of how the target list of 586 G4Jy sources was derived from the original catalogue of 1 863 sources (White et al. 2020a,b). This summary also accounts for the 299 redshifts presented in this paper. Note that 12 of the 98 host-galaxy positions provided by Sejake et al. (2023) were confirmation of existing identifications in the G4Jy catalogue.

| **Description** | **Tally of redshifts** | **Tally of SALT targets** |
|---|---|---|
| 1863 sources in the G4Jy catalogue | – | – |
| 1606 with AllWISE identifications (IDs) | – | – |
| SALT-accessible sources ($-76^{\circ} <$ Dec. $< 11^{\circ}$) | – | 1 253 |
| Apply the $R$-band cut, aiming for completeness | – | 706 |
| 136 sources in 6dFGS DR3 (Jones et al. 2009) | 136 | 573 |
| 104 sources in SDSS DR12 (Alam et al. 2015) | 240 | 548 |
| 98 IDs via MeerKAT images (Sejake et al. 2023) | – | – |
| 38 of these sources are at $-76^{\circ} <$ Dec. $< 11^{\circ}$ | – | 586 |
| 53 sources in SDSS DR16 (Ahumada et al. 2020) | 293 | – |
| 36 of these overlap with SDSS DR12 | 257 | – |
| 100 SALT spectra reduced as part of monitoring | 299 | 586 |

Targetting $(706 - 158 + 38 =)$ 586 G4Jy sources[b] enables investigation of a subsample that is over 10 times larger than the number of 3CRR sources at $R < 20.0$. This allows us to determine more-robust statistics for comparing with galaxy-evolution simulations, making this an excellent legacy dataset for studying powerful active galaxies as a function of redshift and environment. The all-sky distribution of our sources means that we have relaxed scheduling restraints with respect to Right Ascension and Declination, whilst the broad range of $R$-band magnitudes (Fig. 2) means that we can take advantage of SALT time that is available for different moon conditions; 'Bright' time is suitable for the 316 targets that are at $R < 18.5$ mag, and 'Grey' time is selected for the 270 targets that are at $18.5 \leq R/\text{mag} < 20.0$. Thanks to the sources with brighter magnitudes, we can also accommodate more-difficult seeing conditions (up to 2.5 arcsec).

We observe the sources via the pg0900 grating on the SALT Robert Stobie Spectrograph (RSS), at the standard grating-angle of 15.875°, and with the pc03850 Order-Blocking Filter. This provides the resolution (R = $\sim$660–930) and wavelength coverage ($\sim$4 500 – 7 500Å) for maximising redshift completeness ($z \lesssim 4.5$) and broad-line width measurement. Acquisition involves imaging the field using the SALT Imaging CAMera (SALTICAM, also abbreviated to SCAM), and (typically) identifying an acquisition object that allows fainter targets to enter the longslit via the 'blind-offset' method (see Appendix B). Also, we split the observations into 2 to 3 exposures per target (in order to enable cosmic-ray removal), and perform wavelength calibrations using the RSSMOSPipeline software[c] (Hilton et al. 2018), facilitated by arc-observations of a Xenon lamp. This Python-based pipeline automatically detects sources that are present along the 2-arcsec-width longslit of the RSS, uses data from the intervening regions

[b]These 586 targets include 124 of the 264 G4Jy sources belonging to the G4Jy-3CRE subset (Section 1).

[c]https://github.com/mattyowl/RSSMOSPipeline.

for sky subtraction, and performs stacking of spectra from multiple exposures.

## 3. Results and discussion

The (homogenised, SuperCOSMOS-scale) $R$-band magnitudes that informed our observations are shown in Fig. 2(a). This dataset not only aids redshift determination but also provides further understanding of host-galaxy properties, which are essential for understanding the role of AGN in galaxy evolution. 124 of the 586 G4Jy sources belong to the G4Jy-3CRE subset (Massaro et al. 2023a), and 429 G4Jy sources have a (less than) 1-arcsec cross-match with DR4 of the SkyMapper survey (Onken et al. 2024)[d] We note that the application of an $R = 20.0$ mag cut to the homogenised magnitudes corresponds to an $r \simeq 21.0$ limit in the SkyMapper-PSF magnitudes. A scatter-plot comparison of the magnitudes is shown in Fig. 2(b), and indicates that there is a large degree of scatter, irrespective of source brightness. As such, our multisemester campaign needed to be extended so that underexposed host-galaxies could be reobserved (in order to achieve better signal-to-noise ratios). This is crucial as it facilitates the determination of redshifts for a larger sample of radio sources, which is essential for understanding their evolution and the environments that they inhabit.

### 3.1. SALT spectra

The first batch of 42 G4Jy sources that already have good signal-to-noise ratios for deriving redshifts are shown in figure B1, with the redshifts obtained being summarised in Table 2. Twenty-six of the sources listed belong to the G4Jy-3CRE subset, and so supplement the 42 (entirely-new) redshifts obtained by García-Pérez et al. (2024) and via ongoing observing campaigns. In addition, we note that six of the observations presented here are made possible by the host-galaxy positions determined through MeerKAT imaging (Sejake et al. 2023).

If the target spectrum is dominated by emission lines, then we attempt to fit it with a ‘quasar’ template[e]; if, instead, absorption lines dominate the spectrum, then we employ an ‘early-type galaxy’ template[f] (Yip et al. 2004). In sources that appear to have an even mixture of emission- and absorption-lines, we fit the target spectrum with a ‘galaxy’ template that incorporates (for example) both a strong H$\alpha$ emission-line and the Calcium-II H+K doublet[e]. As a reminder, the aim is to obtain redshifts rather than complete emission-/absorption-line characterisation (for which flux calibration would be needed). The error in the redshift is estimated by applying lower and upper limits to the template, and seeing (by eye) how well-aligned the spectral features remain. Such template-shifting allows us to get an idea of the ‘tolerance’ of the fit, and also encompasses any error in the wavelength calibration (which is usually $< 1$ Å).

Like Mauch & Sadler (2007), we see a mixture of emission-line (‘e’), absorption-line (‘a’), and absorption-/emission-line (‘ae’) spectra, and no correspondence between $r$-band magnitude and the spectroscopic redshift (Table 2). However, we note that the brightness of emission lines allows for the redshift to be more-readily determined than via fitting absorption lines (which are more difficult to differentiate from a noisy continuum), and therefore our first set of results are biased towards the former spectral-type. For the interested reader, Sejake et al. (in prep.) will present the relative fractions for a more-completely-defined subsample of G4Jy sources.

### 3.2. Comments on individual sources

G4Jy 530, G4Jy 590, and G4Jy 1819 overlap with the sources analysed by García-Pérez et al. (2024). Our redshifts are in agreement for G4Jy 530 and G4Jy 1819, but differ for G4Jy 590 (where they estimate the redshift as $z = 0.5384 \pm 0.0027$). Following correspondence with the authors, we provide the corrected spectroscopic-redshift of $z = 0.529 \pm 0.002$, in disagreement with the photometric redshift ($z = 0.58$) provided by Burgess & Hunstead (2006).

The largest redshift error in the present work is for G4Jy 541 ($z = 0.200 \pm 0.007$), on account of the CaII H+K doublet being tentatively detected (towards the edge of the wavelength coverage) in addition to strong [OIII] emission-lines. We note that there are not emission lines in the template that align with the distinctive peaks at ∼6 650 and ∼6 825Å, but caution that the target’s continuum is affected by suboptimal sky-background subtraction.

The ‘double’ radio morphologies of G4Jy 1704 and G4Jy 1705 were first published by White et al. (2020a,b), based on the PhD thesis of Haigh (2001), and show extended emission within the Abell-3785 cluster. They are referred to as ‘the dancing ghosts’ by Norris et al. (2021) and Velović et al. (2023), who present ASKAP and MeerKAT imaging of these radio sources, respectively. Our spectroscopic redshifts are in agreement with the photometric redshifts previously acquired (Bilicki et al. 2014), with G4Jy 1704 at $z = 0.078 \pm 0.001$ and G4Jy 1705 at $z = 0.076 \pm 0.002$. These values were determined via fitting their SALT spectra with an ‘early-type galaxy’ template, as shown in figure B1.

Within the SALT spectra we identify the MgII emission-line for 8 of the 42 sources, these being: G4Jy 672, G4Jy 706, G4Jy 901, G4Jy 909, G4Jy 1511, G4Jy 1665, G4Jy 1698, and G4Jy 1709. This is of special interest for follow-up, as the black-hole mass can be estimated via measuring the linewidth of this line (e.g. McLure & Jarvis 2002).

We recheck how many G4Jy sources have spectra in SDSS (Appendix C), and find that 53 overlap with SDSS DR16 (Ahumada et al. 2020), which includes 36 that have had their SDSS-DR12 entries updated. We retain the remaining ($104 - 36 =$) 68 SDSS-DR12 sources (Alam et al. 2015) that are cross-identified with the G4Jy Sample, and combine these with the 136 redshifts from 6dFGS (Jones et al. 2009). The redshift distributions of these different datasets are shown in Fig. 3.

The highest of these redshifts is $z = 2.17827 \pm 0.00015$ from SDSS DR12, which is corroborated by Hewett & Wild (2010), who measure $z = 2.17920 \pm 0.00055$. The G4Jy overlay for this source, G4Jy 1065 (also known as 4C +11.45), is shown in Fig. 4. The large MWA beam means that the GLEAM contours are affected by confusion with a nearby unrelated source towards the south-west, whilst the TGSS contours show extended radio morphology at 150 MHz. The latter (‘double’ morphology) was first presented by Barthel et al. (1988) in a VLA map at 5 GHz.

### 3.3. Intrinsic radio properties

The combination of SALT, 6dFGS, and SDSS redshifts allows us to calculate intrinsic radio-properties for a sample of 299 G4Jy

[d] Earlier data releases were not of the required optical depth for obtaining a large fraction of crossmatches with the G4Jy Sample. (White et al. 2020a,b).

[e] This template was provided courtesy of P.C. Hewett, and is for a ‘pure quasar’, with no host-galaxy contribution towards the emission. A refined version of the quasar template, and a link to Python code, is now available in Temple, Hewett & Banerji (2021).

[f] https://classic.sdss.org/dr5/algorithms/spectemplates/.

**Table 2.** Spectroscopic redshifts for 42 G4Jy sources, as determined via SALT optical-spectroscopy (Appendix B). Sources belonging to the G4Jy-3CRE subset (Massaro et al. 2023a) and the MeerKAT-2019 subset (Sejake et al. 2023) are indicated with a flag of '1' in the respective columns. The point-spread function (PSF) *r*-band magnitude provided via DR4 of SkyMapper (Onken et al. 2024) is also presented, where available. The night of observation is given in the format of YYYY-MM-DD.

| Source name | G4Jy-3CRE subset | MeerKAT-2019 subset | Host-galaxy R.A. (h:m:s, J2000) | Host-galaxy Dec. (d:m:s, J2000) | SuperCOSMOS *R*-band magnitude | SkyMapper PSF *r*-band magnitude | Night of observation | Spectral type | Spectroscopic redshift |
|---|---|---|---|---|---|---|---|---|---|
| G4Jy 43 | 1 | 0 | 00:23:08.86 | −25:02:29.6 | 19.78 | $19.9458 \pm 0.2633$ | 2023-08-06 | absorption-line | $0.354 \pm 0.001$ |
| G4Jy 45 | 1 | 0 | 00:24:30.15 | −29:28:54.3 | 17.65 | $17.6794 \pm 0.0249$ | 2021-09-08 | emission-line | $0.407 \pm 0.002$ |
| G4Jy 98 | 0 | 0 | 00:54:08.43 | −03:33:55.2 | 17.81 | $17.9300 \pm 0.0774$ | 2021-11-18 | emission-line | $0.211 \pm 0.001$ |
| G4Jy 172 | 0 | 0 | 01:33:33.18 | −44:44:17.7 | 16.56 | $17.0252 \pm 0.1376$ | 2021-07-23 | emission-line | $0.091 \pm 0.003$ |
| G4Jy 192 | 1 | 0 | 01:50:35.94 | −29:31:55.3 | 19.16 | $19.8300 \pm 0.0378$ | 2020-10-24 | emission-line | $0.413 \pm 0.003$ |
| G4Jy 373 | 1 | 0 | 03:38:46.01 | −35:22:52.0 | 17.58 | $17.7767 \pm 0.0676$ | 2020-12-28 | emission-line | $0.113 \pm 0.001$ |
| G4Jy 492 | 1 | 0 | 04:44:37.70 | −28:09:54.3 | 16.67 | $17.1836 \pm 0.0837$ | 2021-08-25 | emission-line | $0.148 \pm 0.001$ |
| G4Jy 530 | 1 | 1 | 05:12:47.41 | −48:24:16.5 | 18.37 | $18.6866 \pm 0.2019$ | 2022-02-13 | emission-line | $0.305 \pm 0.002$ |
| G4Jy 541 | 0 | 0 | 05:23:20.72 | −48:16:30.6 | 17.35 | $17.8014 \pm 0.0524$ | 2024-03-19 | emission-line | $0.200 \pm 0.007$ |
| G4Jy 590 | 1 | 0 | 06:03:12.22 | −34:26:32.6 | 19.97 | $20.6115 \pm 0.2703$ | 2022-08-27 | emission-line | $0.529 \pm 0.002$ |
| G4Jy 604 | 0 | 0 | 06:18:13.03 | −48:44:58.3 | 14.35 | $14.7611 \pm 0.1184$ | 2021-12-19 | absorption-line | $0.049 \pm 0.003$ |
| G4Jy 611 | 1 | 1 | 06:26:20.46 | −53:41:35.1 | 17.21 | $14.6096 \pm 0.1107$ | 2021-11-18 | absorption-line | $0.055 \pm 0.001$ |
| G4Jy 641 | 0 | 1 | 07:05:32.94 | −45:13:08.8 | 17.62 | $18.0829 \pm 0.1151$ | 2023-11-24 | emission-line | $0.128 \pm 0.002$ |
| G4Jy 672 | 1 | 1 | 07:43:31.61 | −67:26:25.5 | 16.18 | $16.3000 \pm 0.0233$ | 2022-11-07 | emission-line | $1.510 \pm 0.001$ |
| G4Jy 692 | 0 | 0 | 08:15:27.81 | −03:08:26.6 | 16.65 | $17.4301 \pm 0.1780$ | 2021-11-18 | absorption-/emission-line | $0.198 \pm 0.001$ |
| G4Jy 706 | 1 | 0 | 08:27:17.41 | −20:26:24.8 | 17.70 | $17.1893 \pm 0.0446$ | 2021-01-25 | emission-line | $0.828 \pm 0.004$ |
| G4Jy 901 | 0 | 1 | 11:11:19.43 | −40:30:51.9 | 19.18 | $17.8111 \pm 0.0852$ | 2024-03-19 | emission-line | $0.725 \pm 0.002$ |
| G4Jy 909 | 0 | 0 | 11:18:26.95 | −46:34:14.9 | 17.52 | $17.0930 \pm 0.0098$ | 2024-03-13 | emission-line | $0.714 \pm 0.001$ |
| G4Jy 957 | 1 | 0 | 11:49:06.68 | −12:04:33.4 | 16.62 | $17.5547 \pm 0.0843$ | 2022-06-02 | absorption-/emission-line | $0.119 \pm 0.002$ |
| G4Jy 965 | 1 | 0 | 11:54:21.78 | −35:05:29.0 | 17.83 | $18.1017 \pm 0.0637$ | 2021-05-25 | emission-line | $0.258 \pm 0.001$ |
| G4Jy 1112 | 0 | 0 | 14:02:31.57 | +02:15:46.5 | 17.16 | $17.3919 \pm 0.0666$ | 2022-04-19 | emission-line | $0.180 \pm 0.003$ |
| G4Jy 1135 | 1 | 0 | 14:16:33.15 | −36:40:53.7 | 16.22 | $16.9136 \pm 0.1238$ | 2021-07-25 | emission-line | $0.075 \pm 0.001$ |
| G4Jy 1157 | 1 | 0 | 14:24:16.47 | −38:26:47.6 | 16.92 | – | 2021-08-22 | emission-line | $0.406 \pm 0.001$ |
| G4Jy 1203 | 1 | 0 | 14:54:28.22 | −36:40:04.7 | 19.37 | $19.5418 \pm 0.1037$ | 2021-06-11 | emission-line | $0.420 \pm 0.001$ |
| G4Jy 1306 | 0 | 0 | 16:06:12.69 | +00:00:27.2 | 15.16 | $15.9605 \pm 0.1408$ | 2022-04-21 | absorption-line | $0.058 \pm 0.001$ |
| G4Jy 1343 | 1 | 0 | 16:31:41.60 | −26:56:51.3 | 18.96 | $18.8668 \pm 0.1255$ | 2023-05-01 | emission-line | $0.168 \pm 0.001$ |
| G4Jy 1364 | 0 | 0 | 16:45:42.39 | +02:11:45.0 | 16.25 | $16.3795 \pm 0.1673$ | 2022-04-21 | absorption-/emission-line | $0.094 \pm 0.001$ |
| G4Jy 1487 | 1 | 0 | 18:30:58.92 | −36:02:30.7 | 16.24 | $15.9748 \pm 0.1575$ | 2021-07-24 | emission-line | $0.078 \pm 0.001$ |
| G4Jy 1511 | 1 | 0 | 18:57:10.80 | −30:19:40.1 | 17.94 | $17.7130 \pm 0.0215$ | 2020-09-15 | emission-line | $1.554 \pm 0.003$ |
| G4Jy 1518 | 1 | 1 | 19:15:48.68 | −26:52:57.4 | 18.94 | $18.3254 \pm 0.1774$ | 2024-04-04 | absorption-/emission-line | $0.231 \pm 0.003$ |

**Table 2.** Continued.

| Source name | G4Jy-3CRE subset | MeerKAT-2019 subset | Host-galaxy R.A. (h:m:s, J2000) | Host-galaxy Dec. (d:m:s, J2000) | SuperCOSMOS *R*-band magnitude | SkyMapper PSF *r*-band magnitude | Night of observation | Spectral type | Spectroscopic redshift |
|---|---|---|---|---|---|---|---|---|---|
| G4Jy 1555 | 1 | 0 | 19:33:25.00 | −39:40:20.7 | 15.09 | $15.3791 \pm 0.1508$ | 2021-06-20 | absorption-line | $0.074 \pm 0.001$ |
| G4Jy 1565 | 1 | 0 | 19:41:15.02 | −15:24:31.2 | 19.91 | $19.3288 \pm 0.0846$ | 2020-10-14 | emission-line | $0.454 \pm 0.001$ |
| G4Jy 1581 | 0 | 0 | 19:52:15.79 | +02:30:24.1 | 15.68 | $15.7368 \pm 0.1094$ | 2021-08-29 | absorption-/emission-line | $0.059 \pm 0.001$ |
| G4Jy 1660 | 0 | 0 | 20:52:02.35 | −57:04:07.5 | 12.01 | – | 2022-05-14 | absorption-/emission-line | $0.012 \pm 0.001$ |
| G4Jy 1665 | 0 | 0 | 20:56:16.36 | −47:14:47.6 | 18.48 | $17.1078 \pm 0.0525$ | 2021-08-04 | emission-line | $1.488 \pm 0.003$ |
| G4Jy 1698 | 1 | 0 | 21:31:01.47 | −20:36:56.1 | 19.54 | $19.5835 \pm 0.0954$ | 2020-10-12 | emission-line | $1.630 \pm 0.002$ |
| G4Jy 1704 | 0 | 0 | 21:34:06.70 | −53:34:18.7 | 14.49 | $15.2795 \pm 0.1473$ | 2022-05-14 | absorption-line | $0.078 \pm 0.001$ |
| G4Jy 1705 | 0 | 0 | 21:34:17.69 | −53:38:11.1 | 14.28 | – | 2021-08-29 | absorption-line | $0.076 \pm 0.002$ |
| G4Jy 1709 | 1 | 0 | 21:37:50.00 | −20:42:31.7 | 19.61 | $19.7336 \pm 0.0280$ | 2022-05-24 | emission-line | $0.638 \pm 0.004$ |
| G4Jy 1781 | 1 | 0 | 22:29:18.63 | −40:51:32.8 | 19.11 | $19.0730 \pm 0.0298$ | 2023-11-10 | emission-line | $0.448 \pm 0.002$ |
| G4Jy 1795 | 1 | 0 | 22:53:03.11 | −40:57:46.7 | 18.49 | $19.1680 \pm 0.1630$ | 2020-12-02 | emission-line | $0.307 \pm 0.003$ |
| G4Jy 1819 | 1 | 0 | 23:19:56.26 | −27:28:07.4 | 16.68 | – | 2022-11-07 | emission-line | $0.174 \pm 0.001$ |

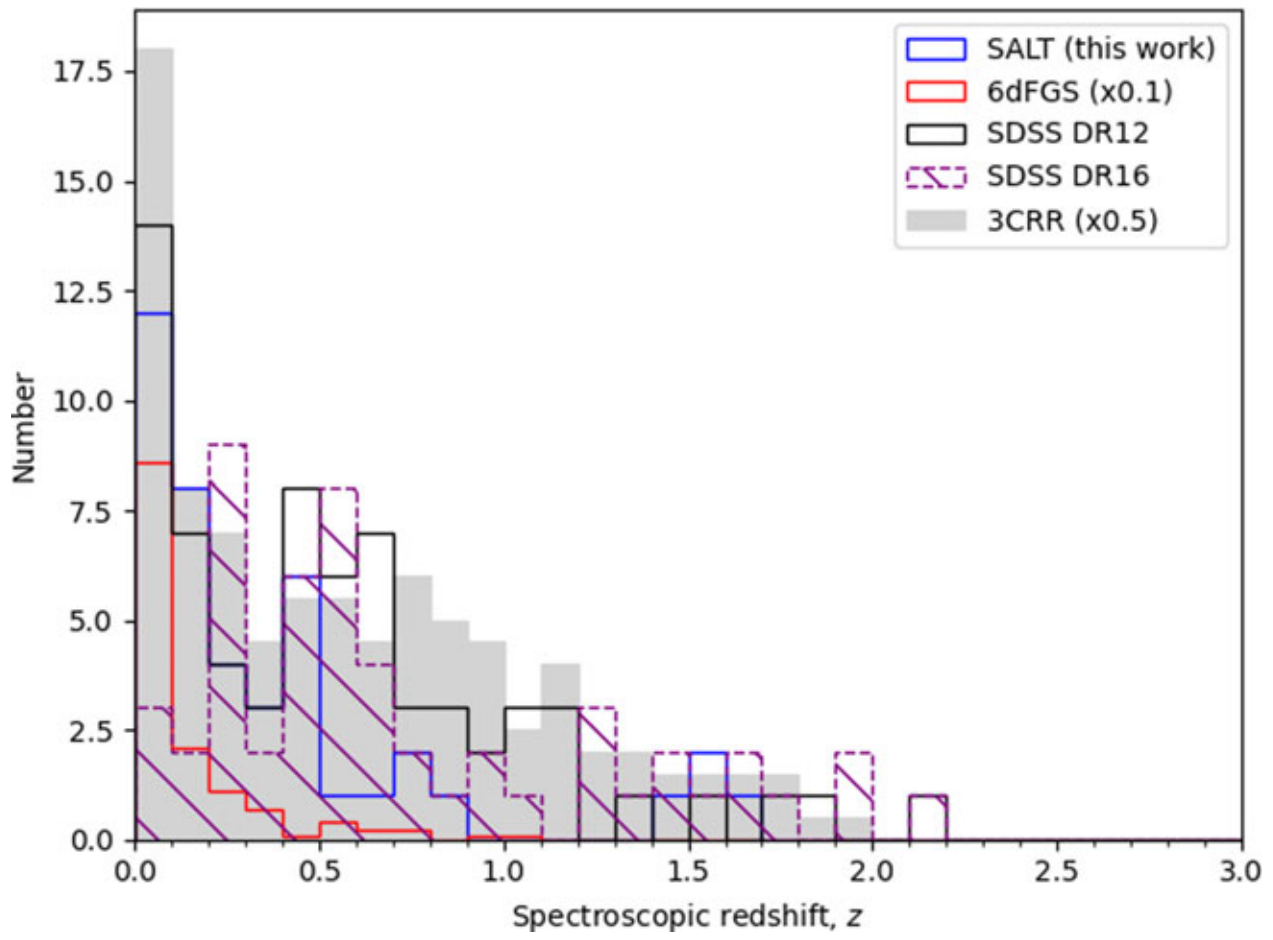

**Figure 3.** Distributions of the redshifts (Section 3.2) for different subsets of the G4Jy Sample (with no restrictions based on Declination). The redshift distribution for 3CRR sources (Laing et al. 1983) is added for comparison (grey histogram, scaled by 0.5), and the 6dFGS distribution has been scaled by 0.1.

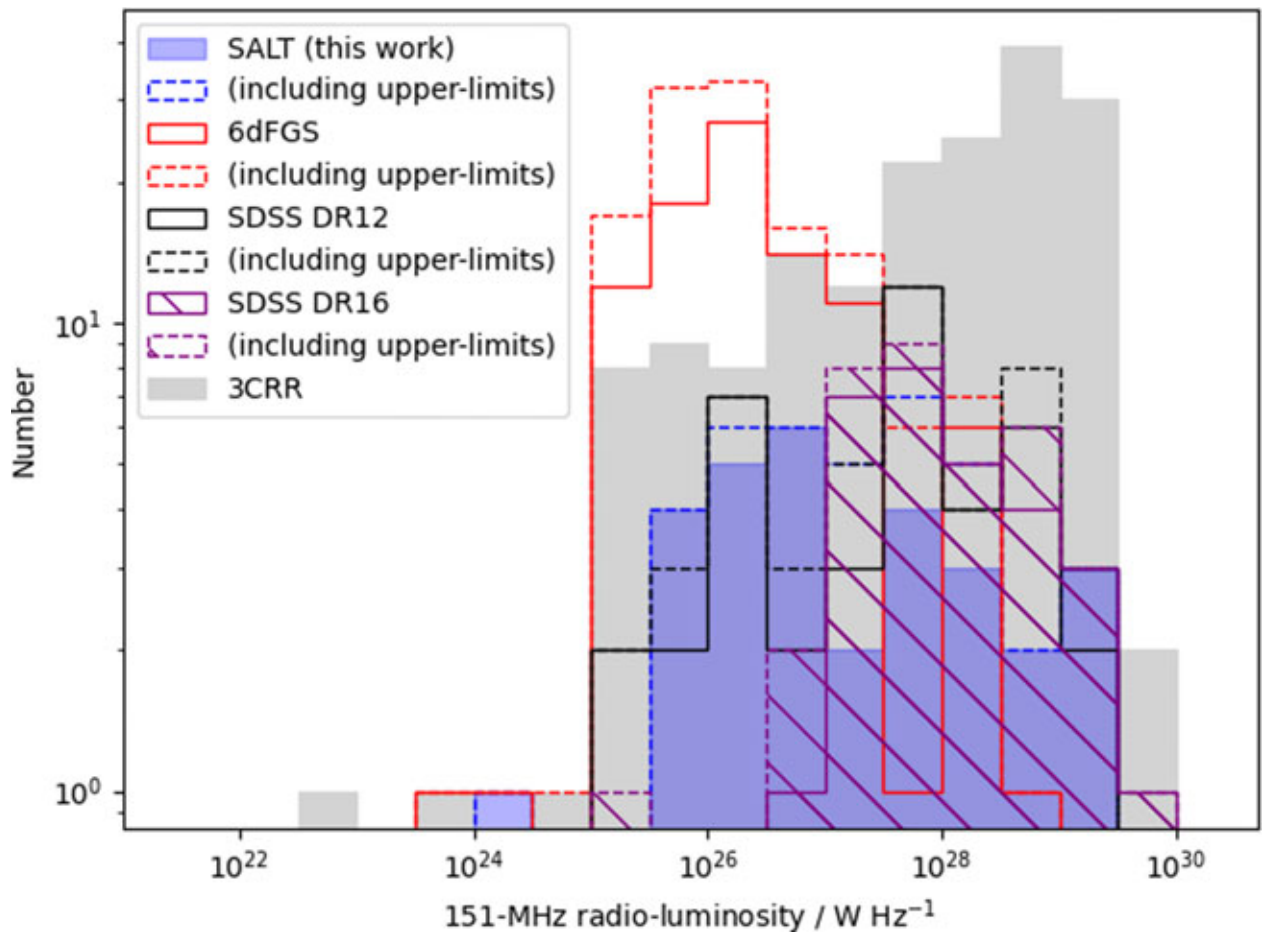

**Figure 5.** Distributions of the 151-MHz radio-luminosities (Section 3.3) for different subsets of the G4Jy Sample (with no restrictions based on Declination). The luminosity distribution for 3CRR sources (Laing et al. 1983) is added for comparison (grey histogram).

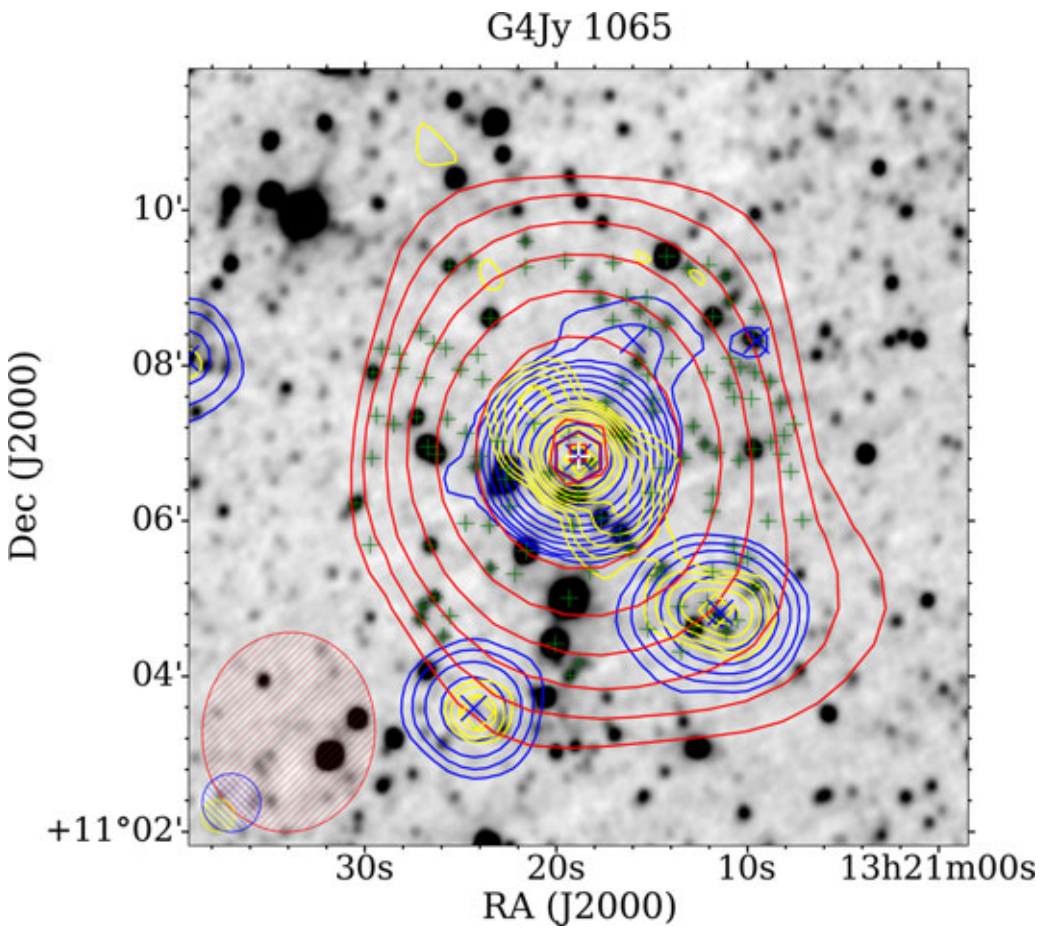

**Figure 4.** Radio contours (GLEAM [200 MHz] in red, NVSS [1 400 MHz] in blue, and TGSS [150 MHz] in yellow) for G4Jy 1065, with the respective beam-sizes of the different radio surveys shown in the bottom left-hand corner. The inverted-greyscale image is from the W1 band of AllWISE, with green plusses ('+') marking AllWISE catalogue positions within 3 arcmin of the radio-centroid position (purple hexagon). The host galaxy of the radio emission is indicated by a white '+', in close alignment with the radio positions from the different radio surveys (a red square, a blue '×', and a yellow diamond, respectively).

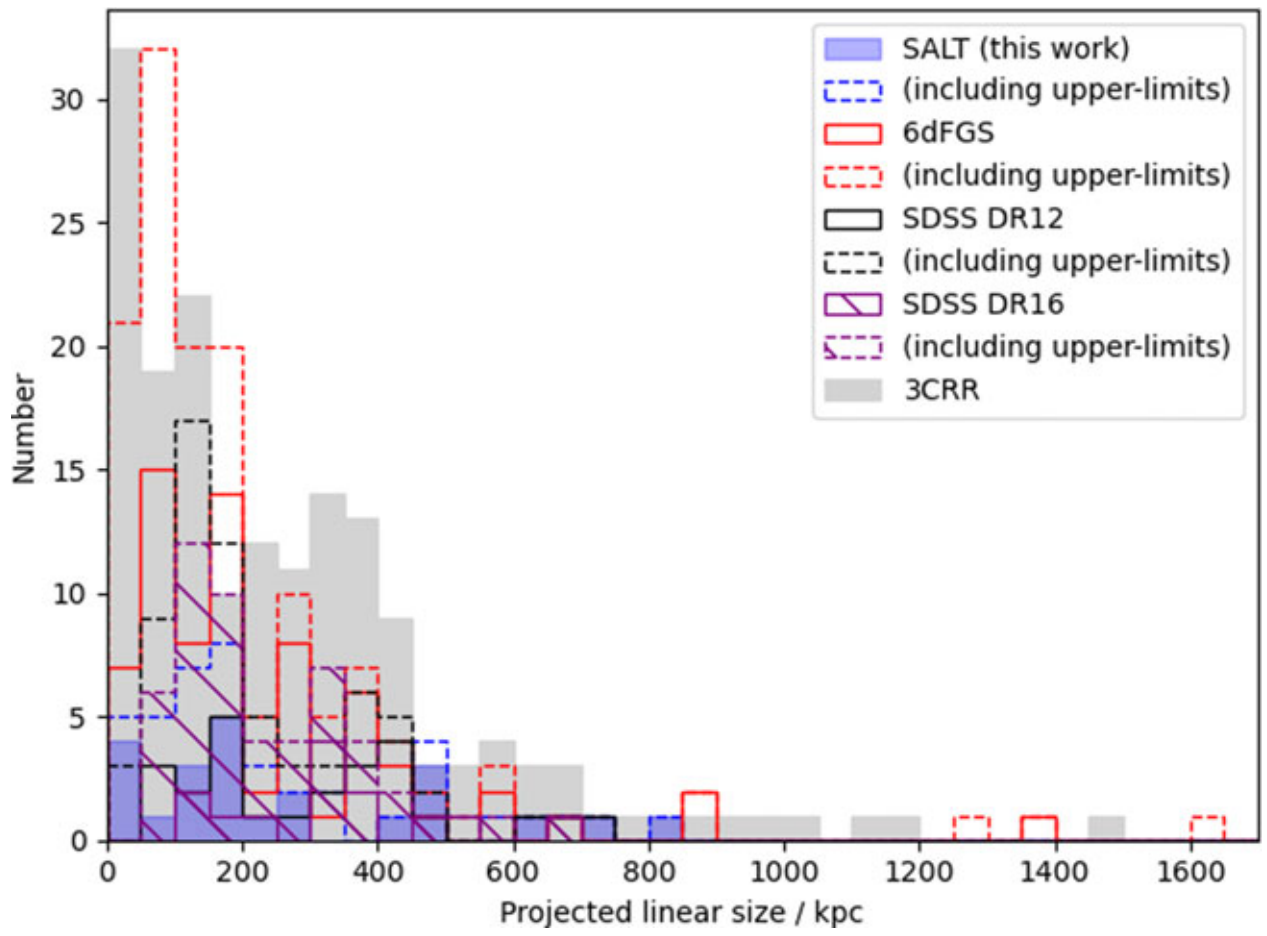

**Figure 6.** Distributions of the (projected) linear sizes (Section 3.3) for different subsets of the G4Jy Sample (with no restrictions based on Declination). The size distribution for 3CRR sources (Laing et al. 1983) is added for comparison (grey histogram), with three of these sources (NGC 6251, 3C 326 = G4Jy 1282, and 3C 236) having linear sizes that are beyond the plot range (i.e. 1 900–4 530 kpc).

sources. For this work, we focus on radio luminosities (based upon measurements at 151 MHz) and projected linear-sizes (based upon angular sizes determined through NVSS and SUMSS catalogue positions, at 1.4 GHz and 843 MHz, respectively; see section 6.3.1 by White et al. 2020a). These values are provided in Table C1 (Appendix C). We note that 49 sources do not have spectral indices because the G4Jy flux-densities (20 measurements across 72–231 MHz, provided in the G4Jy catalogue) are not well-fit by a power-law function ($S \propto \nu^{\alpha}$, where $\alpha$ is the spectral index within the GLEAM band). As a result, we do not calculate their radio luminosities for Fig. 5. This is because these luminosities are K-corrected, *assuming* a power-law description of the radio emission. Future work will present the broadband spectra of these sources, which show significant spectral curvature in the radio (White et al., in preparation).

In order to compare the G4Jy Sample with 3CRR (Laing et al. 1983), we scale the 178-MHz radio-luminosities to 151-MHz radio-luminosities for the 3CRR sources (with the spectral index, $\alpha$, taken from the 3CRR catalogue). We remind the reader that the G4Jy (sub-)sample is not spectroscopically-complete yet, but it is encouraging that new redshifts from SALT are already enabling us to probe a wide parameter space in luminosity (Fig. 5). The five G4Jy sources with the highest luminosities, above $1.4 \times 10^{29}$ W Hz$^{-1}$, are: G4Jy 1511 (SALT, $z = 1.554 \pm 0.003$), G4Jy 1698 (SALT, $z = 1.630 \pm 0.002$), G4Jy 845 (re-fitted $z = 1.706 \pm 0.001$; see figure B3), G4Jy 1065 (SDSS DR12, $z = 2.17827 \pm 0.00015$), and G4Jy 682 (SDSS DR16, $z = 1.96818 \pm 0.00016$). Joining the most-powerful ($\gtrsim 10^{29}$ W Hz$^{-1}$) radio-galaxies in the Universe (e.g. Laing et al. 1983; Saxena et al. 2018; Capetti & Balmaverde 2024), we anticipate that the G4Jy Sample will allow more-robust comparisons with cosmological and galaxy-evolution simulations, for a range of redshifts and environments.

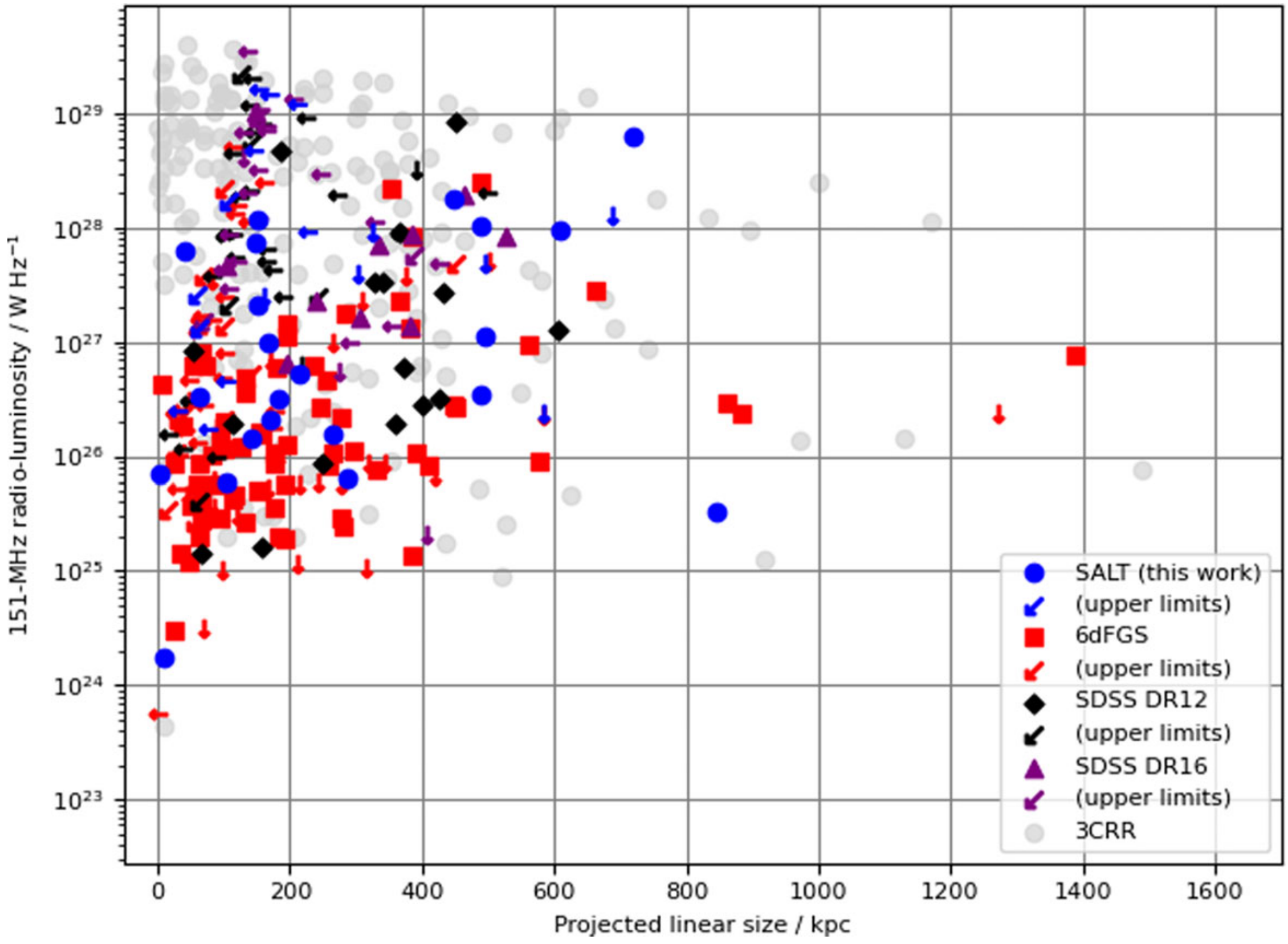

**Figure 7.** The distribution of 151-MHz radio-luminosities against the (projected) linear sizes of the G4Jy Sample (section 3.3), with no restrictions based on Declination. Each set of 'upper limits' is with respect to the linear-size value and/or the radio luminosity. Upper limits in the linear size are represented by horizontal arrows, and are the result of the angular size of the source being an upper limit (due to the resolution of SUMSS/NVSS imaging). Vertical arrows represent upper limits in the radio luminosity, which are a consequence of the 151-MHz flux-density being affected by blended emission from unrelated radio sources. (The affected G4Jy sources are demarcated via a 'confusion flag' of '1'.) Meanwhile, diagonal arrows are used when both the linear size and the radio luminosity are upper limits. The distribution for 3CRR sources (Laing et al. 1983) is added for comparison (grey circles), with three of these sources (NGC 6251, 3C 326 = G4Jy 1282, and 3C 236) having linear sizes that are beyond the plot range (at 1 900–4 530 kpc). In addition, 49 G4Jy sources do not appear in this figure on account of their lack of a spectral-index fit (White et al. 2020a). (This is required for appropriate K-correction of the radio luminosity.)

Connected to the gas-density distribution of the environment is the linear extent of the radio emission associated with a particular host galaxy. Of course, the 2-dimensional imaging of radio sources means that we can only calculate the *projected* linear size, when going from the observed frame to the intrinsic frame. As such, all linear sizes are (in effect) lower limits, but for this work we treat these sizes as absolute values (see Fig. 6). We also present the linear-size distributions when upper limits are included (dashed-line histograms in Fig. 6). These correspond to G4Jy sources that are either unresolved in NVSS/SUMSS imaging or have radio emission at 151 MHz that is blended with an unrelated radio-source. (The former are indicated via an inequality sign, '<', associated with the angular size in the G4Jy catalogue, and the latter are assigned a 'Confusion flag' of '1'; Table C1.).

Again, the linear sizes of 3CRR sources are added for comparison, with these values being obtained from the online 3CRR database[g]. However, they are derived from radio maps at a range of frequencies (typically 1.4–8 GHz) and so the linear sizes will be connected to the different spatial resolutions (and sensitivity to different angular-scales) that are involved. Our linear sizes (for the GJ4y Sample) are more-consistent in that they are from a homogeneous dataset; NVSS or SUMSS, both of which are at 45-arcsec resolution. Either way, another caveat to add, regarding both the G4Jy Sample and the 3CRR sample, is that the linear-size values will be lower limits for FRI ('centre-brightened') sources and better estimates for FRII ('edge-brightened') sources (Fanaroff & Riley 1974). The scope of this work can be extended to consider FRI and FRII sources separately, once a survey of the entire G4Jy Sample is completed at $\lesssim$10 arcsec resolution (allowing such a morphology classification to be performed).

Like White et al. (2020b), we follow Willis & Strom (1978) in defining a giant radio galaxy (GRG) as having a projected, linear size of $> 1$ Mpc. Amongst the 299 G4Jy sources analysed in this work are three GRGs: G4Jy 1079 (1.271 Mpc), G4Jy 133 (1.388 Mpc), and G4Jy 517 (1.620 Mpc). However, the existence of 6dFGS spectra (Jones et al. 2009) means that these are already-known GRGs, as noted in section 4.8 of White et al. (2020b). If we instead relax the size threshold (for defining a GRG) to 0.7 Mpc

[g] https://3crr.extragalactic.info/cgi/database?1=on&2=on&3=on&4=on&8=on&7=on&5=on&6=on&12=on&13=on.

(e.g. Malarecki et al. 2013; Oei et al. 2023), then we can identify G4Jy 1665 (SALT, $z = 1.488 \pm 0.003$), G4Jy 1173 (SDSS DR12, $z = 0.05509 \pm 0.00001$), G4Jy 604 (SALT, $z = 0.049 \pm 0.003$), G4Jy 619 (6dFGS, $z = 0.05501$), and G4Jy 1741 (6dFGS, $z = 0.14519$) as additional sources belonging to this category of radio galaxy.

Of these, G4Jy 604 is a *new* GRG, with a linear size of 0.846 Mpc. Its absorption-line optical spectrum (Figure B1) may explain why, despite its 6-Jy brightness at 151 MHz, it has not been previously classified as a GRG. Meanwhile, G4Jy 1173 is noted by Oei et al. (2023) as a GRG in the LOw-Frequency ARray (LOFAR) Two-metre Sky Survey (LoTSS), and G4Jy 1741 is recorded in the Rapid ASKAP Continuum Survey (RACS) by Andernach et al. (2021). As for G4Jy 619 (aka PKS B0634−205), its giant size has been known since the work of Danziger et al. (1978).

For further interest, in Fig. 7 we present the 151-MHz radio-luminosity against the linear size. It is thought that extended radio-galaxies follow evolutionary tracks across this plane, in what is effectively a power–size ($P$–$D$) diagram (Baldwin 1982), but this is complicated by various factors, such as: (i) the impact of the environment on the radio morphology (e.g. Vardoulaki et al. 2024), (ii) the strength of magnetic fields (e.g. Miley 1980; Jamrozy et al. 2004) in and around the radio galaxy (through its connection to the radio luminosity), (iii) the impact of viewing angle on the observed brightness of the radio emission (although this is ameliorated by the *low-frequency* selection for the G4Jy Sample; Barthel 1989), (iv) assumptions made about the duty cycle of the AGN (e.g. Turner 2018), and (v) how quickly hotspots advance through the intergalactic medium (Alexander & Leahy 1987). We refer the reader to figure 8 of Hardcastle et al. (2019) and the discussion therein for further details. It is expected that as we gather more redshifts for the G4Jy Sample, we will populate a greater area of the $P$–$D$ diagram, with follow-up observations (such as those studying X-rays and polarisation) allowing us to break the degeneracy of some of the aforementioned factors.

For now, we note that the G4Jy Sample shows a similar spread in radio luminosity and linear size as the 3CRR sample (Fig. 7). The lack of a G4Jy source larger than 1 620 kpc (Fig. 6) may be a combined effect of current incompleteness of the spectroscopic follow-up, and the relative shallowness of existing optical data (resulting in a bias towards the small volume of the local Universe). We also emphasise that the linear sizes are limited by the 45-arcsec resolution of the SUMSS/NVSS imaging. Combined with the Malmquist bias towards higher radio-luminosities at higher redshifts, this could explain the clustering (and positive trend) of linear-size upper-limits at $L_{151\,\mathrm{MHz}} \gtrsim 10^{27}$ W Hz$^{-1}$. This is because the upper limit of the linear size will become more discrepant with increasing redshift for unresolved sources (in a given radio survey). We conclude with a note that this is being addressed via improved angular sizes for the full sample (White et al., in preparation).

## 4. Summary

Our ongoing observing campaign with SALT is to obtain optical spectroscopy (PI: White) for a complete sample (SuperCOSMOS $R2 \lesssim 20.0$) of powerful radio-galaxies from the G4Jy Sample (White et al. 2020a,b), which are selected at low radio-frequencies and distributed over the entire southern sky. This enables accurate redshift measurements, with new spectra presented in this paper (for 42 G4Jy sources; Fig. B1) and as part of a larger data release (Sejake et al., in preparation). We also re-fit the spectra for five G4Jy sources that appear in SDSS (Alam et al. 2015; Ahumada et al. 2020), and present the corrected redshifts in Appendix C.

For an initial analysis of 299 G4Jy sources, we combine SALT redshifts with those from 6dFGS (Jones et al. 2009) and SDSS, and find that the sample spans a wide range in radio luminosities ($L_{151\,\mathrm{MHz}} = 5.7 \times 10^{23}$–$3.5 \times 10^{29}$ W Hz$^{-1}$) and linear sizes (1–1 620 kpc). We expect that the lower flux-density threshold for the G4Jy Sample ($S_{151\,\mathrm{MHz}} > 4.0$ Jy), compared to the famous 3CRR sample ($S_{178\,\mathrm{MHz}} > 10.9$ Jy), will allow us to better-populate the radio-power–physical-size ($P-D$ diagram; Fig. 7) and conduct more-detailed investigations of how radio-loud AGN evolve, over a larger fraction of the Universe's history.[h]

The redshift information collated from multiple sources will also enable us to address several key questions on the interaction of AGN with their environment, without the orientation bias that affects both AGN samples selected at high radio-frequencies (due to relativistic beaming) and those selected in the optical (caused by dust obscuration). Furthermore, the spectroscopy will form a valuable *legacy multiwavelength dataset* for future detailed active-galaxy studies. This is because it addresses a critical gap in prior work on powerful AGN in the southern sky, aiding research with facilities like the Square Kilometre Array (SKA) and its precursor telescopes.

**Acknowledgements.** The observations reported in this paper were obtained with the SALT, under program 2020-1-MLT-008 (PI: White). We thank Christian Hettlage for his expertise with SALT schedule-blocks; Lucia Marchetti, Mattia Vaccari, and the SALT Team for helpful comments on the proposal; Christopher White for help with the SALT Finder Charts; and Dan Smith for support. We also thank the anonymous reader for their comments, which improved the scope of the paper, and both Martin Hardcastle and the anonymous referee, regarding the luminosities.

The national facility capability for SkyMapper has been funded through ARC LIEF grant LE130100104 from the Australian Research Council, awarded to the University of Sydney, the Australian National University, Swinburne University of Technology, the University of Queensland, the University of Western Australia, the University of Melbourne, Curtin University of Technology, Monash University and the Australian Astronomical Observatory. SkyMapper is owned and operated by The Australian National University's Research School of Astronomy and Astrophysics. The survey data were processed and provided by the SkyMapper Team at ANU. The SkyMapper node of the All-Sky Virtual Observatory (ASVO) is hosted at the National Computational Infrastructure (NCI). Development and support of the SkyMapper node of the ASVO has been funded in part by Astronomy Australia Limited (AAL) and the Australian Government through the Commonwealth's Education Investment Fund (EIF) and National Collaborative Research Infrastructure Strategy (NCRIS), particularly the National eResearch Collaboration Tools and Resources (NeCTAR) and the Australian National Data Service Projects (ANDS). Funding for the Sloan Digital Sky Survey IV has been provided by the Alfred P. Sloan Foundation, the U.S. Department of Energy Office of Science, and the Participating Institutions. SDSS-IV acknowledges support and resources from the Center for High Performance Computing at the University of Utah. The SDSS website is . SDSS-IV is managed by the Astrophysical Research Consortium for the Participating Institutions of the SDSS Collaboration including the Brazilian Participation Group, the Carnegie Institution for Science, Carnegie Mellon University, Center for Astrophysics | Harvard & Smithsonian, the Chilean Participation Group, the French Participation Group, Instituto de Astrofísica de Canarias, The Johns Hopkins University, Kavli Institute for the Physics and Mathematics of the Universe (IPMU)/University of Tokyo, the Korean Participation Group, Lawrence Berkeley National Laboratory, Leibniz Institut für Astrophysik Potsdam (AIP), Max-Planck-Institut für Astronomie (MPIA Heidelberg), Max-Planck-Institut für Astrophysik (MPA Garching), Max-Planck-Institut für Extraterrestrische Physik (MPE), National Astronomical Observatories of China, New Mexico State University, New York University, University of Notre Dame, Observatário Nacional/MCTI, The Ohio State University, Pennsylvania State University, Shanghai Astronomical Observatory, United Kingdom Participation Group, Universidad Nacional Autónoma de México, University of Arizona, University of Colorado Boulder, University of Oxford, University of Portsmouth, University of Utah, University of Virginia, University of Washington, University of Wisconsin, Vanderbilt University, and Yale University.

[h]SDSS DR12 (Alam et al. 2015) provides the highest redshift for the G4Jy Sample *so far* (geddit?), at $z = 2.17827 \pm 0.00015$ (for G4Jy 1 065), whilst 3C 9 has the highest redshift in the 3CRR sample (Laing et al. 1983), at $z = 2.012$. See Bridle et al. (1994) for a deep, 5-GHz VLA map of the latter.

**Data availability statement.** The DOI for the SkyMapper DR4 release is 10.25914/5M47-S621, and the SALT spectra are available through the Zenodo repository for the G4Jy Sample: https://zenodo.org/communities/g4jy/.

**Funding statement.** This work is based on the research supported in part by the National Research Foundation of South Africa (Grant Number 151060). The financial assistance of the South African Radio Astronomy Observatory (SARAO) towards this research is also hereby acknowledged.

**Competing interests.** None.

**Ethical standards.** The research meets all ethical guidelines, including adherence to the legal requirements of the study country.

## Appendix A. Collating *R*-band magnitudes

We collated magnitudes for the (cross-identified) G4Jy sources, as observed in the following filters by various surveys:

• the $R2$ filter for SuperCOSMOS[i] (Hambly et al. 2001),

• the $r$ filter for the National Optical Astronomy Observatory (NOAO) Source Catalog (NSC)[j] DR1 (Nidever et al. 2018),

• the $r$ filter for the Dark Energy Spectroscopic Instrument (DESI) Legacy Imaging Surveys (LS)[k] DR8 (Dey et al. 2019),

• the $r$ filter for SDSS DR12 (Alam et al. 2015),

• the Petrosian $r$ filter[l] (Bessell et al. 2011) for SkyMapper (SM) DR1.1 (Wolf et al. 2018),

• the $r$ filter[m] (Tonry et al. 2012) for PanSTARRS (PS; Flewelling et al. 2020),

• and Visual Survey Telescope $r$-band photometry (Kuijken et al. 2015), newly extracted using ProFound (PF; Robotham et al. 2018).

The greatest coverage was provided by SuperCOSMOS (i.e. ∼64% of the 586 G4Jy targets), and so this magnitude scale was chosen to be the 'anchor' to which the other magnitudes were 'homogenised'. This was done by applying the following $y = mx + c$ equations (based upon linear-regression analysis), as appropriate:

$$r^{R2}_{\rm NSC} = (r_{\rm NSC} - c^{R2}_{\rm NSC})/m^{R2}_{\rm NSC} = (r_{\rm NSC} + 0.493)/1.022 \qquad \text{(A1)}$$

$$r^{R2}_{\rm LS} = (r_{\rm LS} - c^{R2}_{\rm LS})/m^{R2}_{\rm LS} = (r_{\rm LS} + 0.729)/1.029 \qquad \text{(A2)}$$

$$r^{R2}_{\rm SDSS} = (r_{\rm SDSS} - c^{R2}_{\rm SDSS})/m^{R2}_{\rm SDSS} = (r_{\rm SDSS} + 1.349)/1.067 \qquad \text{(A3)}$$

$$r^{R2}_{\rm SM} = (r_{\rm SM} - c^{R2}_{\rm SM})/m^{R2}_{\rm SM} = (r_{\rm SM} - 2.600)/0.845 \qquad \text{(A4)}$$

$$r^{R2}_{\rm PS} = (r_{\rm PS} - c^{R2}_{\rm PS})/m^{R2}_{\rm PS} = (r_{\rm PS} - 2.908)/0.866 \qquad \text{(A5)}$$

$$r^{R2}_{\rm PF} = (r_{\rm PF} - c^{R2}_{\rm PF})/m^{R2}_{\rm PF} = (r_{\rm PF} + 0.254)/1.001 \qquad \text{(A6)}$$

## Appendix B. SALT spectra and Finder Charts

Within figure B1, we present optical spectra for 42 G4Jy sources (White et al. 2020a,b), including those that belong to the G4Jy-3CRE subset (Massaro et al. 2023a) and optical spectra for 'the dancing ghosts', G4Jy 1704 and G4Jy 1705.

The corresponding Finder Charts are shown in figure B2, for reference, by way of corroborating the host-galaxy position used for optical follow-up. Note that, for G4Jy 1511, Massaro et al. (2023a) suggest that the AllWISE identification for the host galaxy (White et al. 2020a,b) is 'confused' with a (much fainter) optical source nearby. For our observation, SALT is pointed at the optical source that coincides with the AllWISE position (which has ObjectID = 229802842 within SkyMapper DR4[n]).

Additional spectra will be provided to the community as part of a SALT data release that focuses on G4Jy sources at $-40^\circ < \text{Dec.} < -10^\circ$ (Sejake et al. in preparation).

[i] http://ssa.roe.ac.uk/dboverview.html.
[j] https://datalab.noirlab.edu/nscdr1/index.php.
[k] https://www.legacysurvey.org.
[l] https://datalab.noirlab.edu/skymapper.php.
[m] https://outerspace.stsci.edu/display/PANSTARRS/PS1+Filter+properties#PS1Filterproperties-Filterdescriptions.
[n] https://skymapper.anu.edu.au/object-viewer/dr4/229802842/.

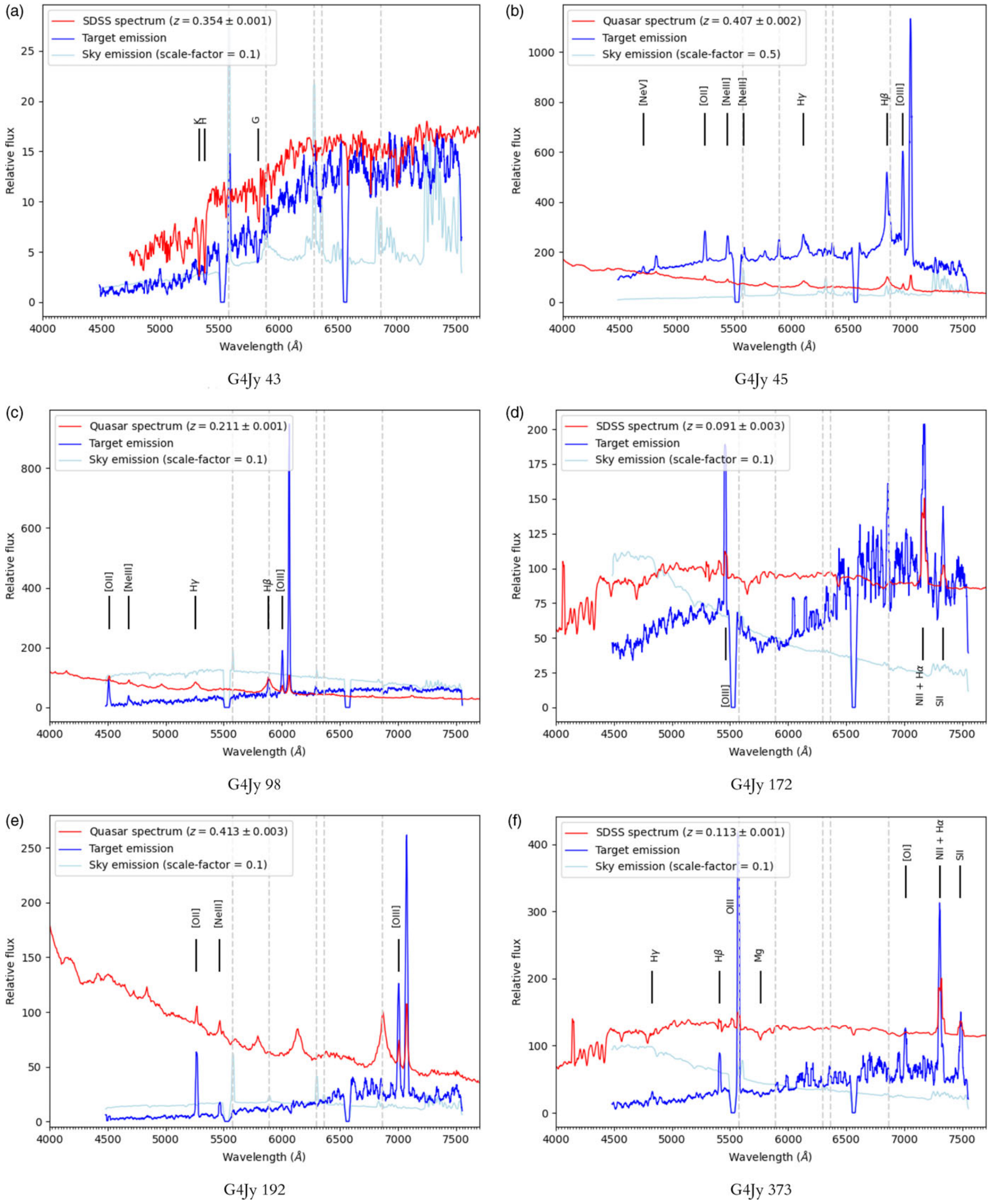

**Figure B1.** SALT spectra (blue lines) of G4Jy sources (Section 3 and Appendix C). CCD chip-gaps are indicated by the blue line dropping to zero relative-flux, whilst the sky-emission spectra are represented by lighter-blue lines. (The latter is scaled to aid comparison with the target emission, and the scale factor that has been applied is noted in each legend.) The dashed, grey, vertical lines indicate the sky-emission that is used to assess the accuracy of the wavelength calibration, and the target spectrum is fitted with the appropriate template spectrum (red line; Section 3).

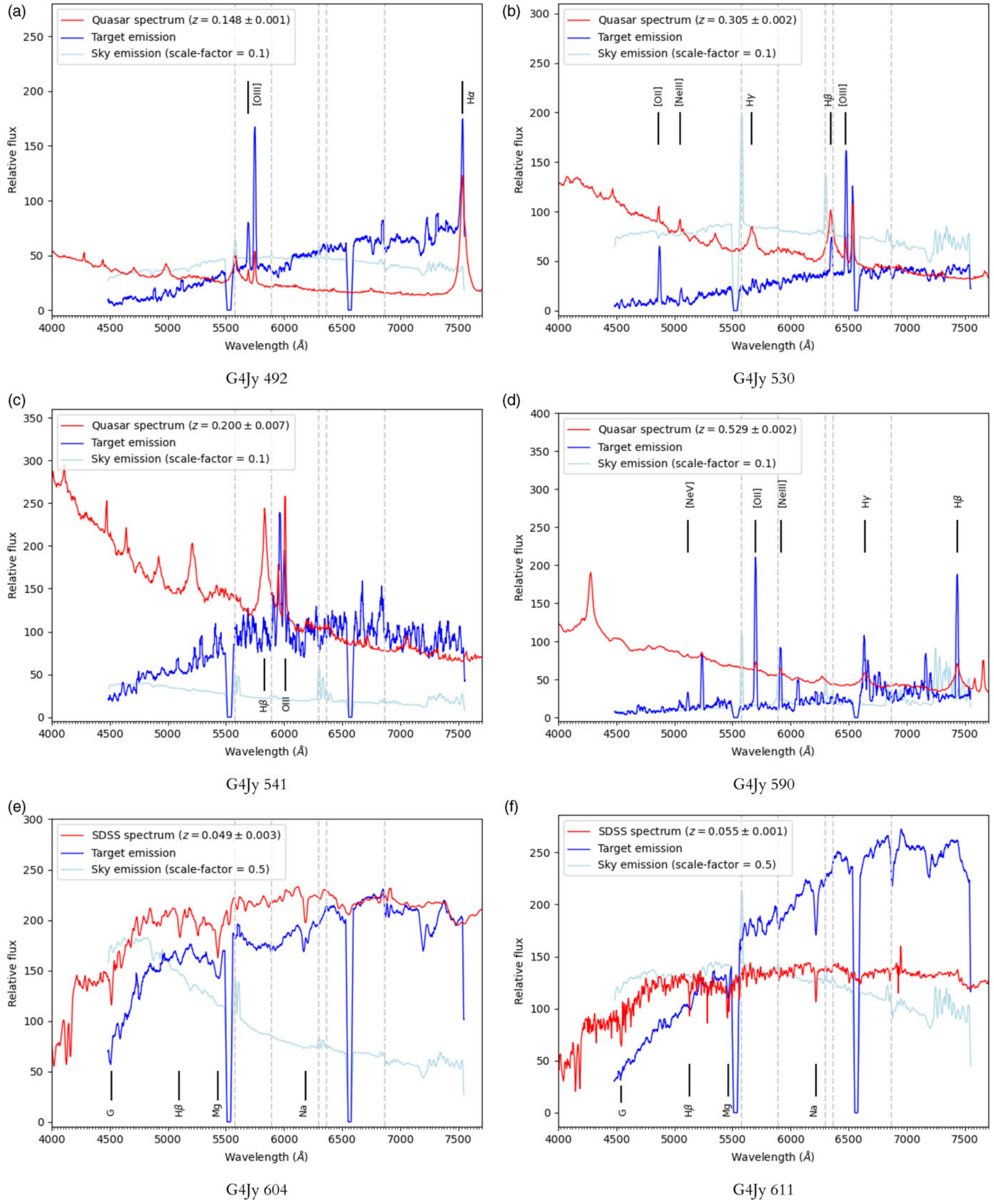

**Figure B1.** Continued.

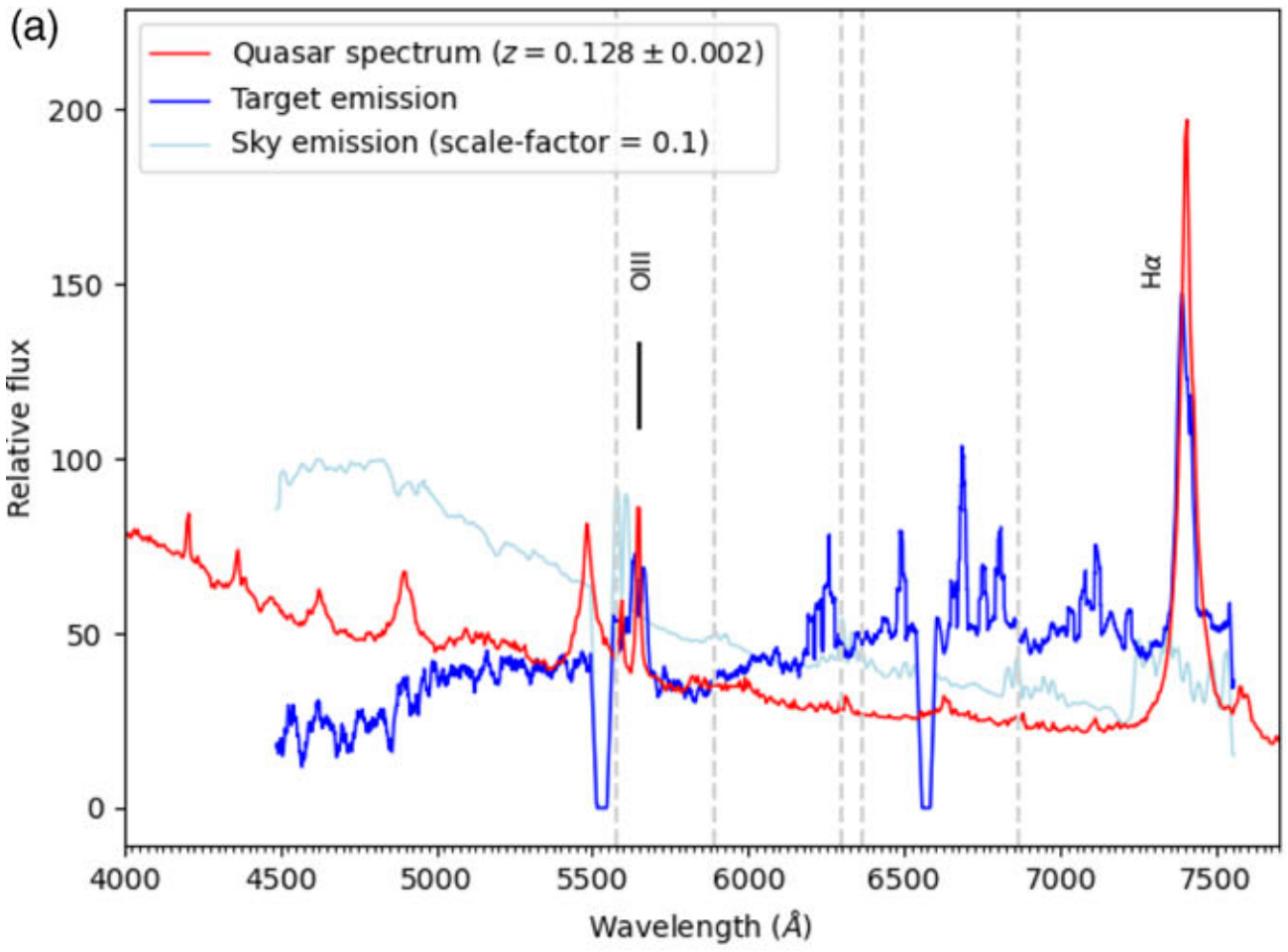

G4Jy 641

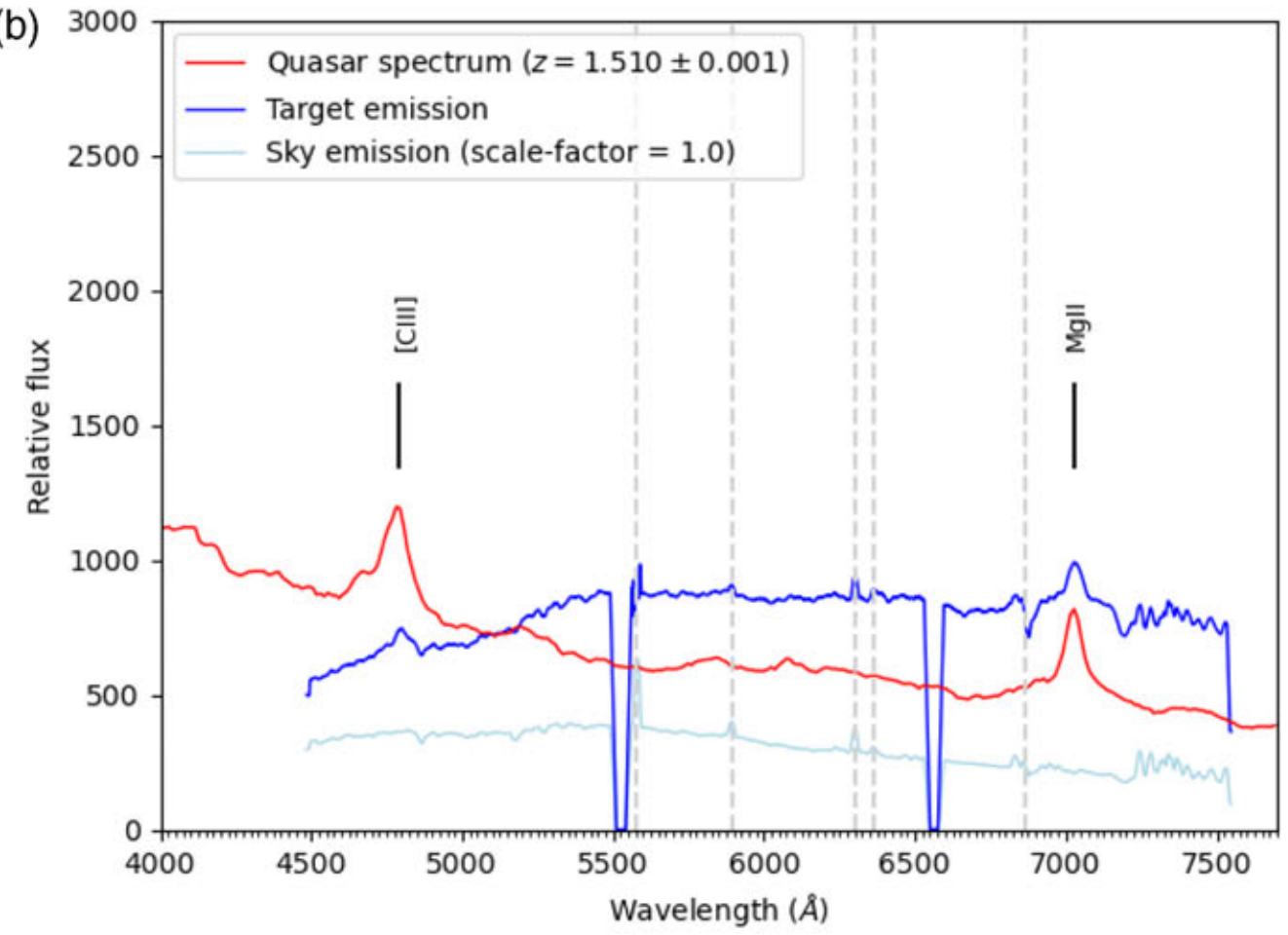

G4Jy 672

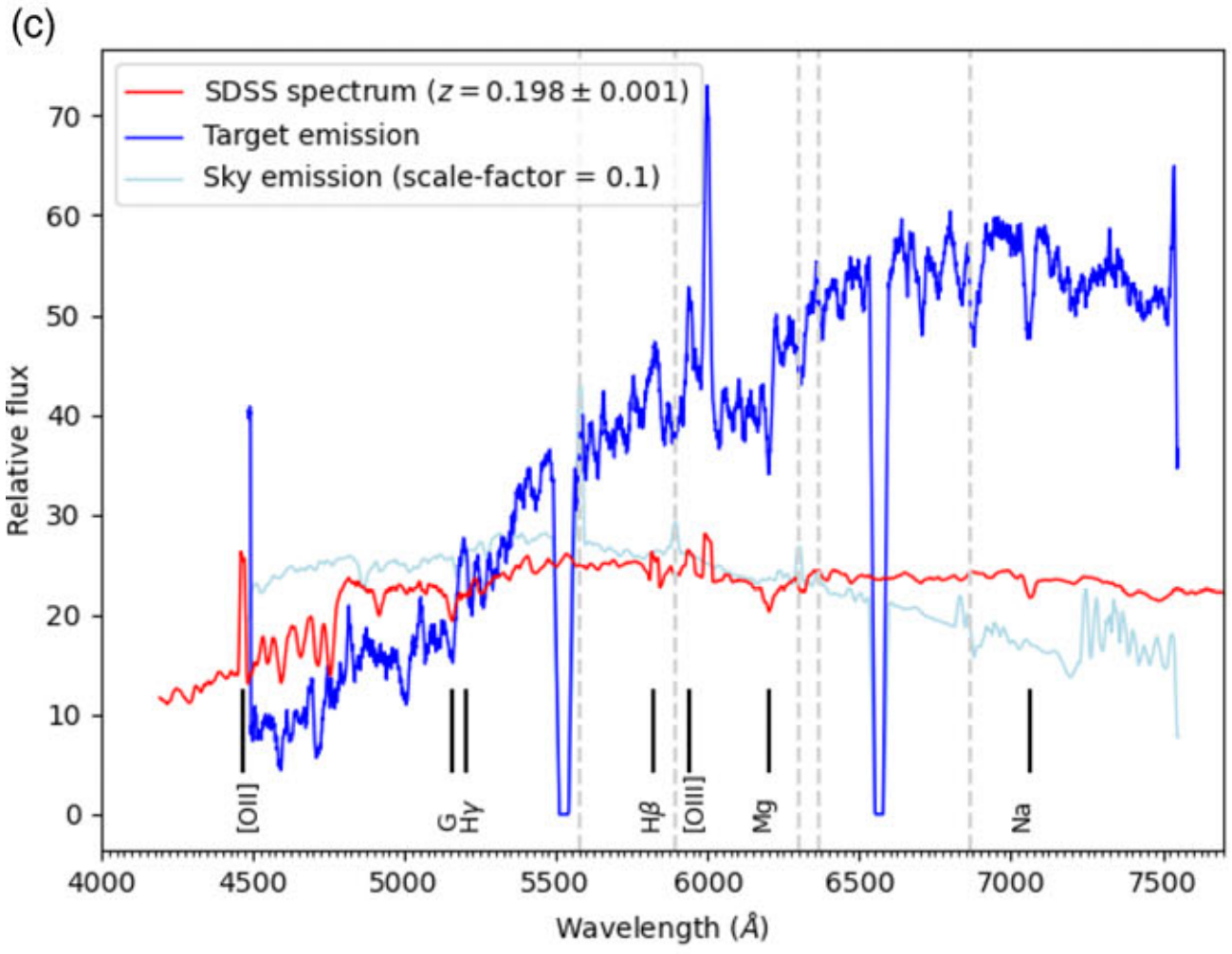

G4Jy 692

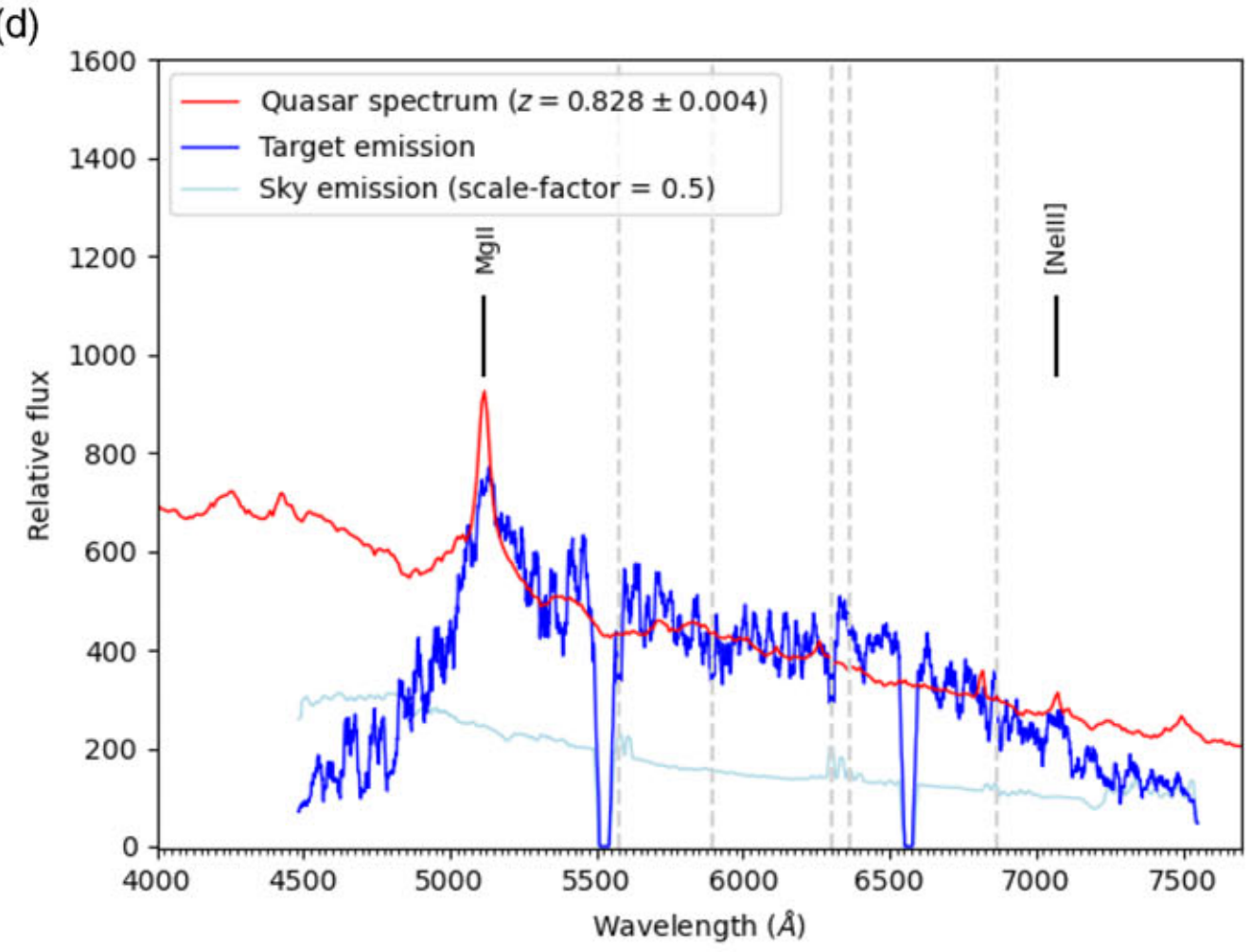

G4Jy 706

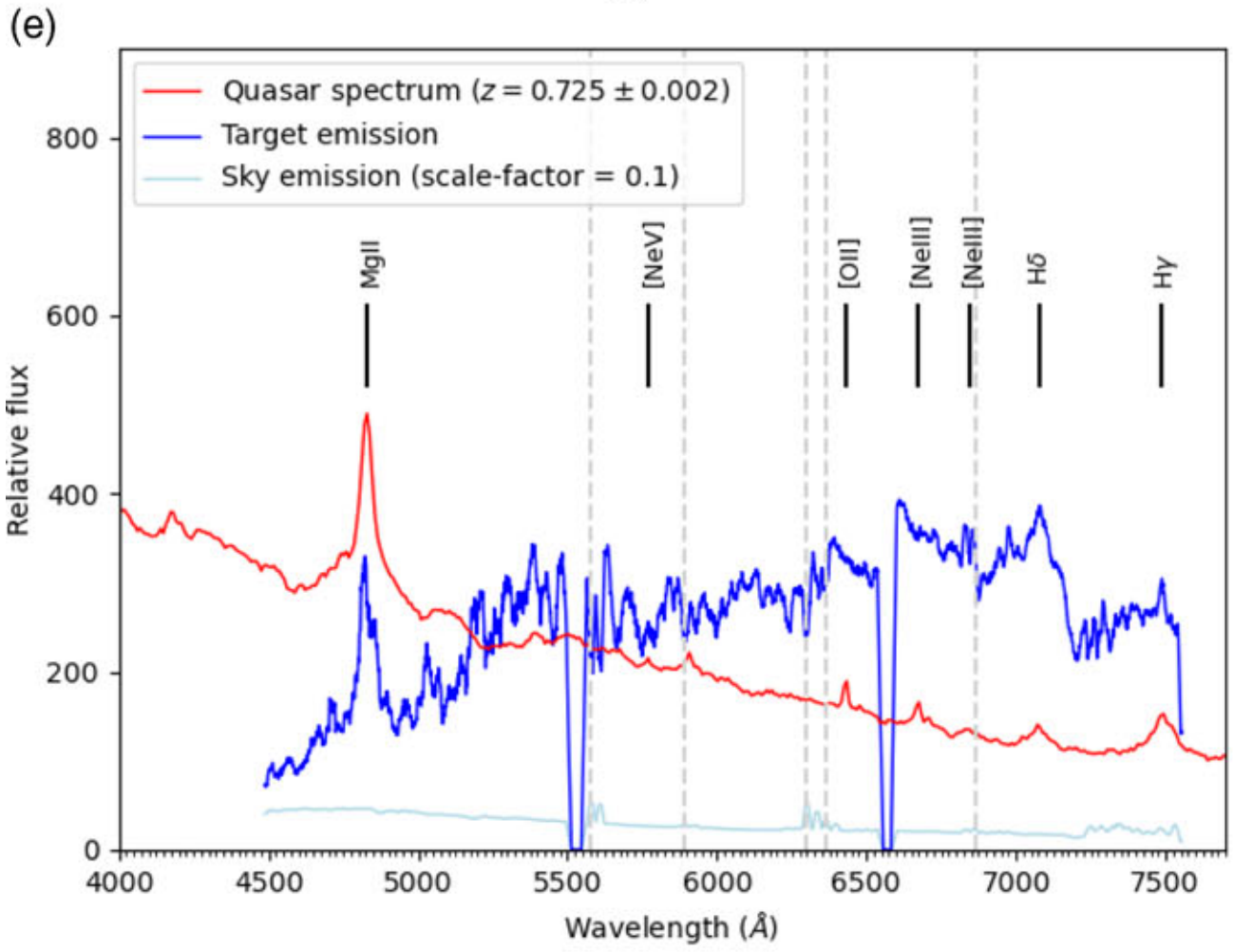

G4Jy 901

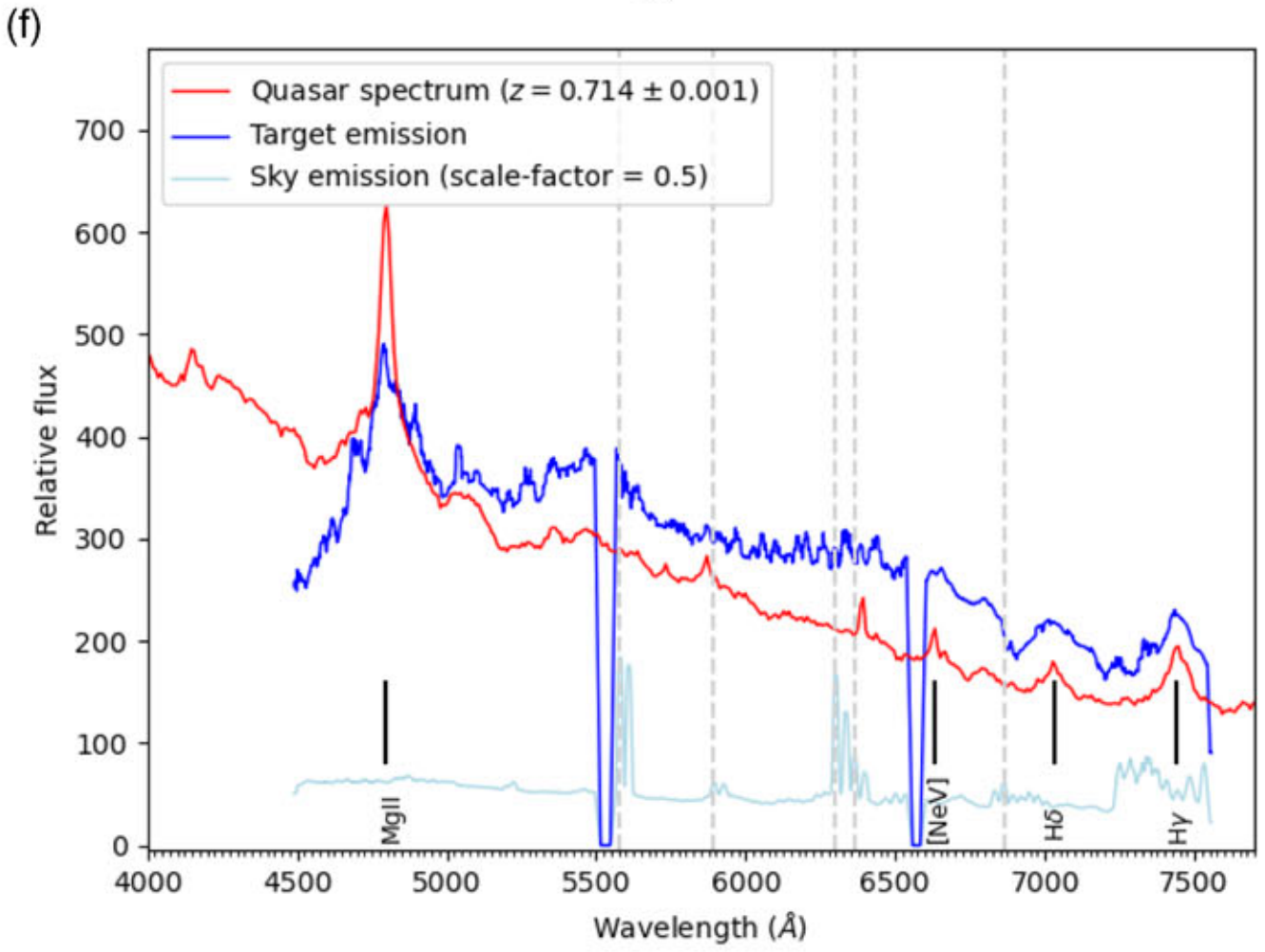

G4Jy 909

**Figure B1.** Continued.

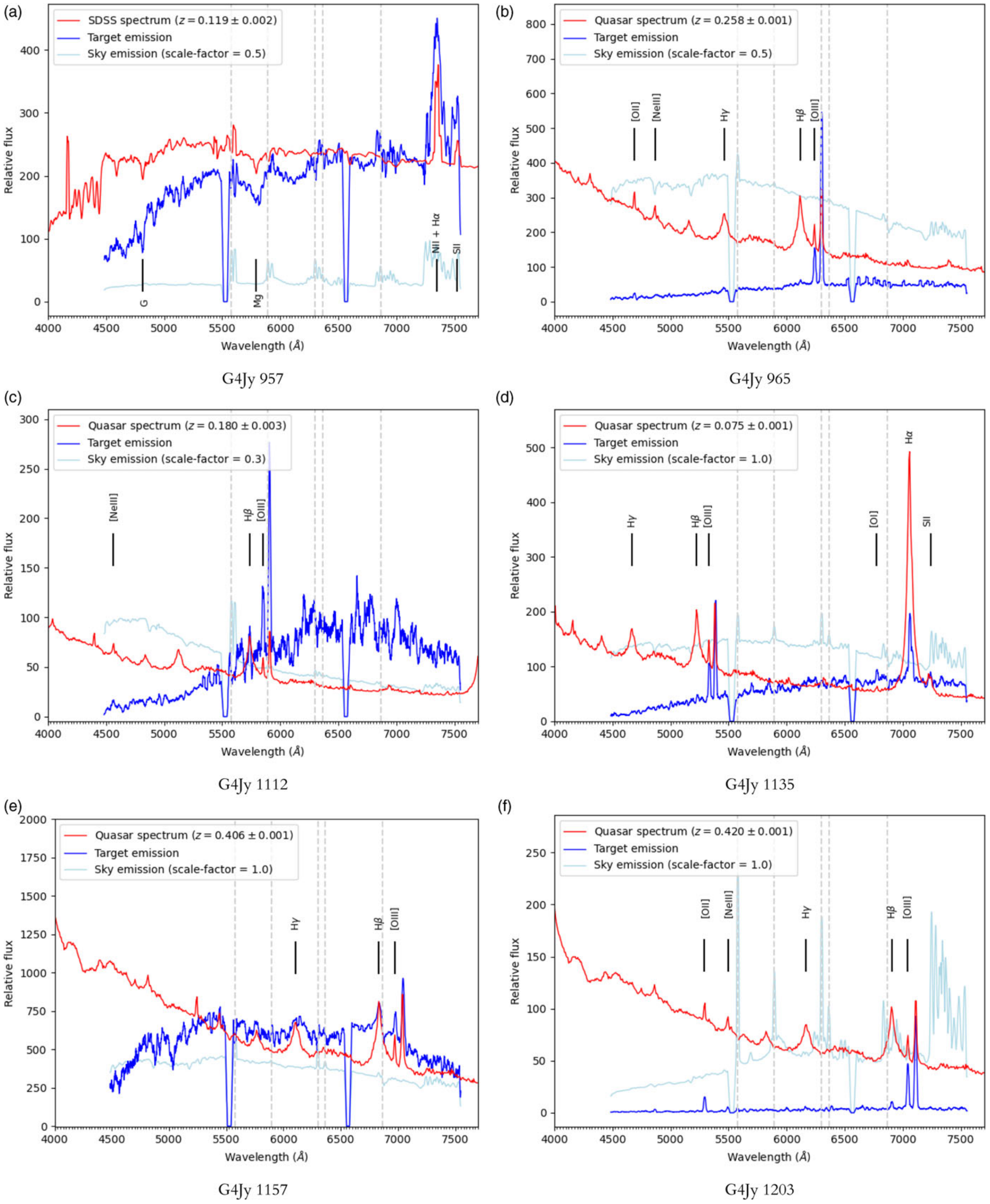

**Figure B1.** Continued.

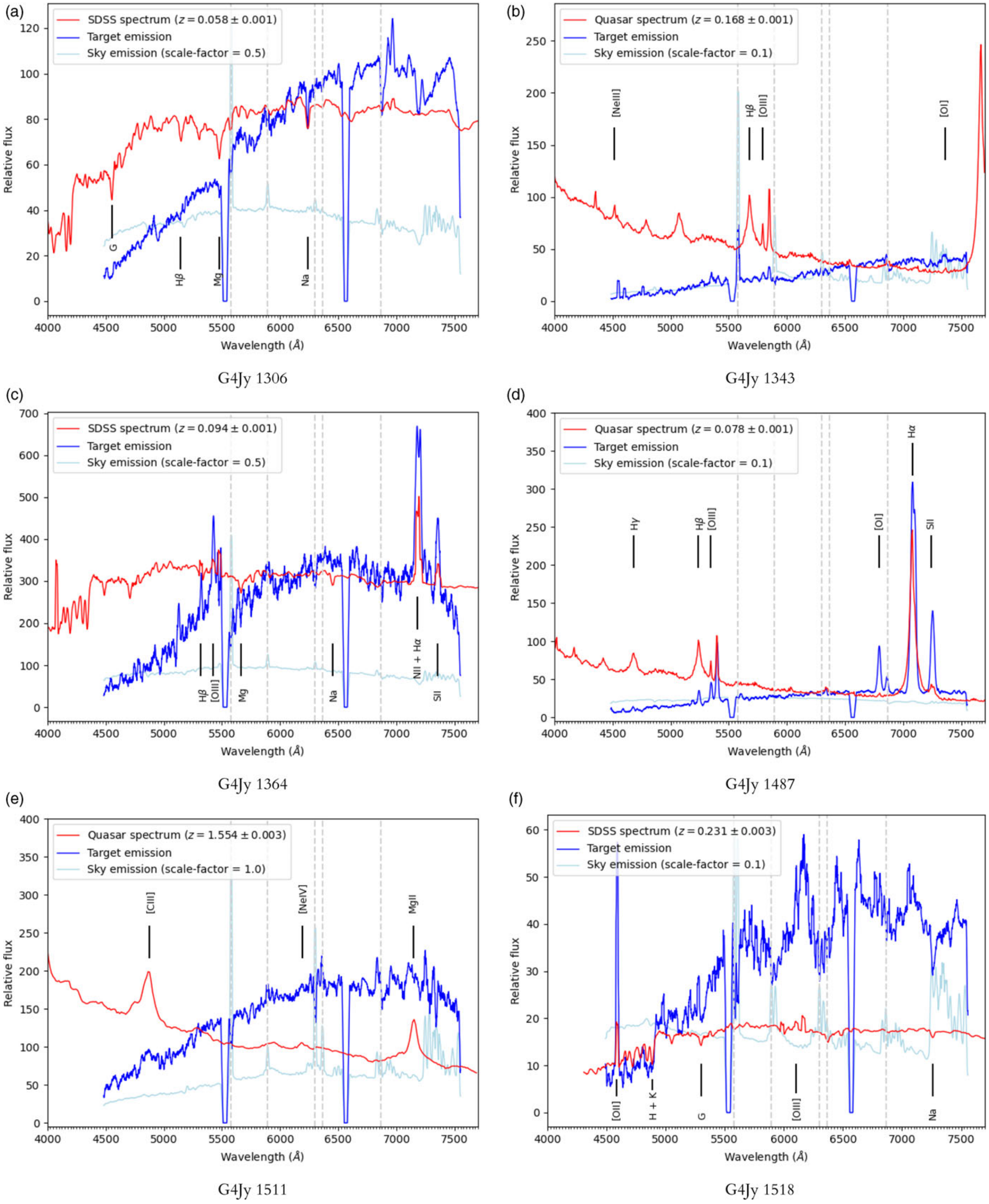

**Figure B1.** Continued.

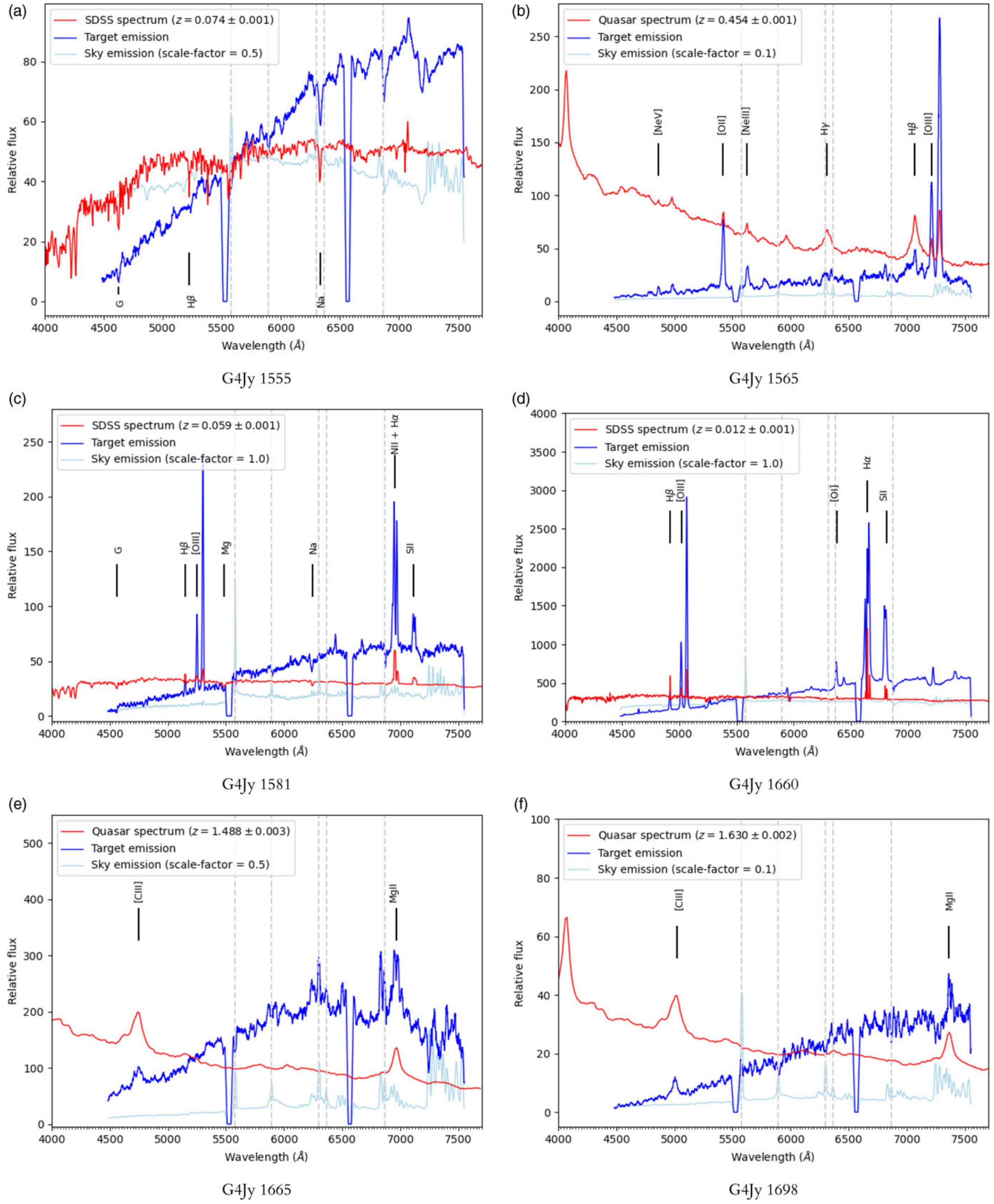

**Figure B1.** Continued.

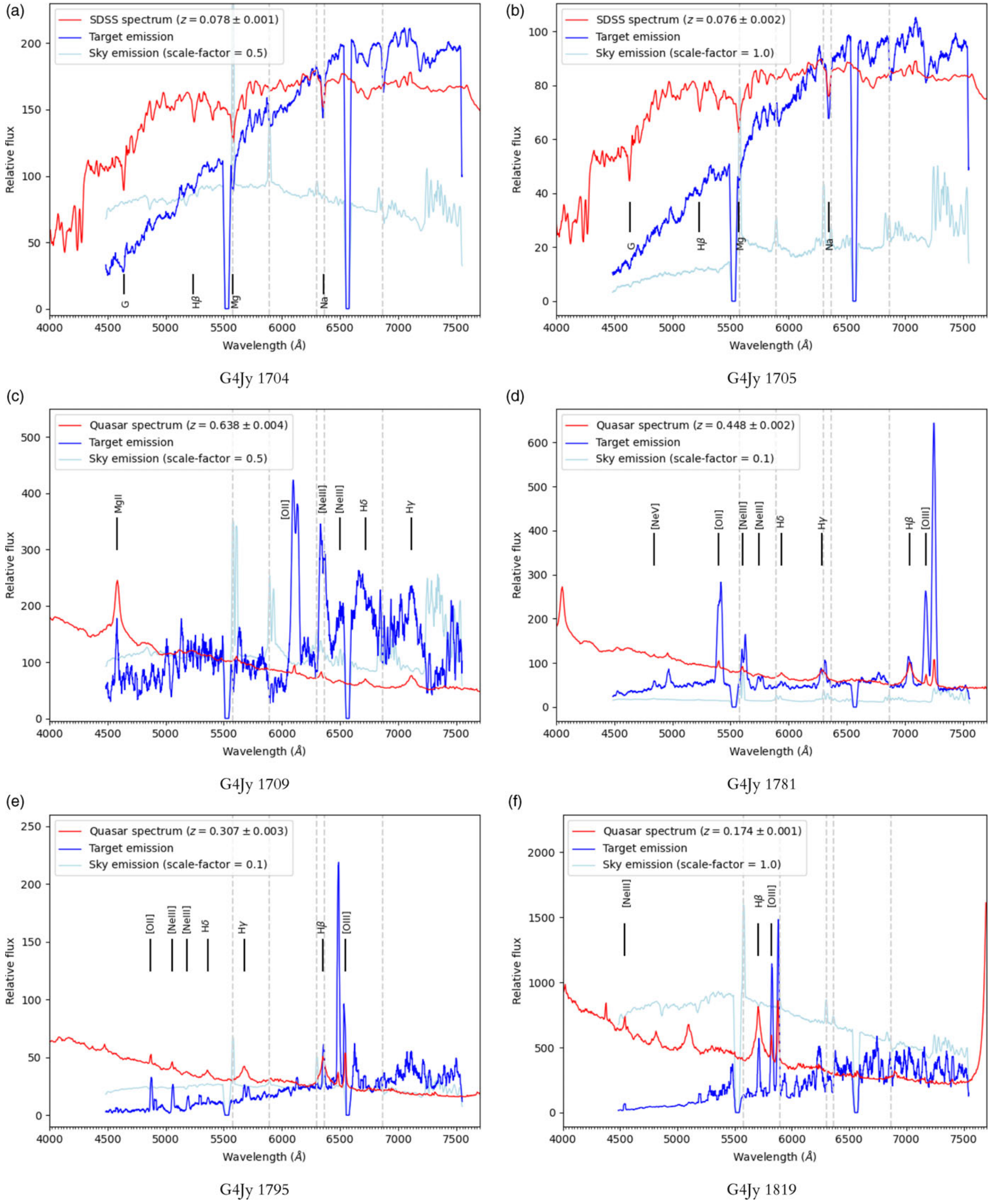

**Figure B1.** Continued.

(a)

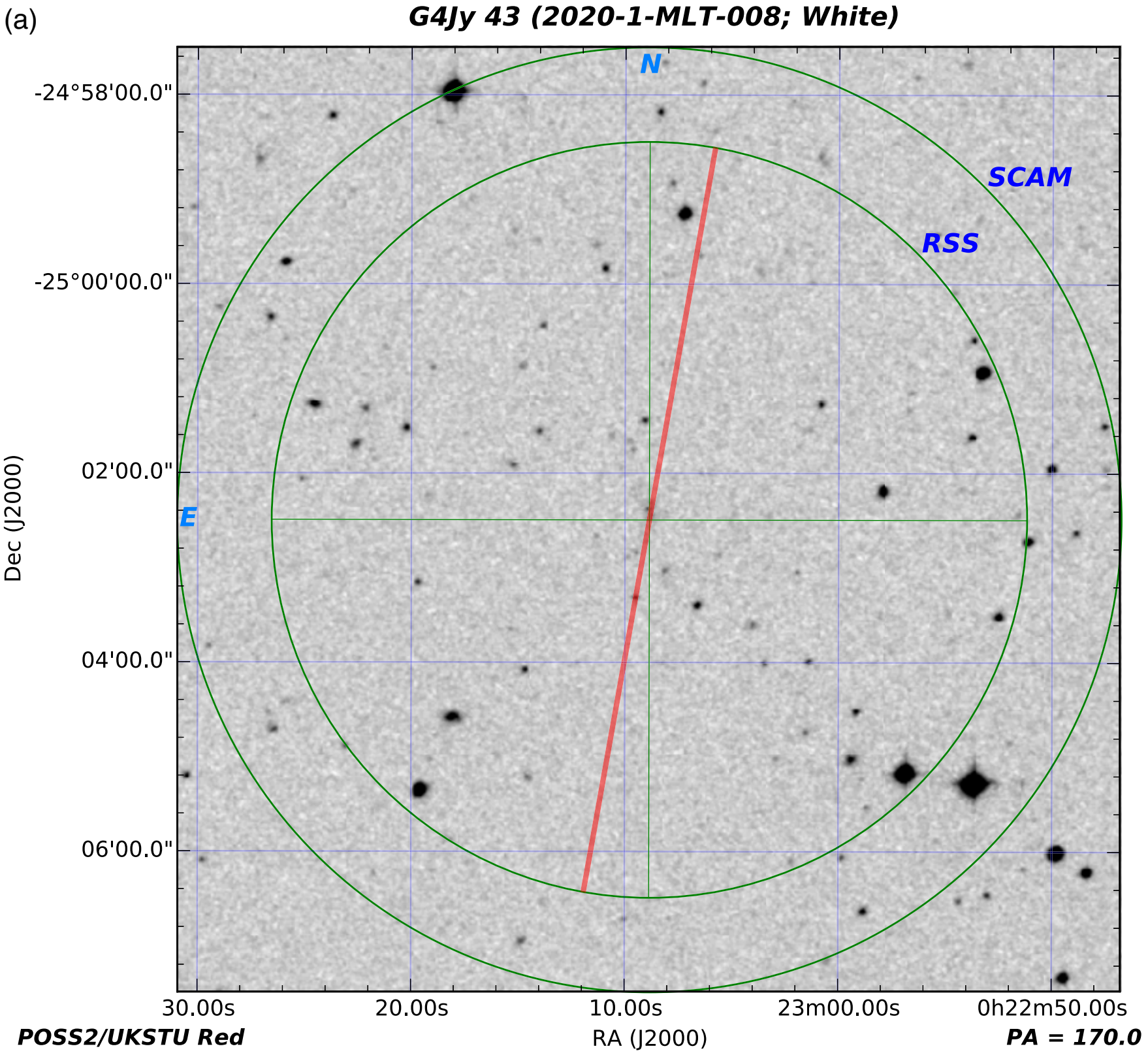

Finder Chart for G4Jy 43

(b)

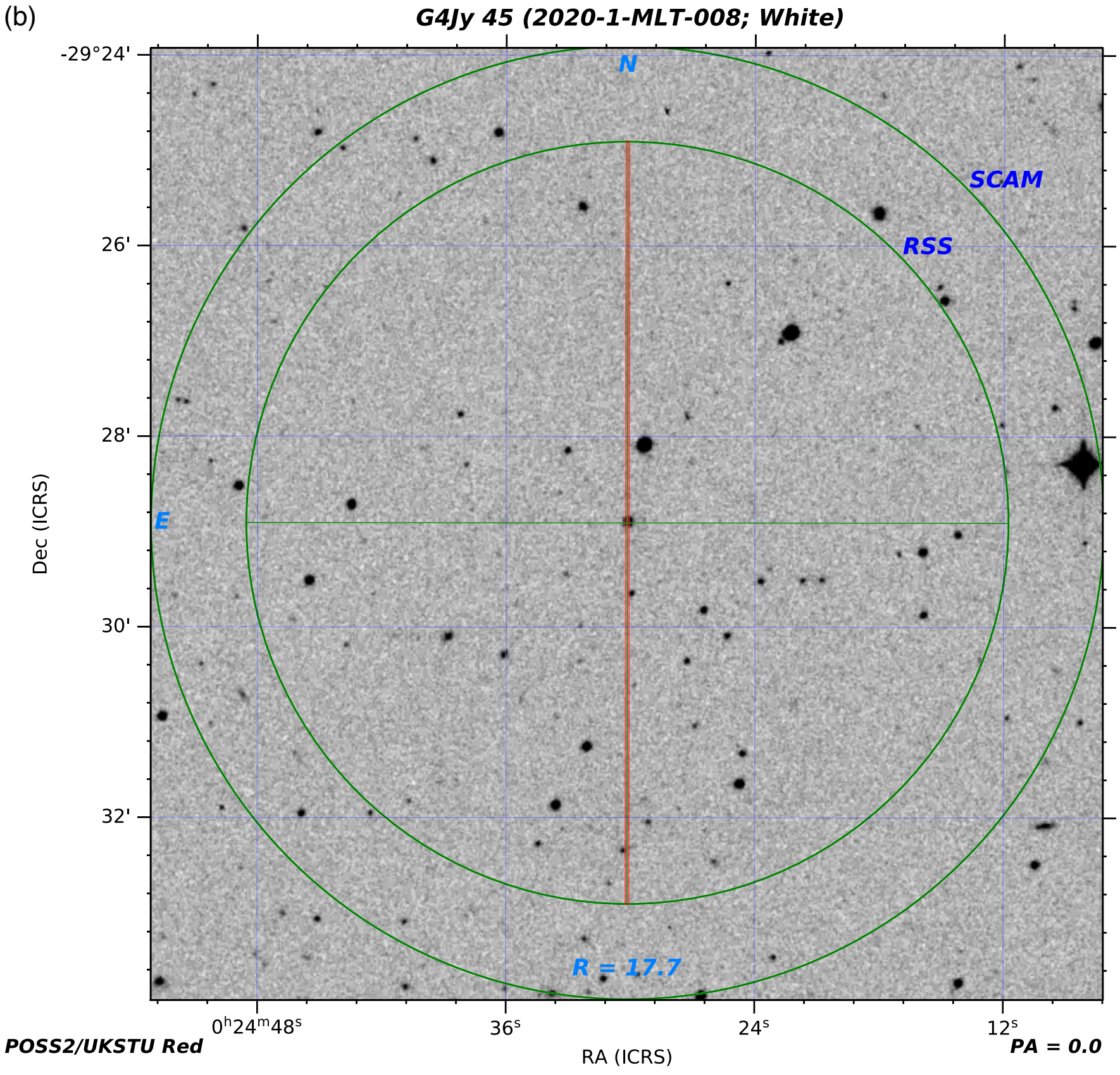

Finder Chart for G4Jy 45

**Figure B2.** SALT Finder Charts, with the target at the centre. A non-zero Position Angle (PA) for the slit indicates that an alignment object was used.

(a)

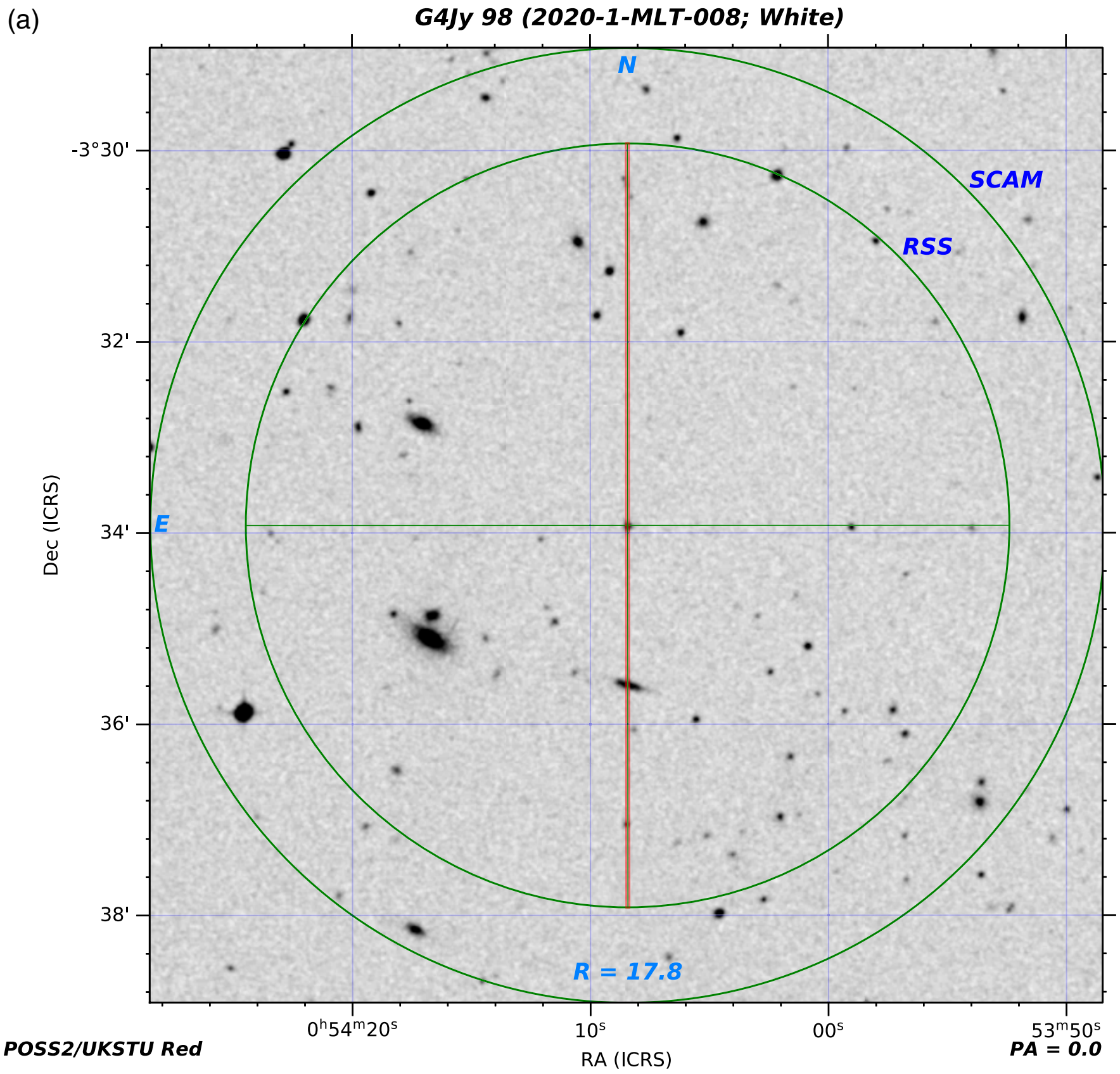

Finder Chart for G4Jy 98

(b)

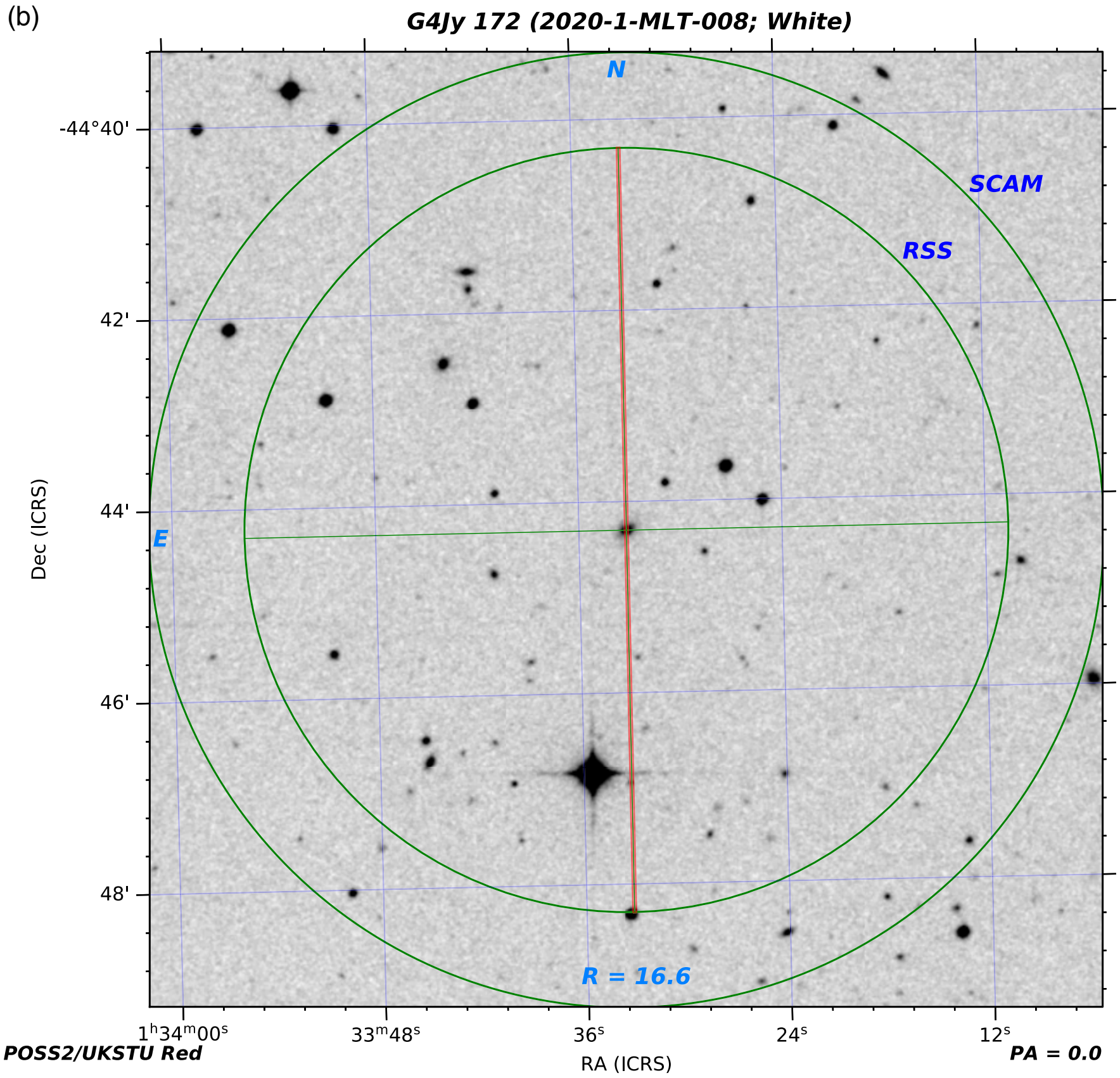

Finder Chart for G4Jy 172

**Figure B2.** Continued.

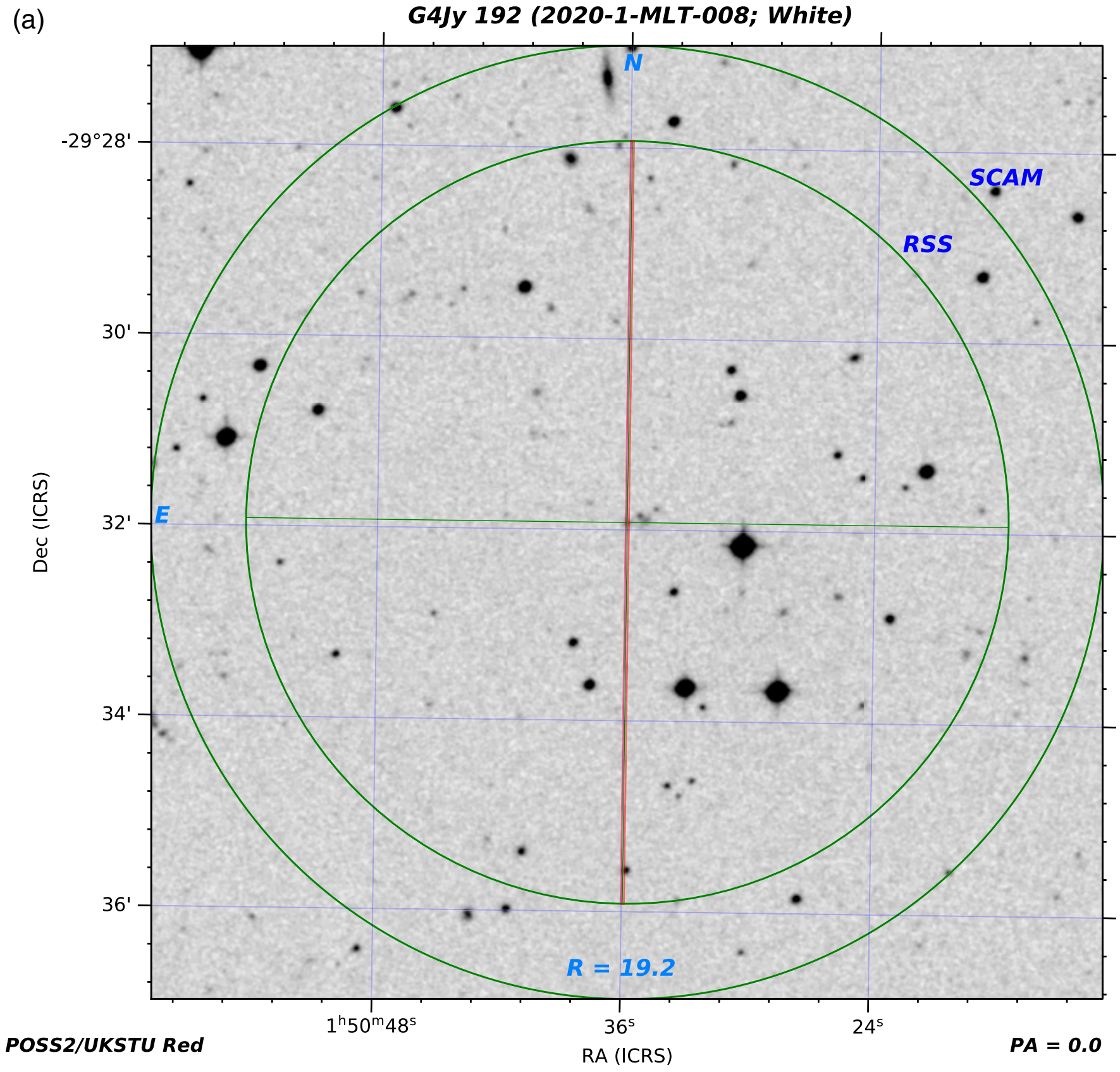

Finder Chart for G4Jy 192

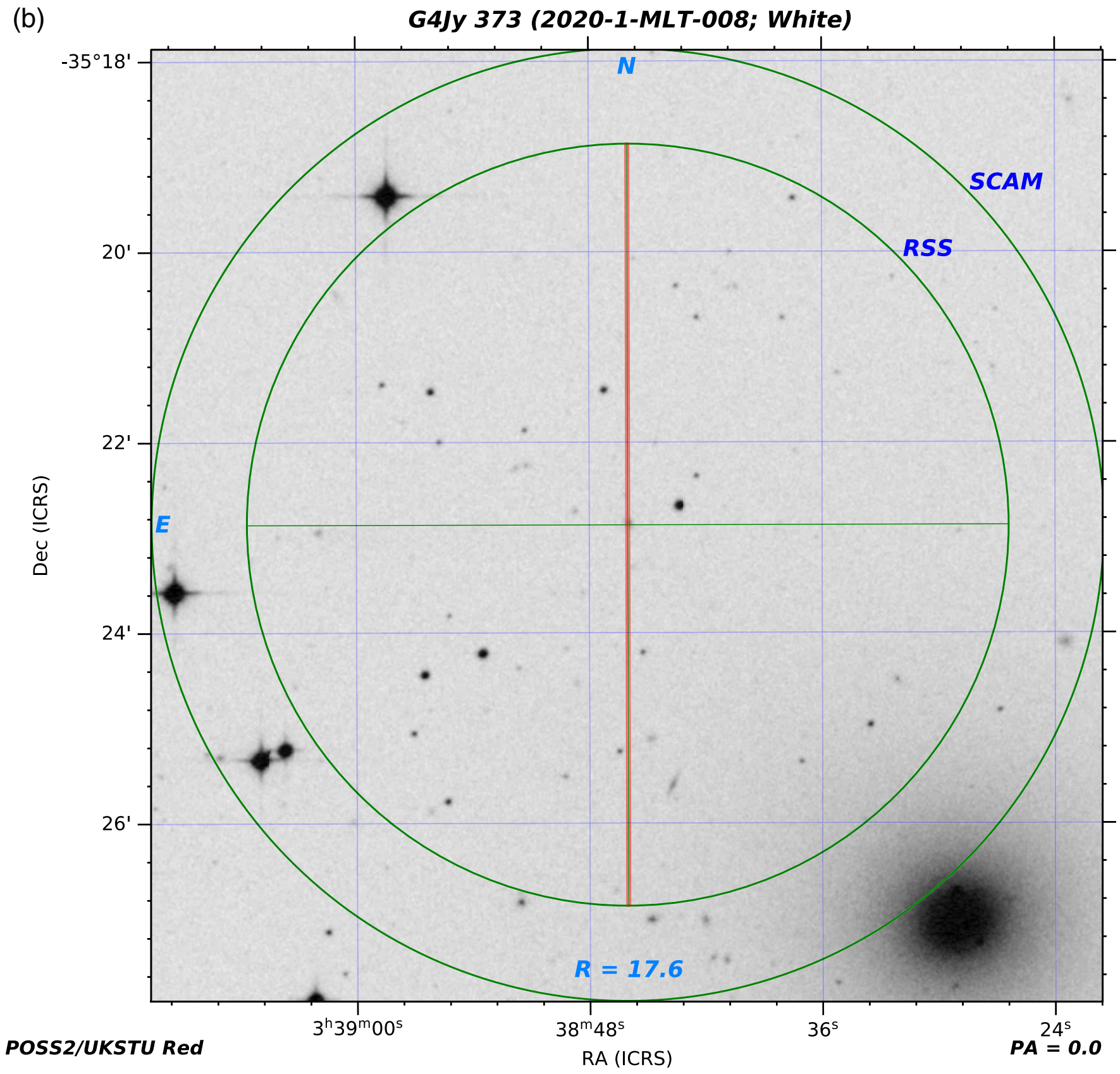

Finder Chart for G4Jy 373

**Figure B2.** Continued.

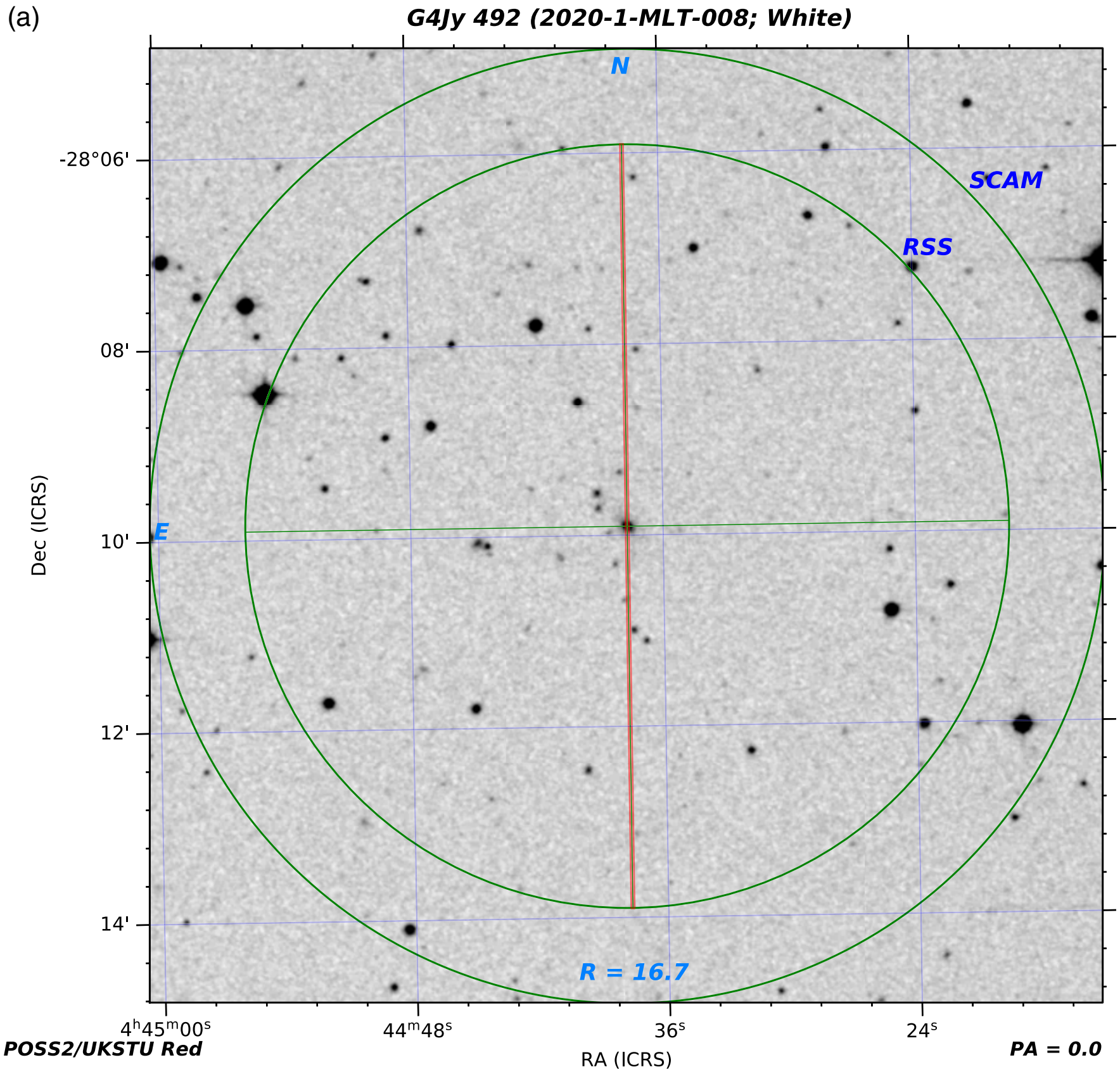

Finder Chart for G4Jy 492

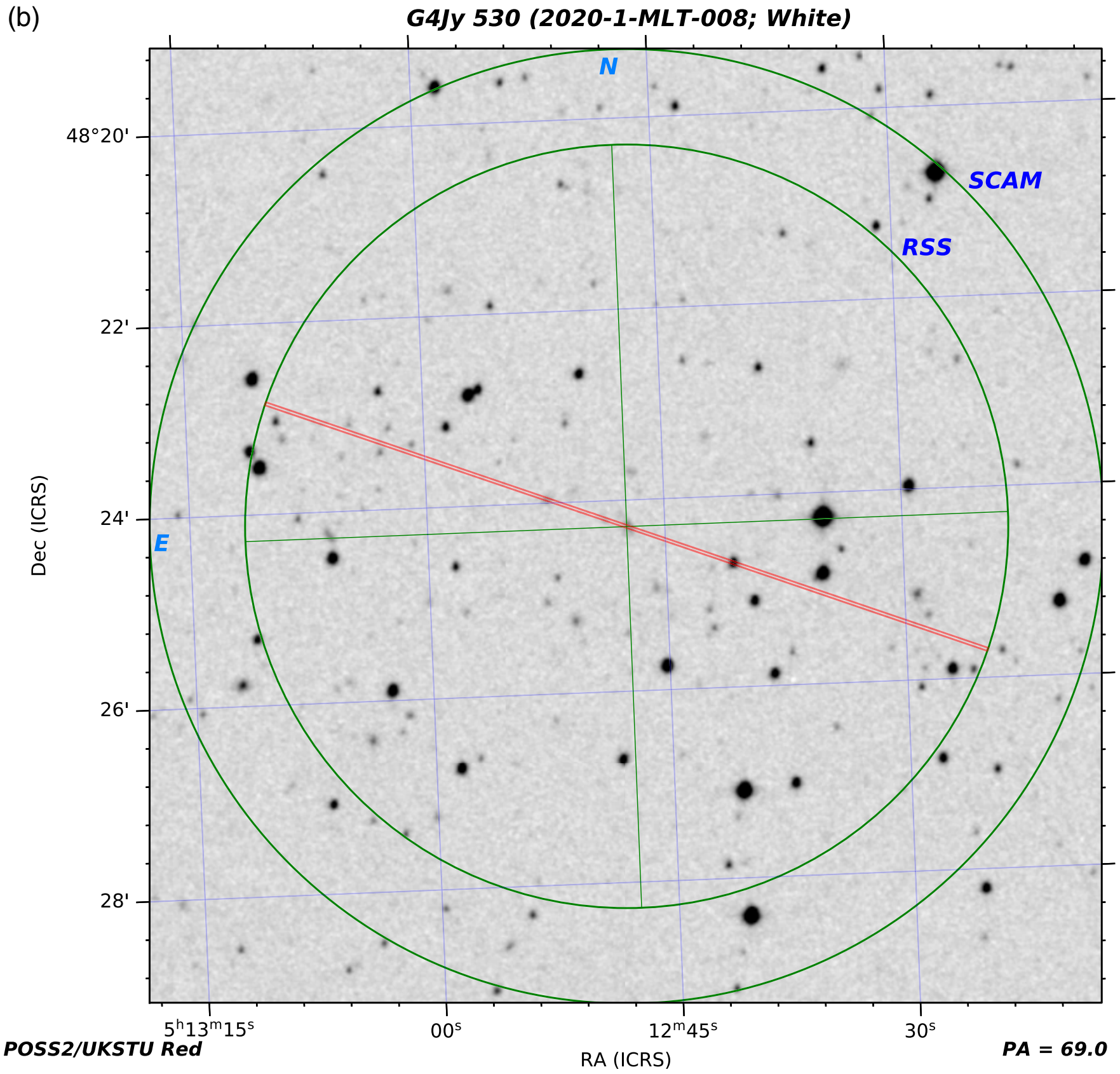

Finder Chart for G4Jy 530

**Figure B2.** Continued.

(a)

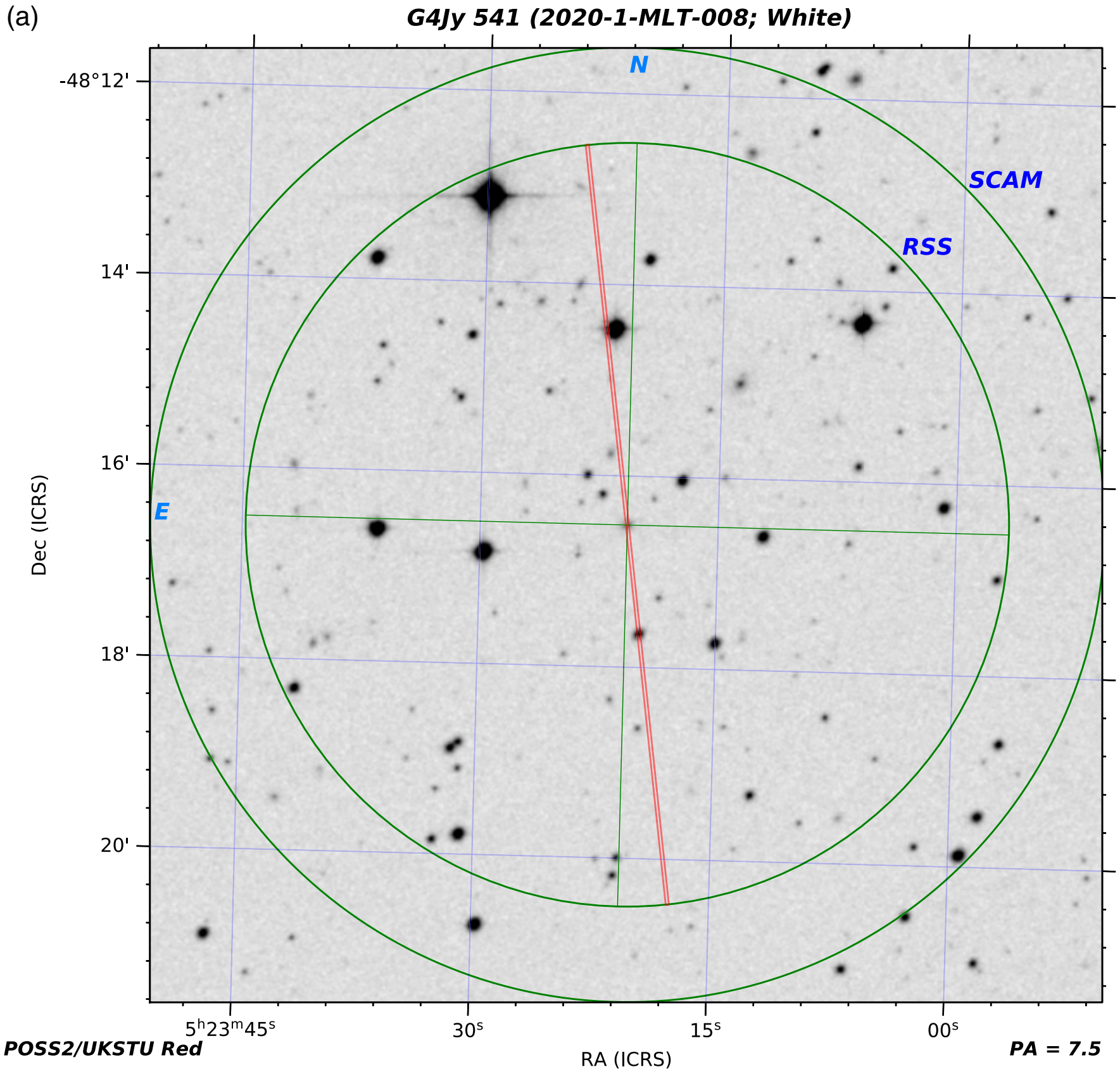

Finder Chart for G4Jy 541

(b)

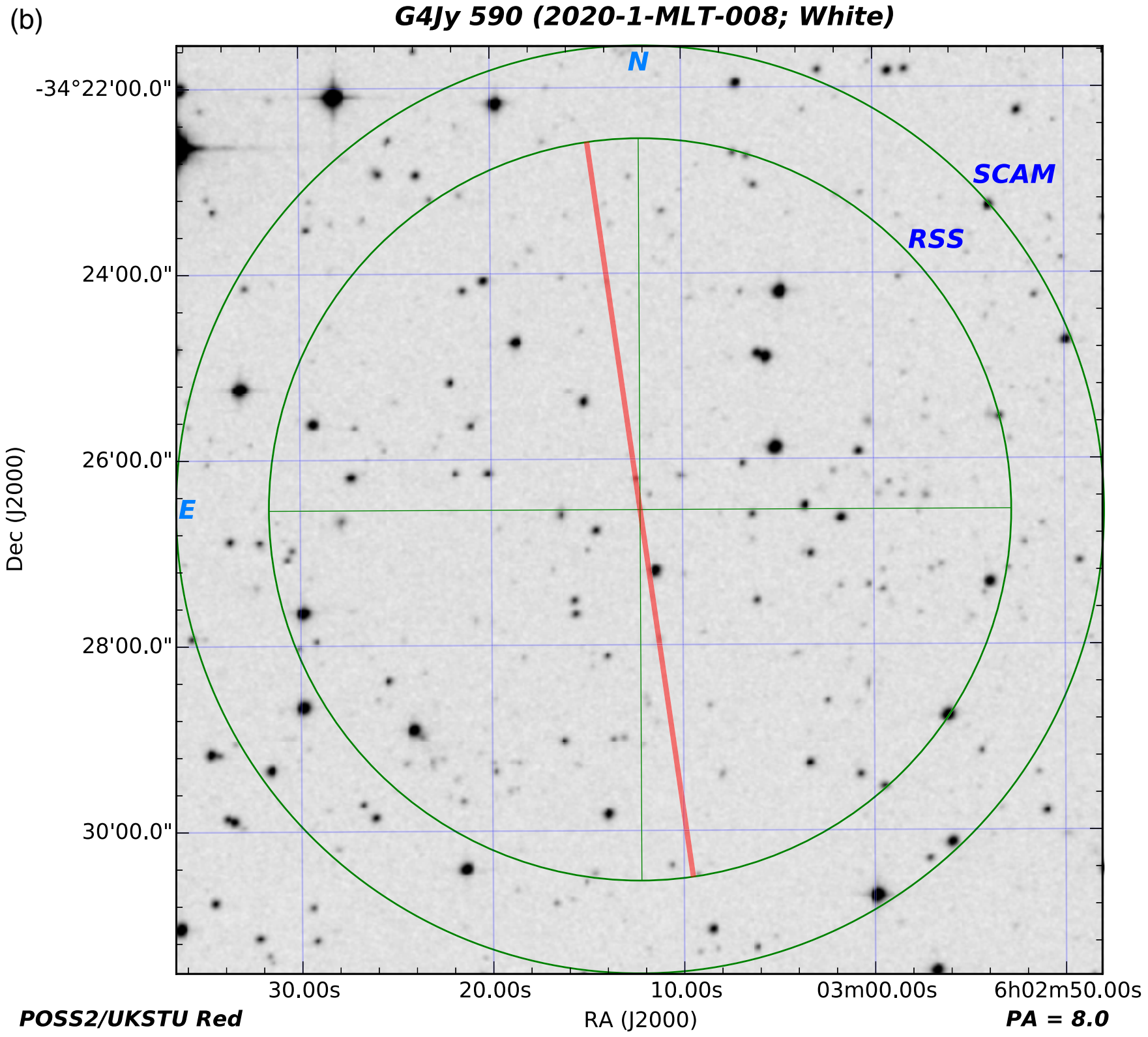

Finder Chart for G4Jy 590

**Figure B2.** Continued.

(a)

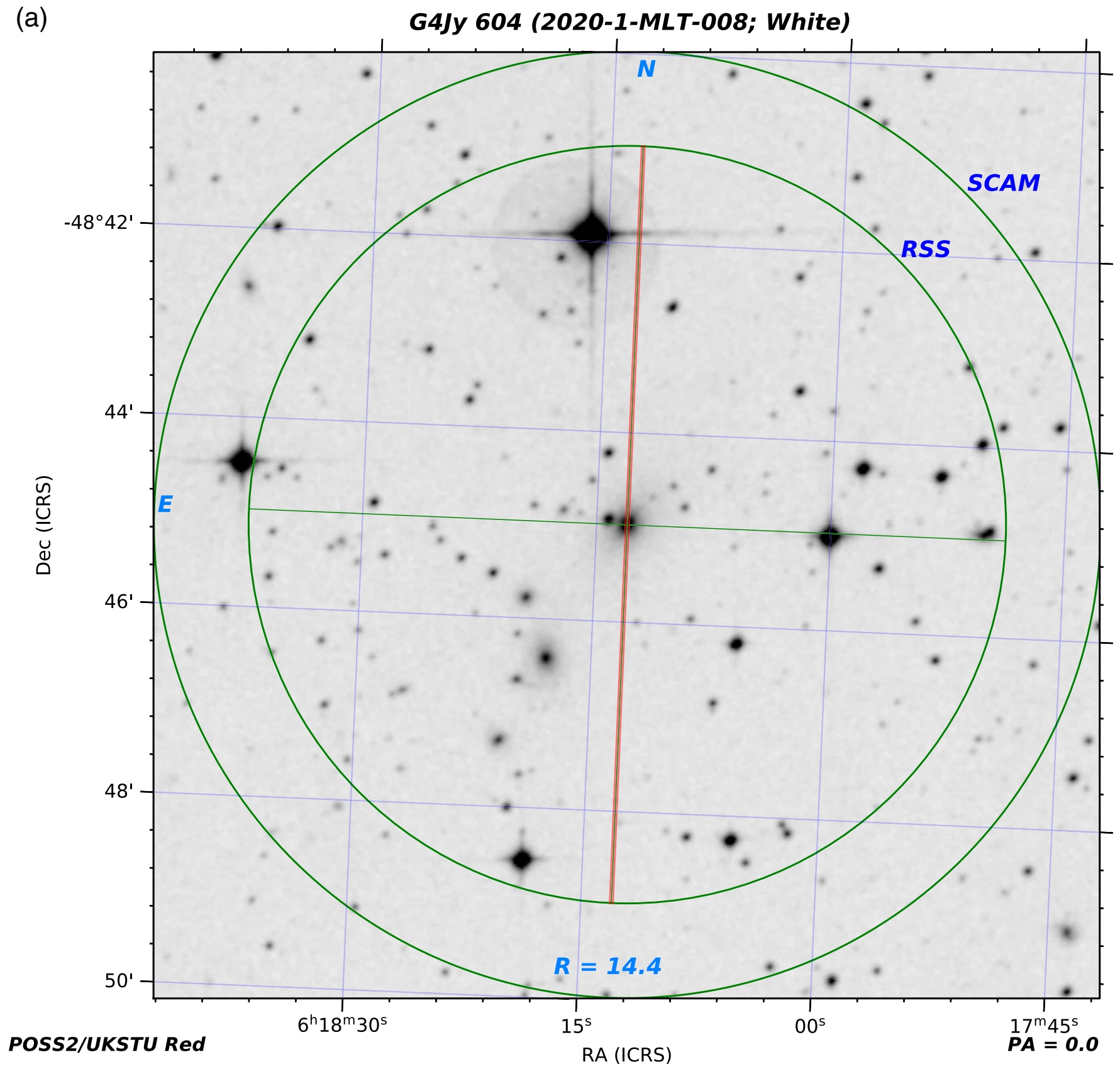

Finder Chart for G4Jy 604

(b)

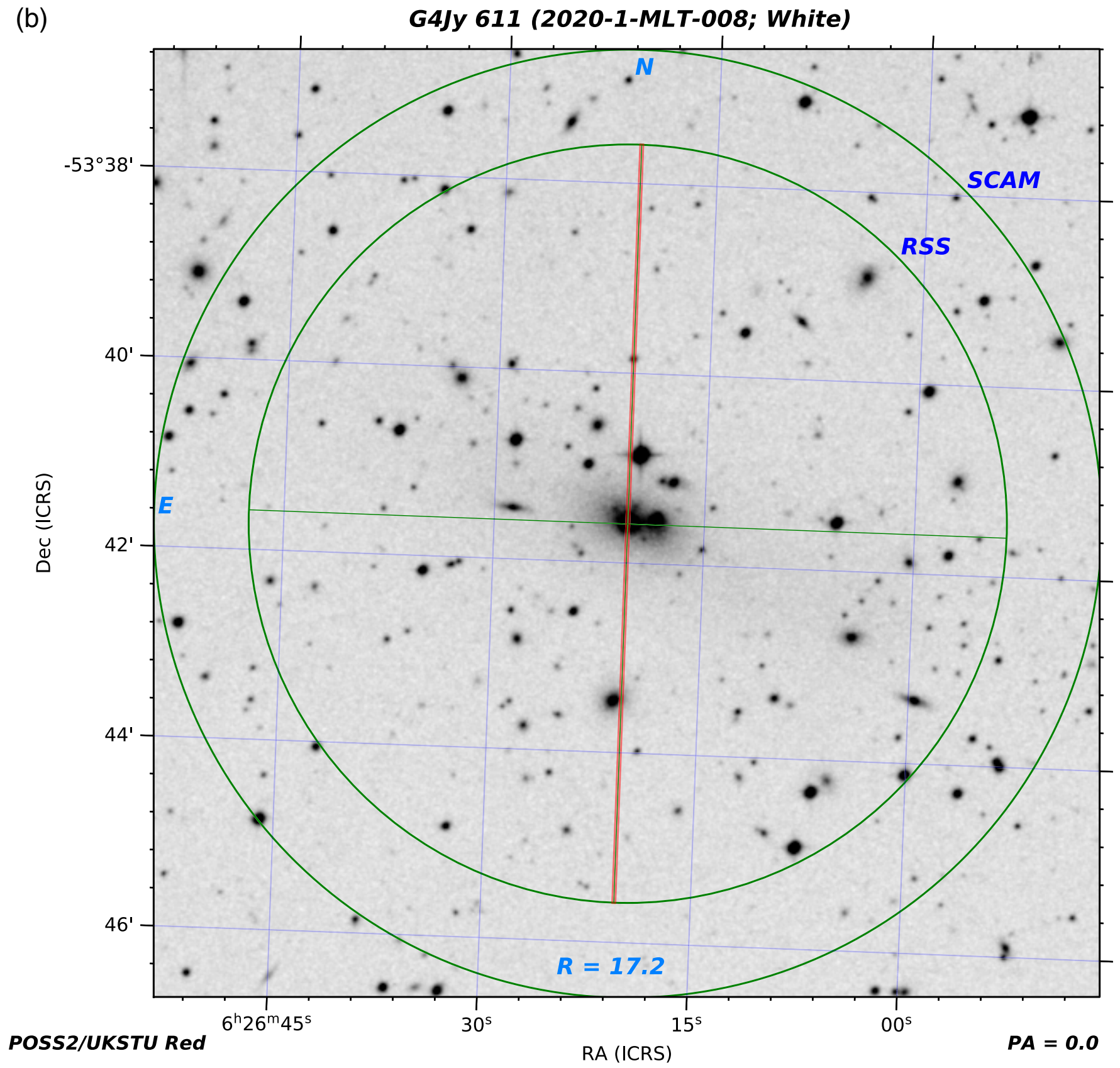

Finder Chart for G4Jy 611

**Figure B2.** Continued.

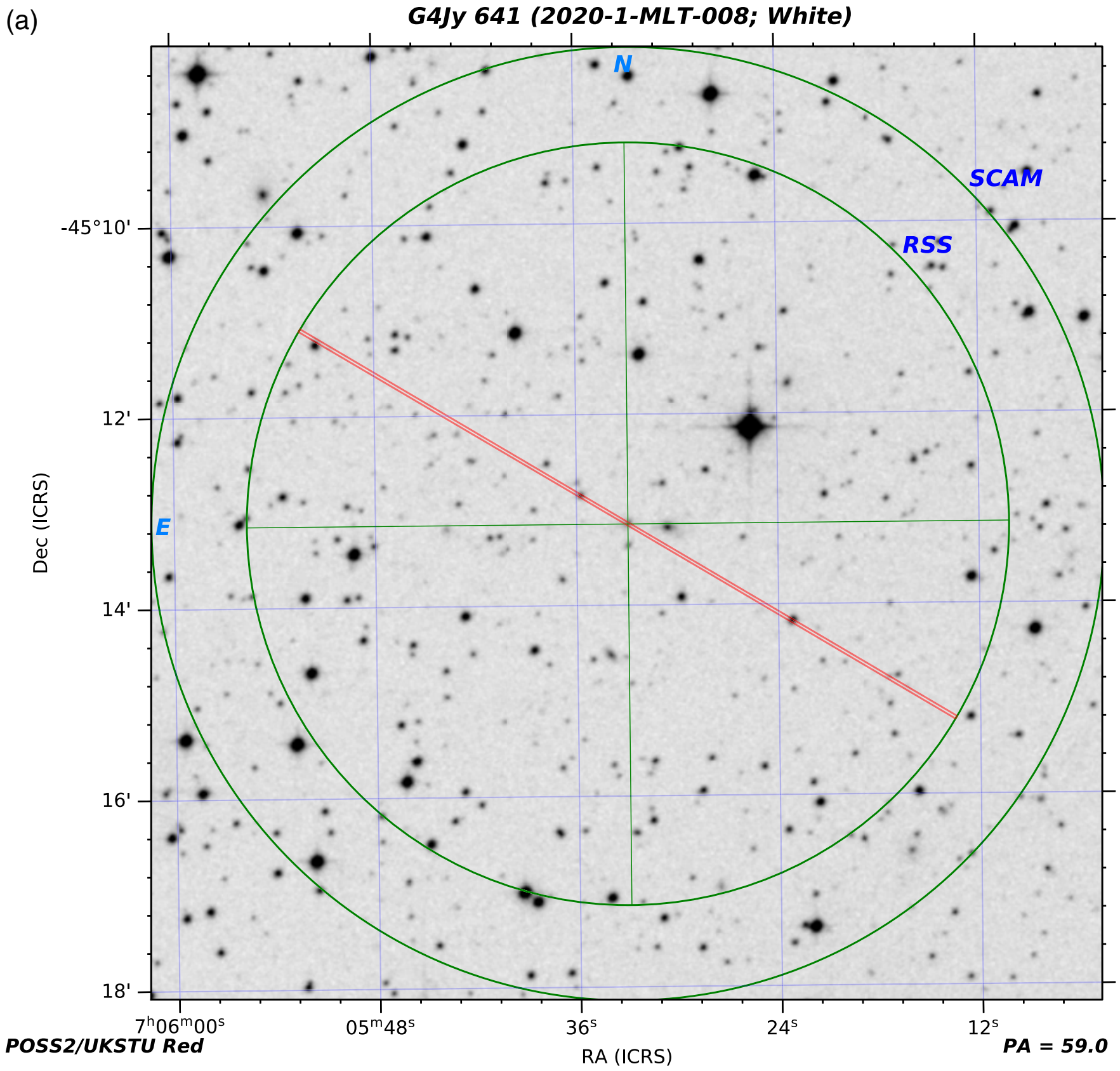

Finder Chart for G4Jy 641

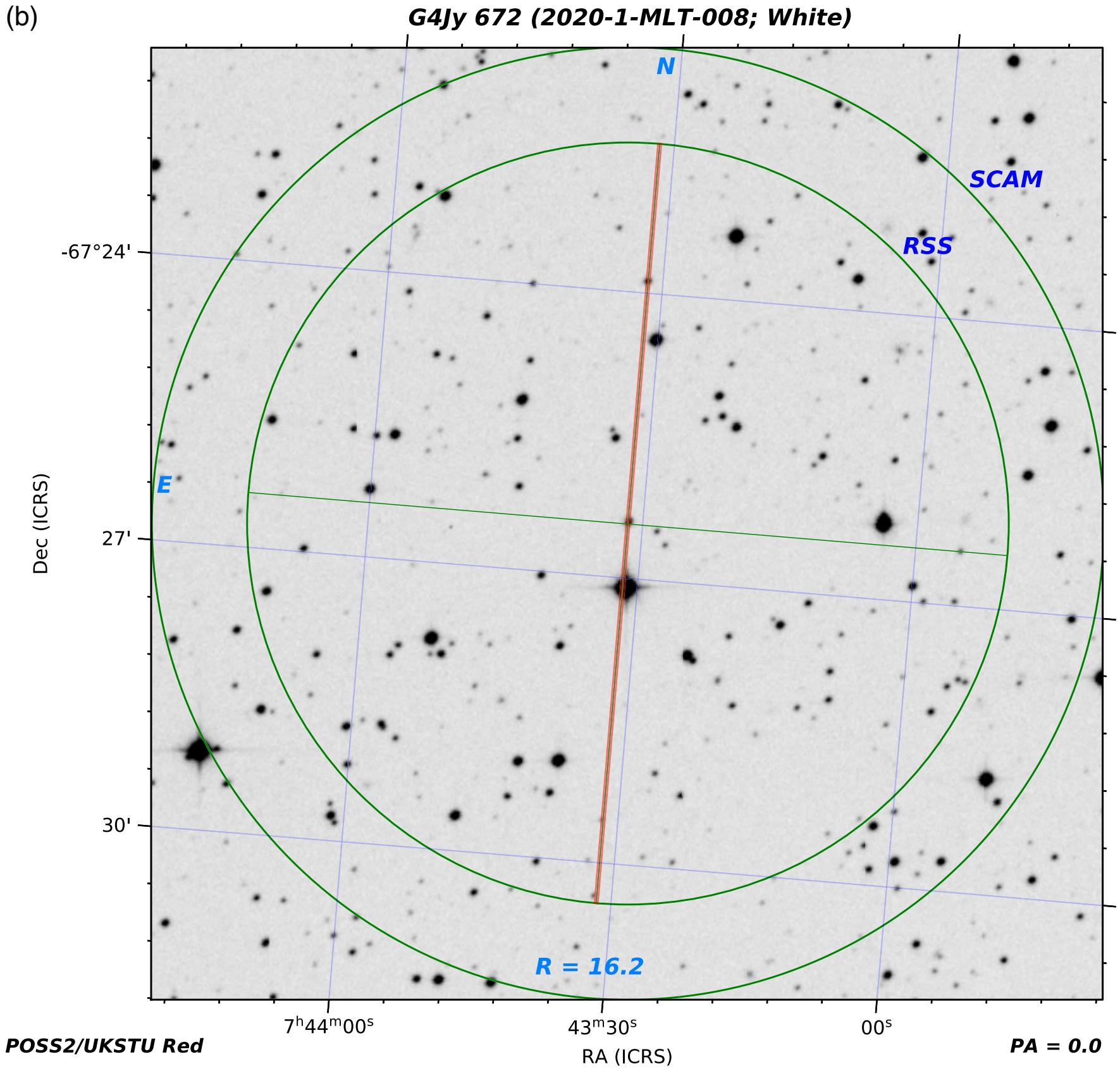

Finder Chart for G4Jy 672

**Figure B2.** Continued.

(a)

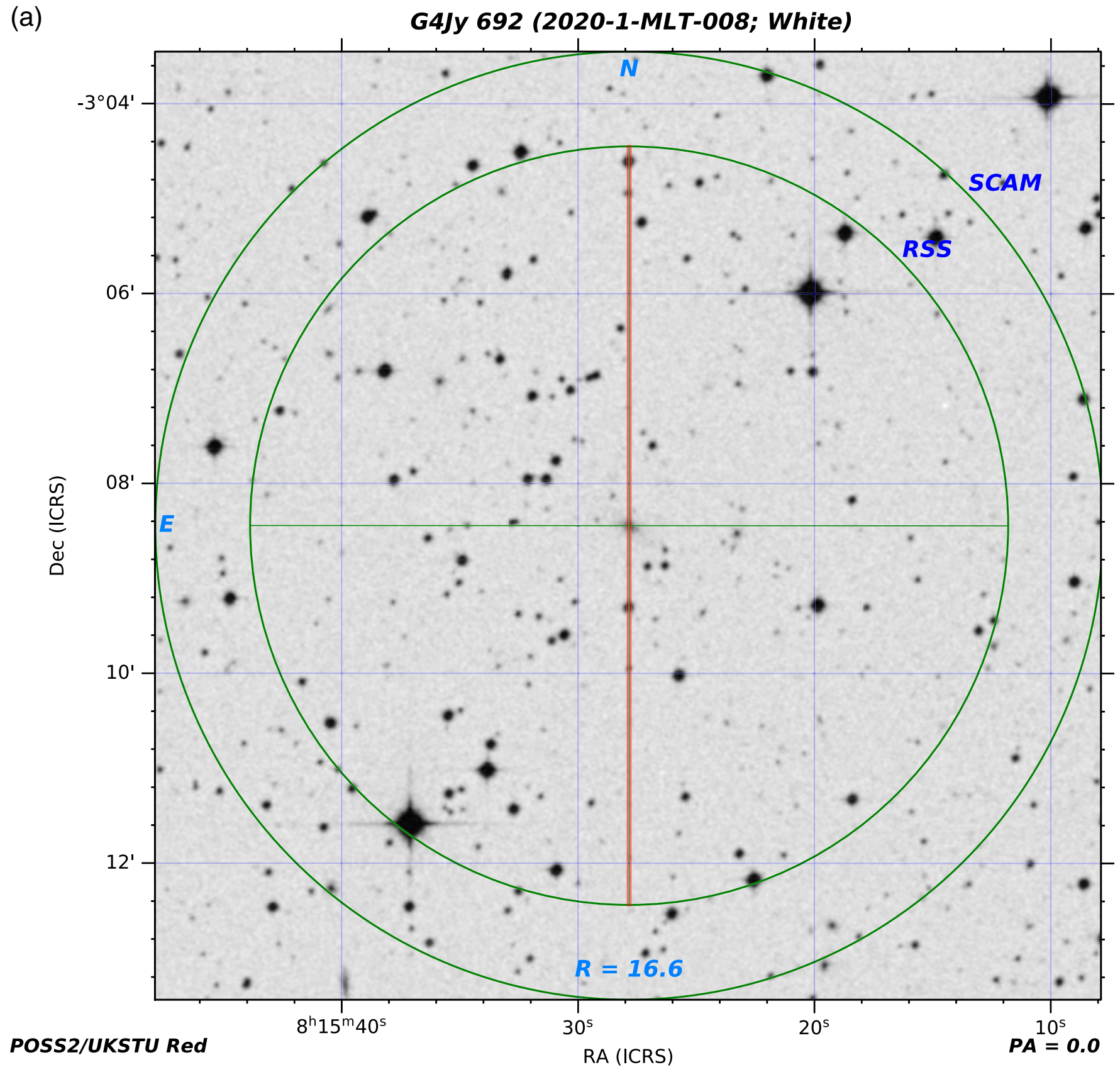

Finder Chart for G4Jy 692

(b)

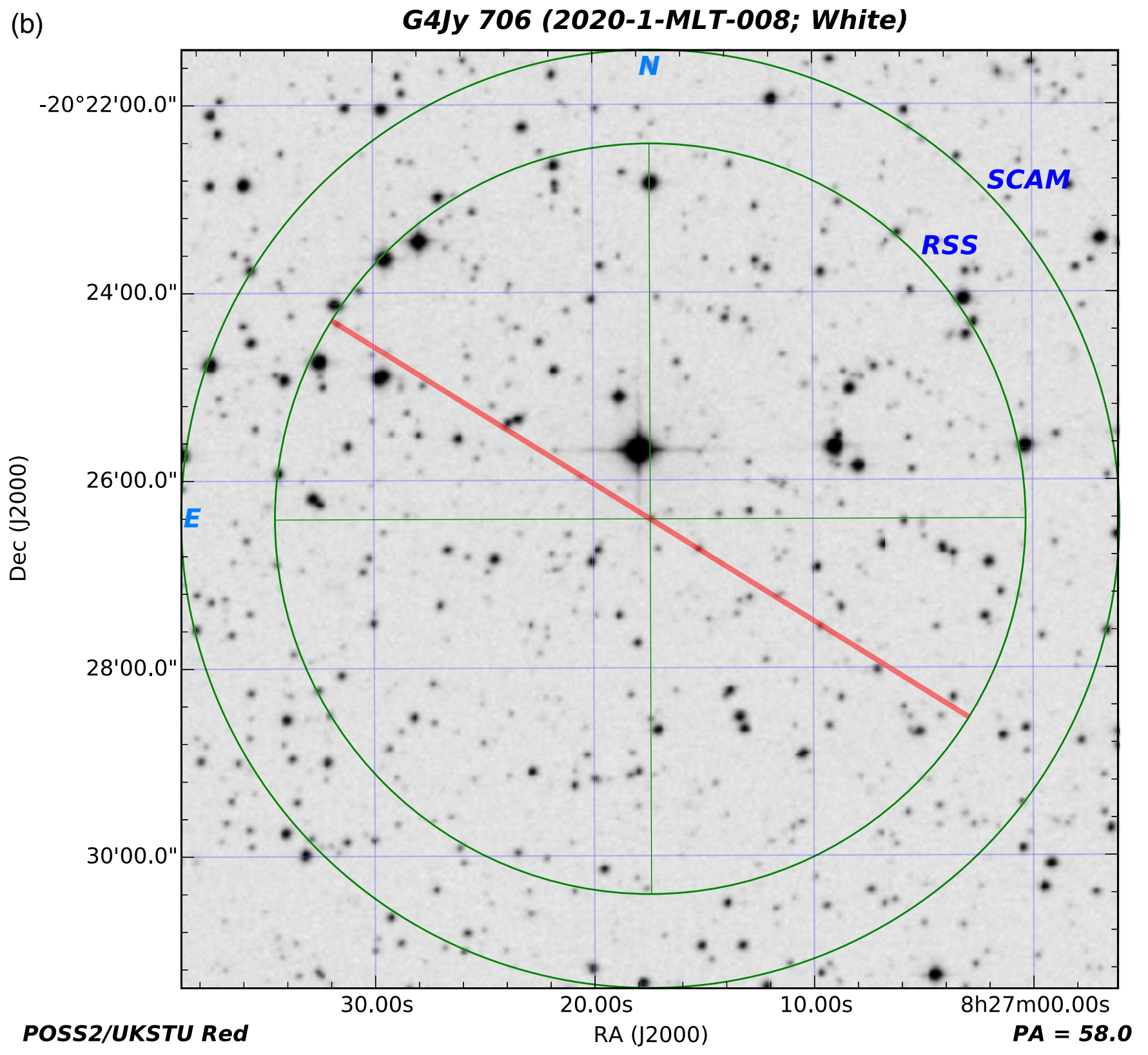

Finder Chart for G4Jy 706

**Figure B2.** Continued.

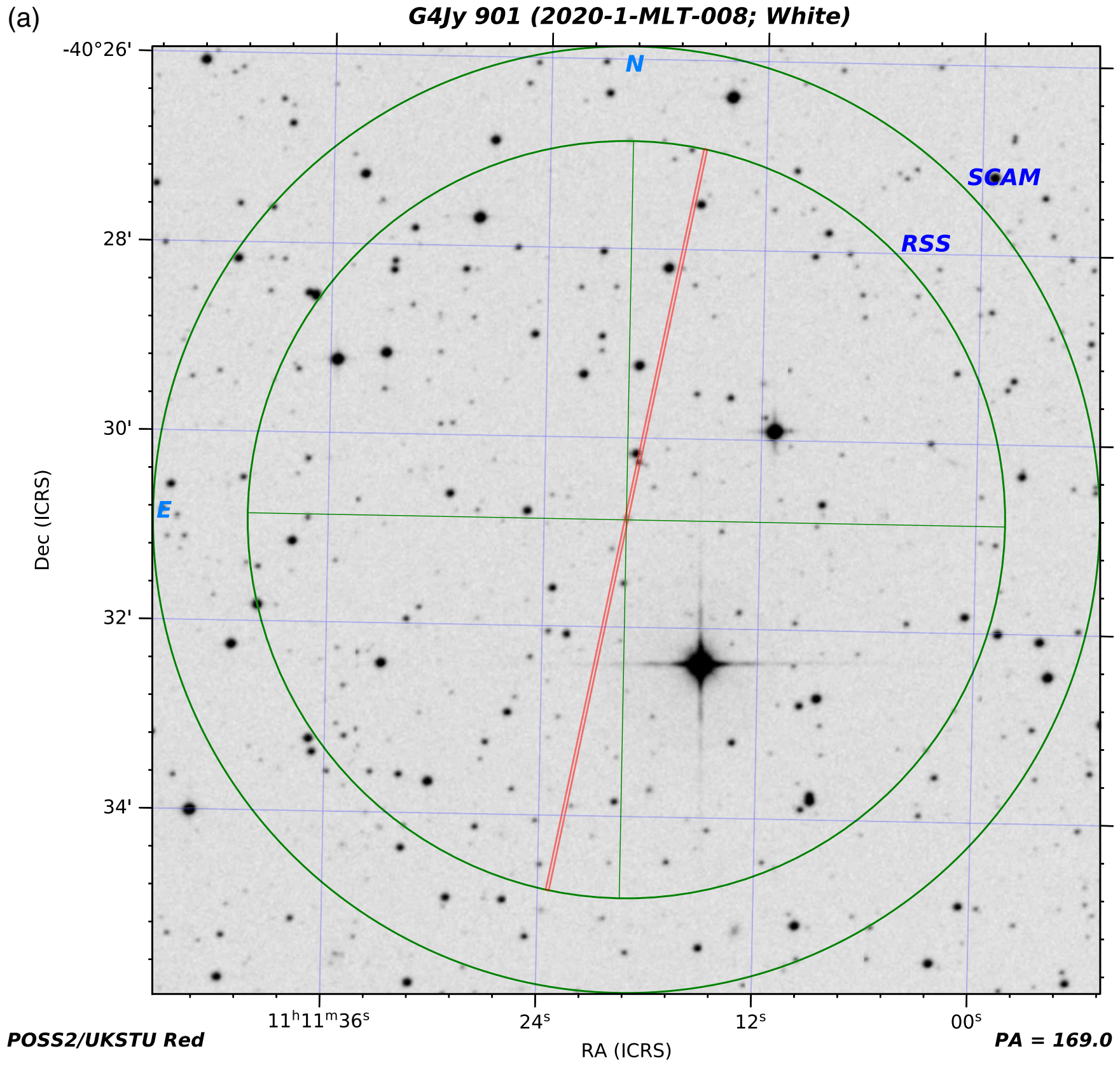

Finder Chart for G4Jy 901

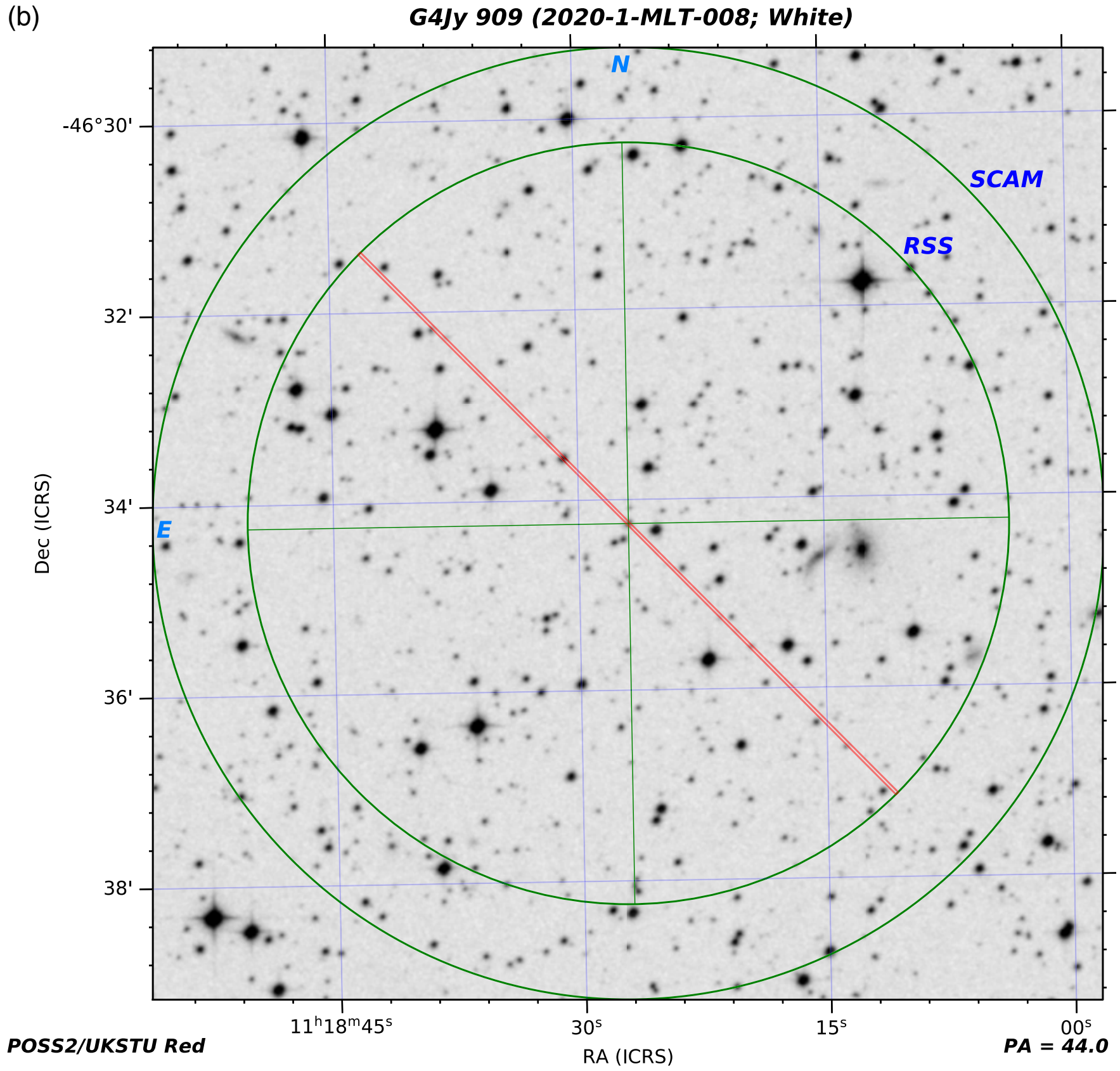

Finder Chart for G4Jy 909

**Figure B2.** Continued.

(a)

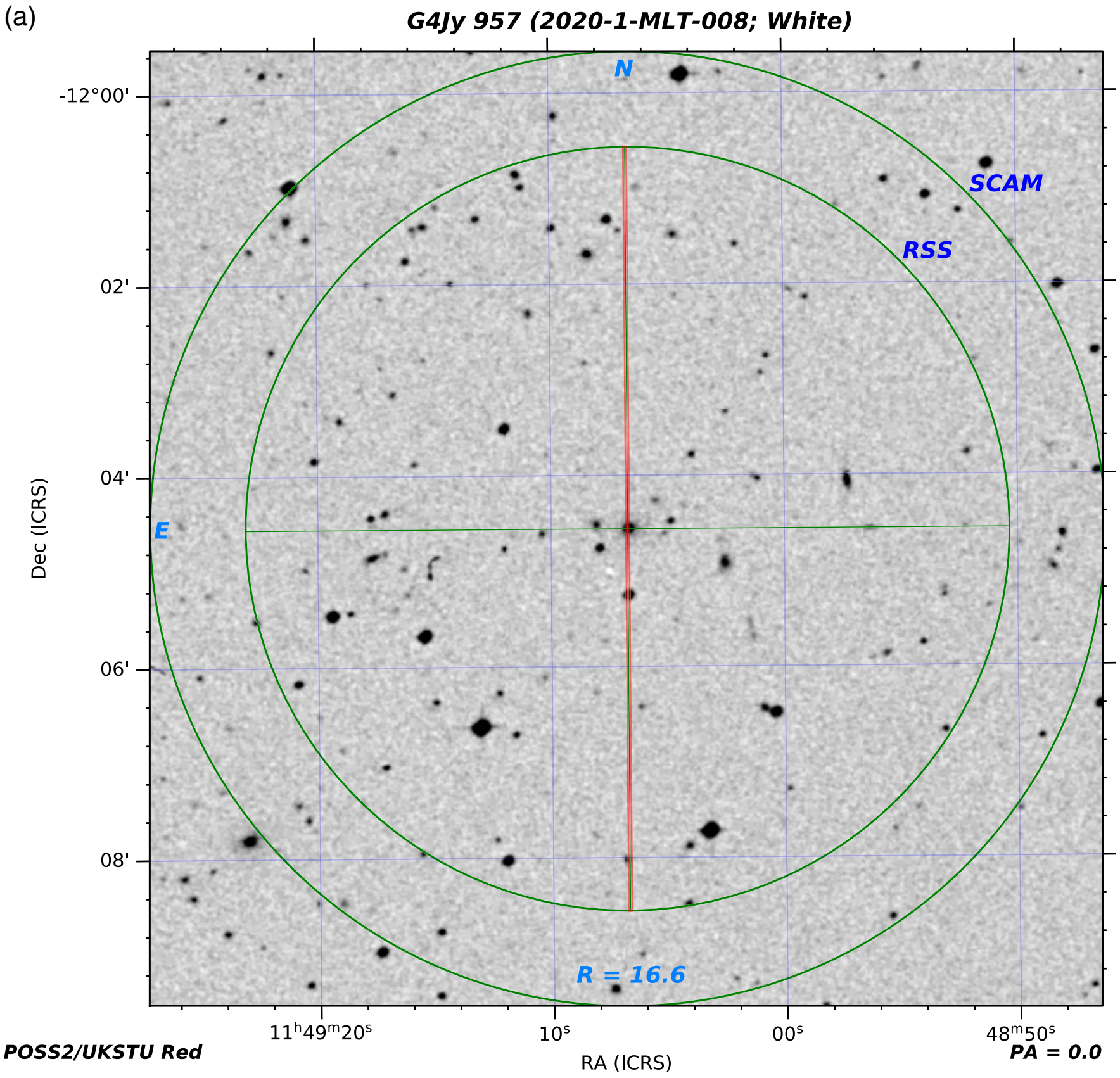

Finder Chart for G4Jy 957

(b)

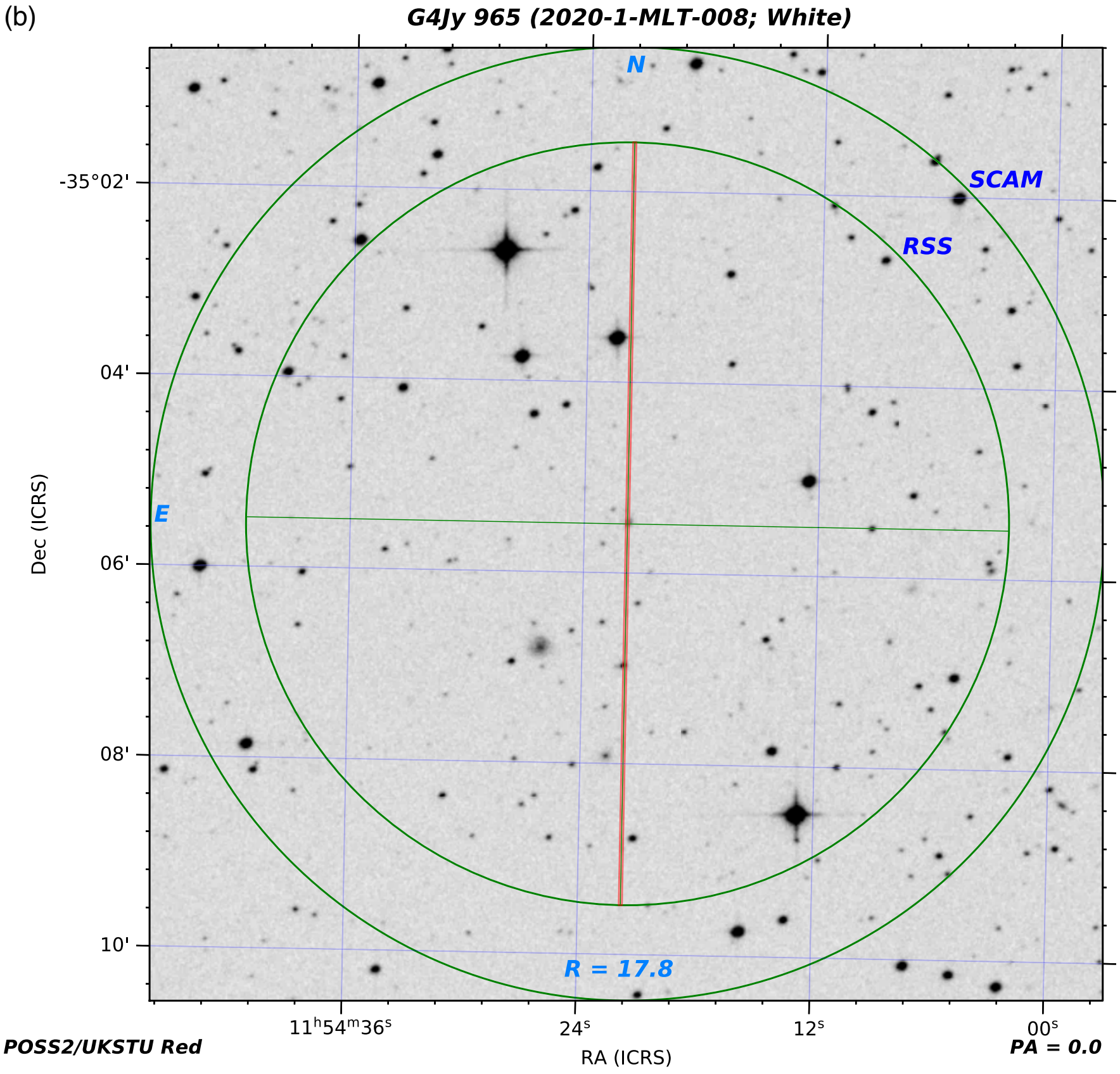

Finder Chart for G4Jy 965

**Figure B2.** Continued.

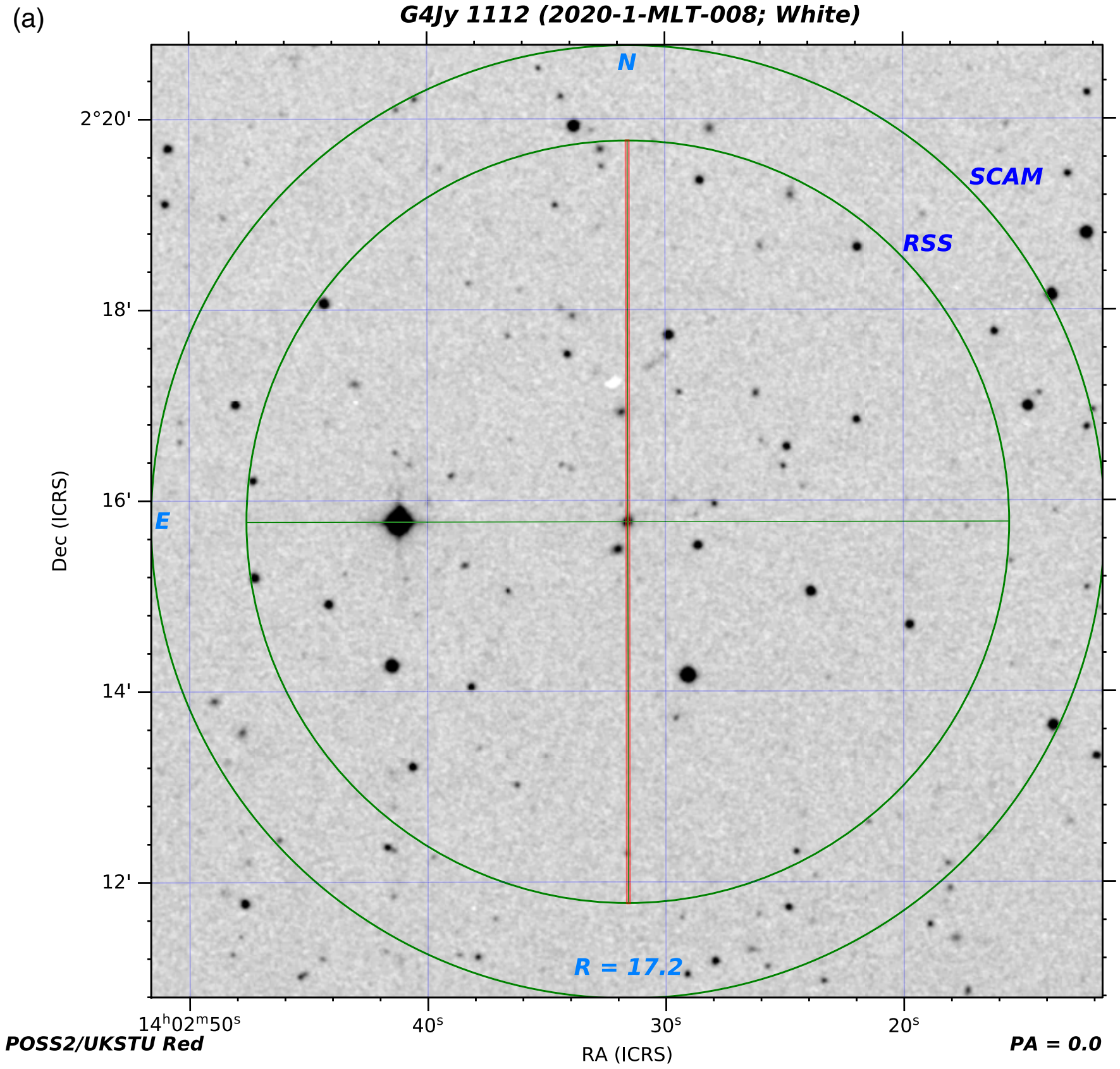

Finder Chart for G4Jy 1112

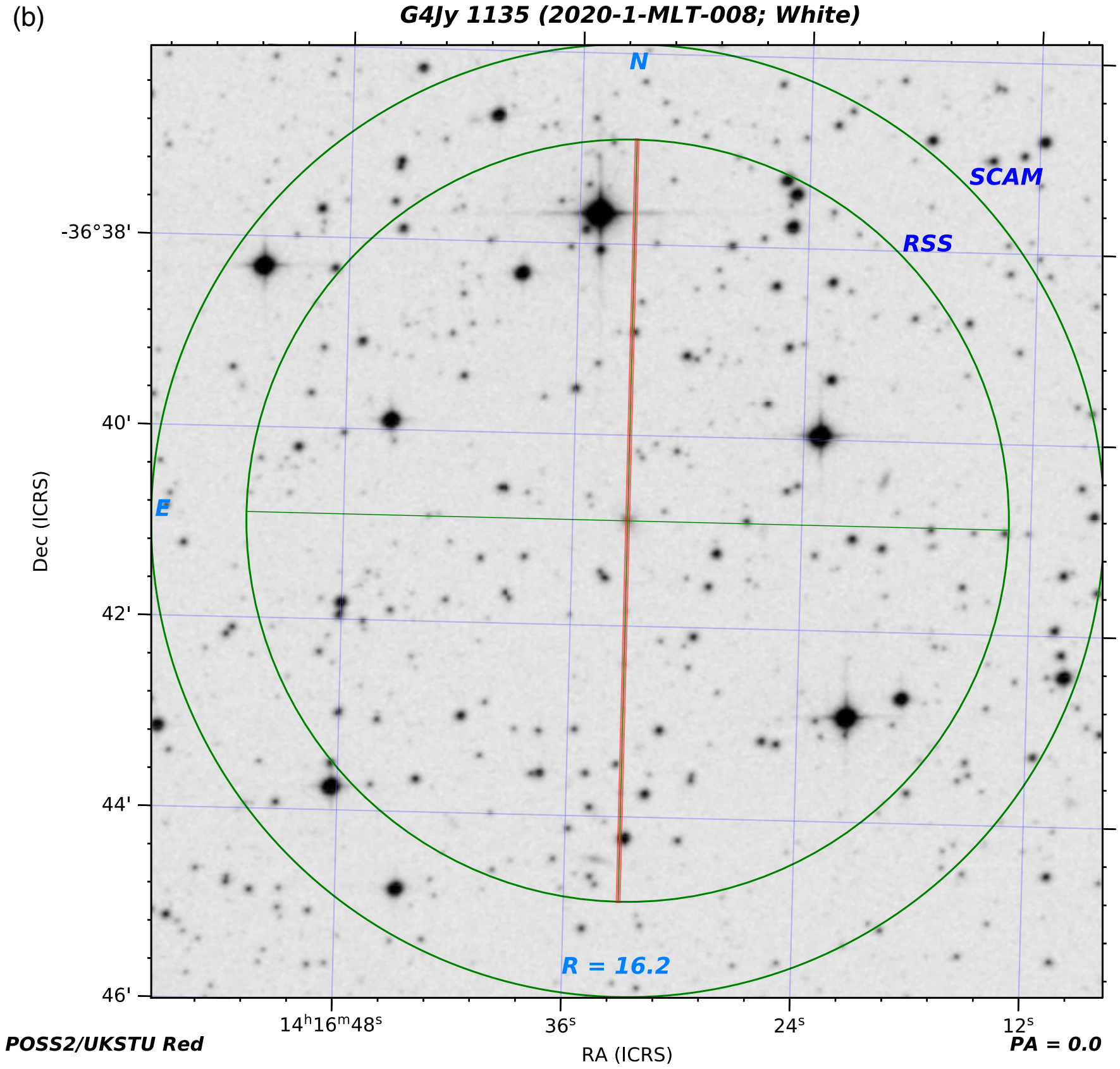

Finder Chart for G4Jy 1135

**Figure B2.** Continued.

(a)

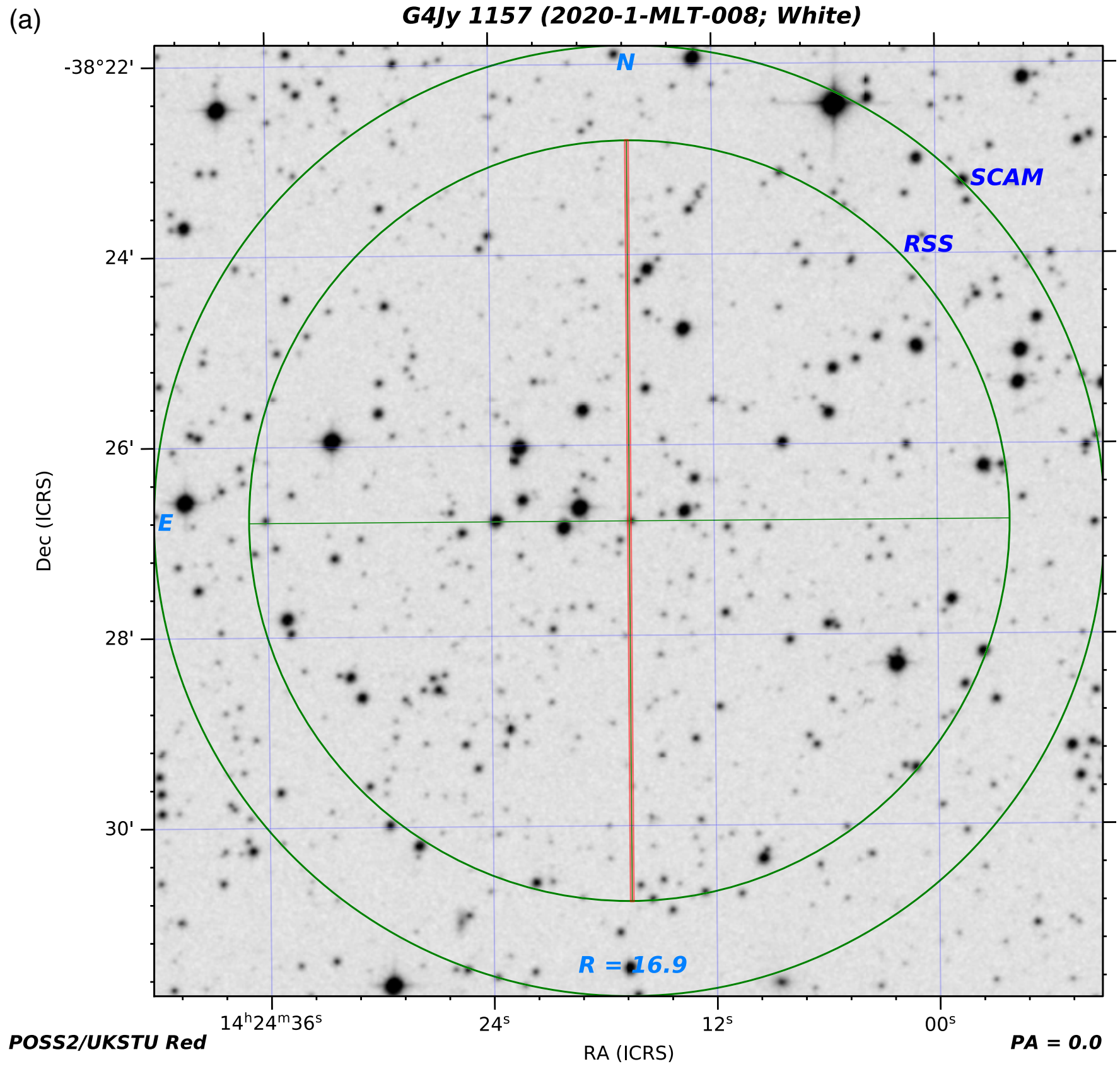

Finder Chart for G4Jy 1157

(b)

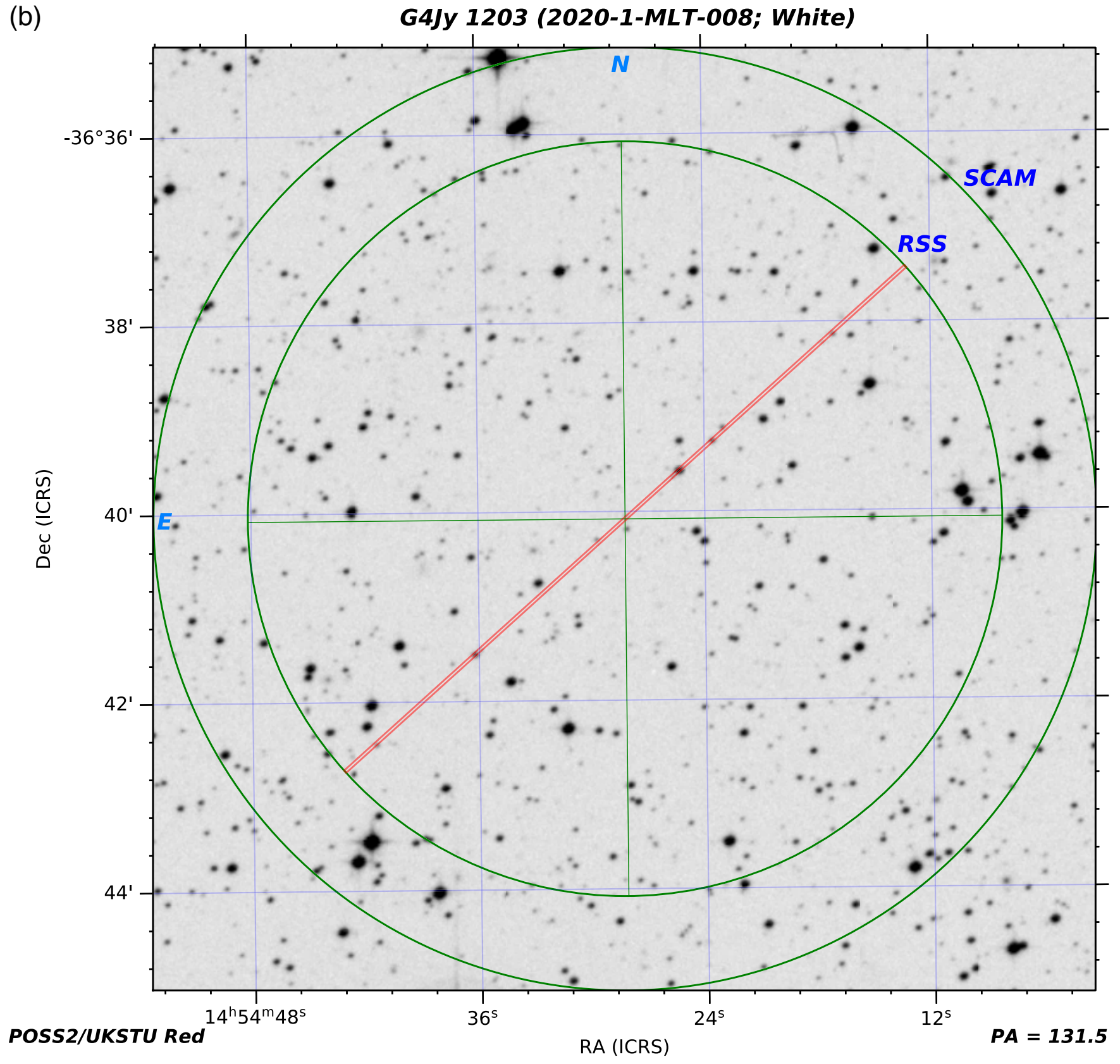

Finder Chart for G4Jy 1203

**Figure B2.** Continued.

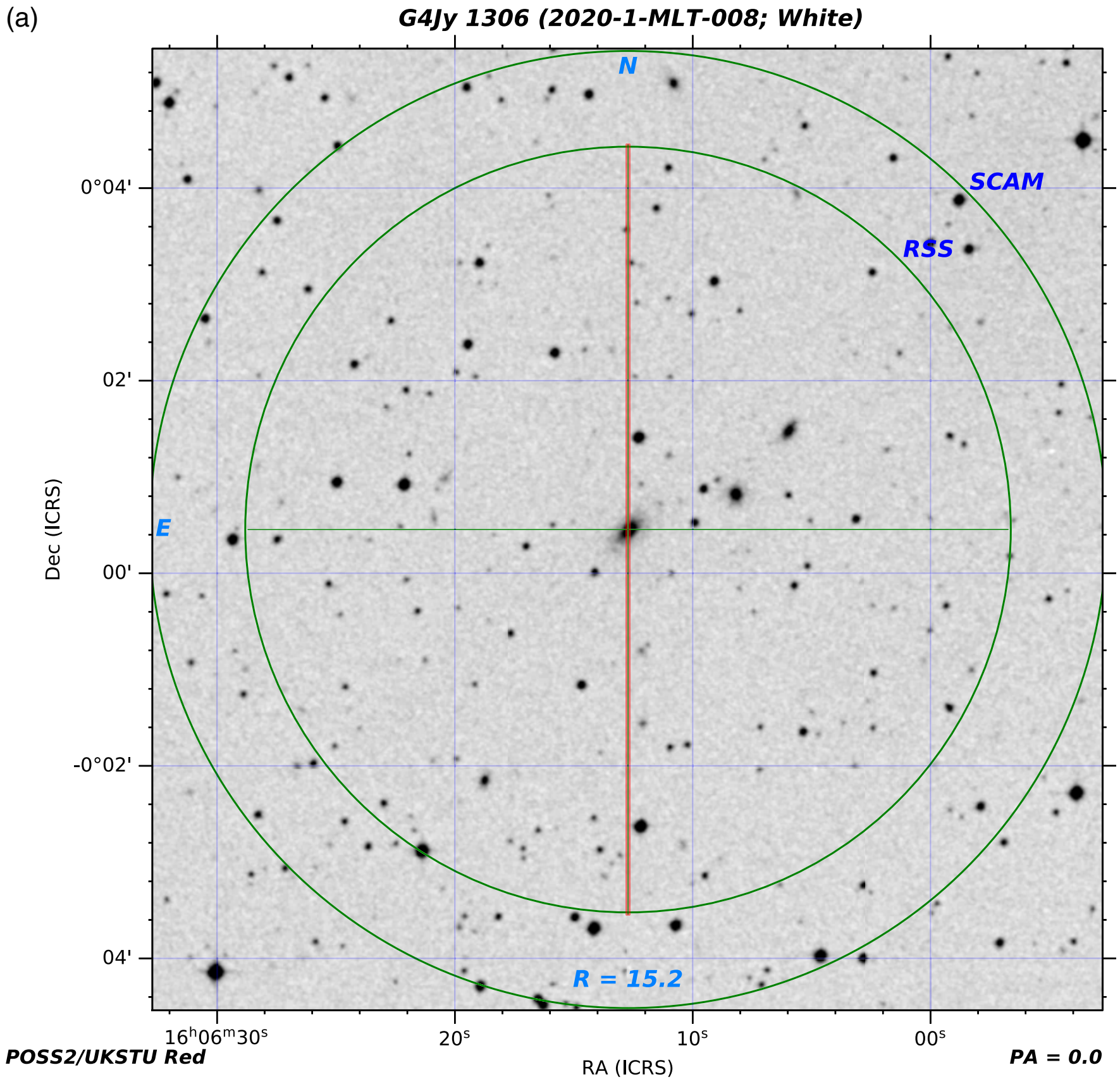

Finder Chart for G4Jy 1306

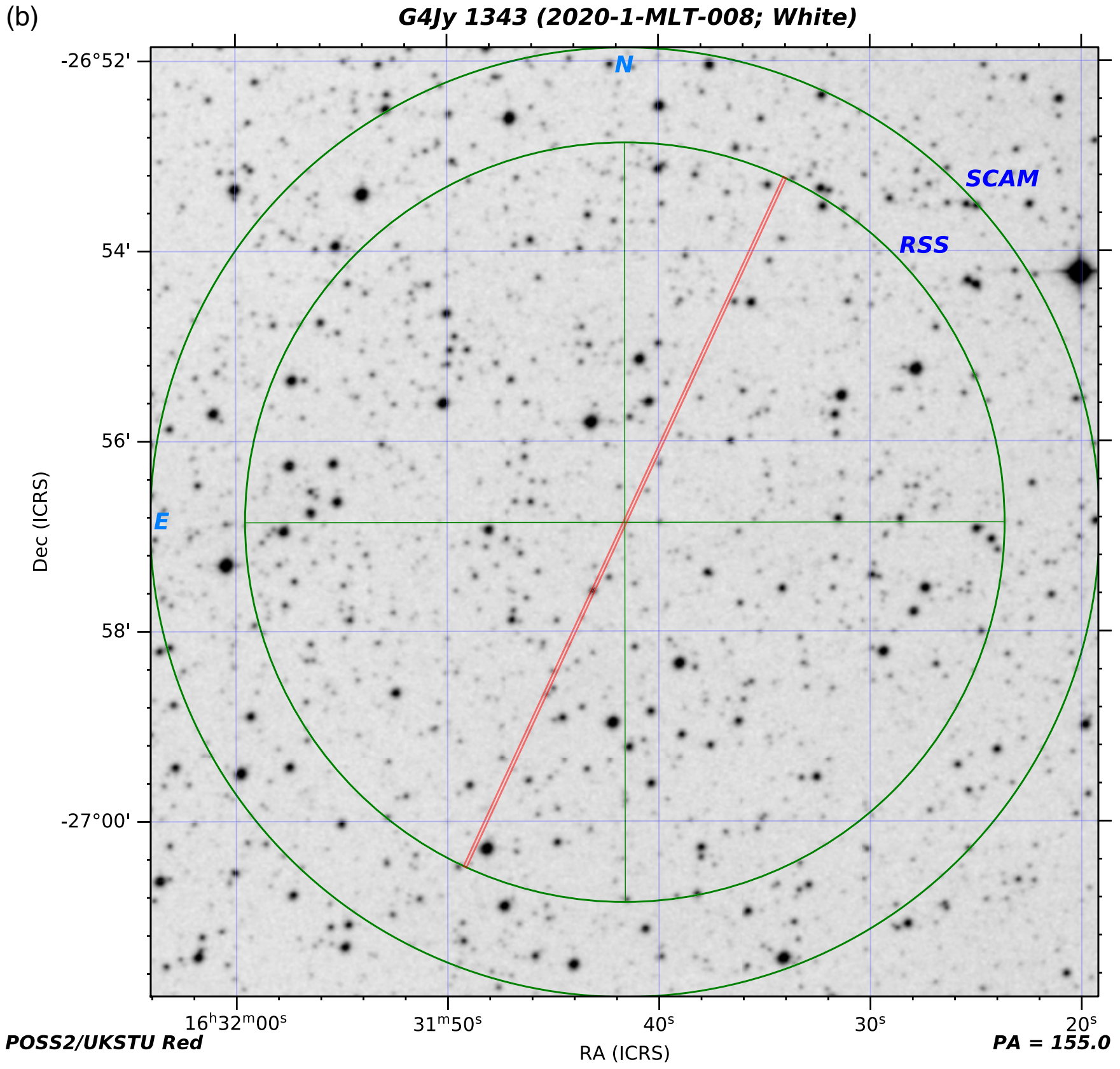

Finder Chart for G4Jy 1343

**Figure B2.** Continued.

(a)
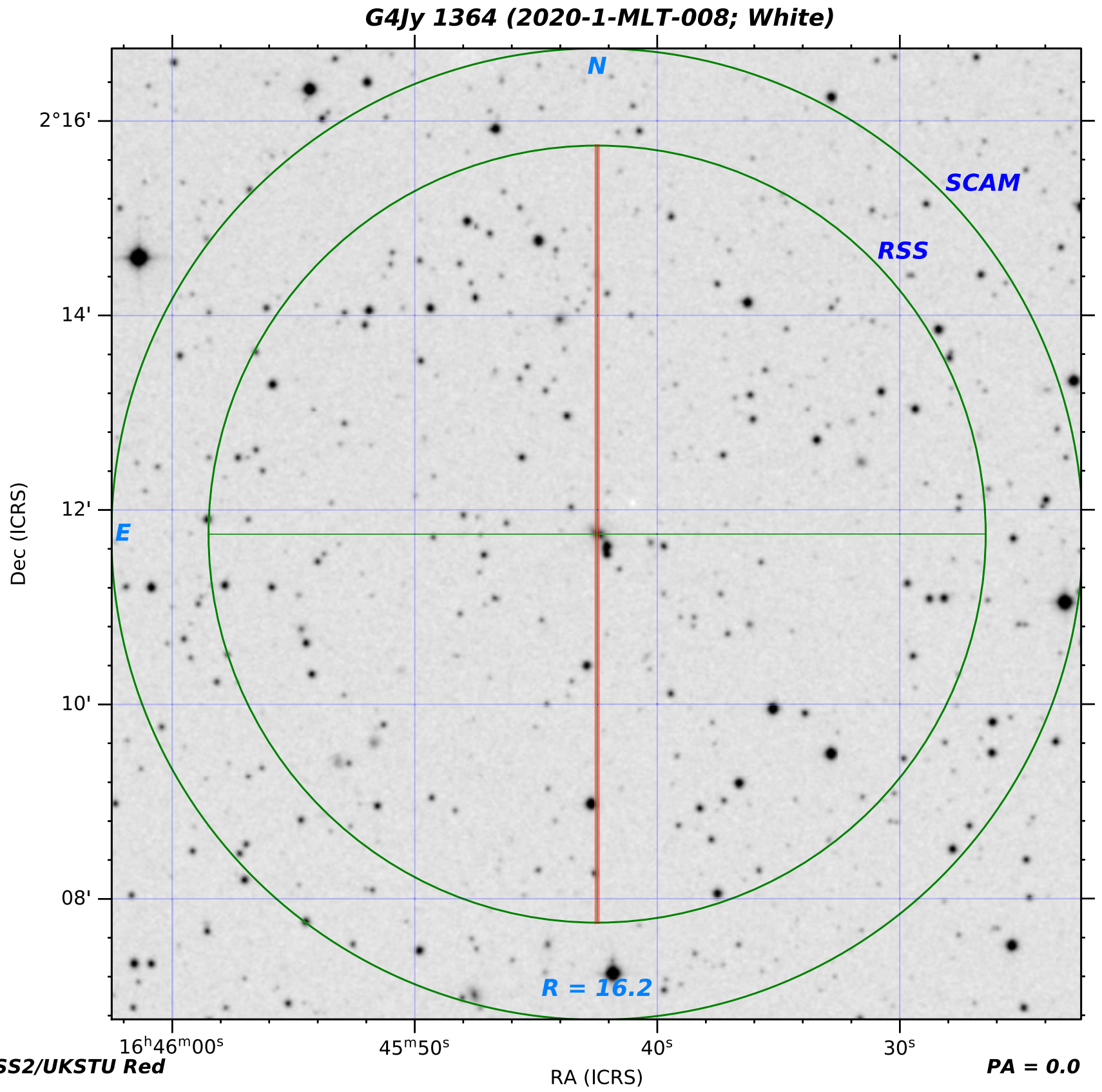

Finder Chart for G4Jy 1364

(b)
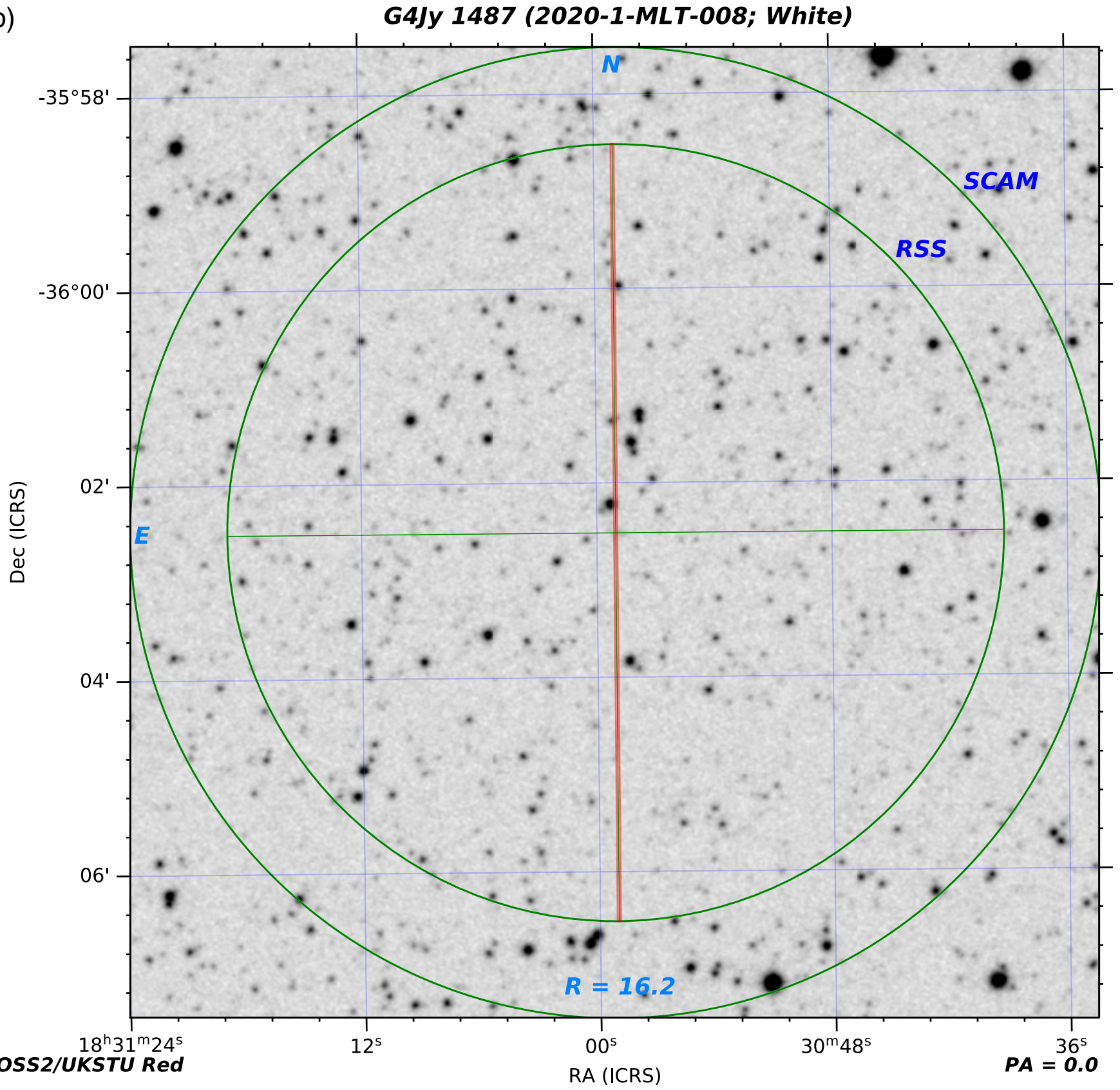

Finder Chart for G4Jy 1487

**Figure B2.** Continued.

(a)

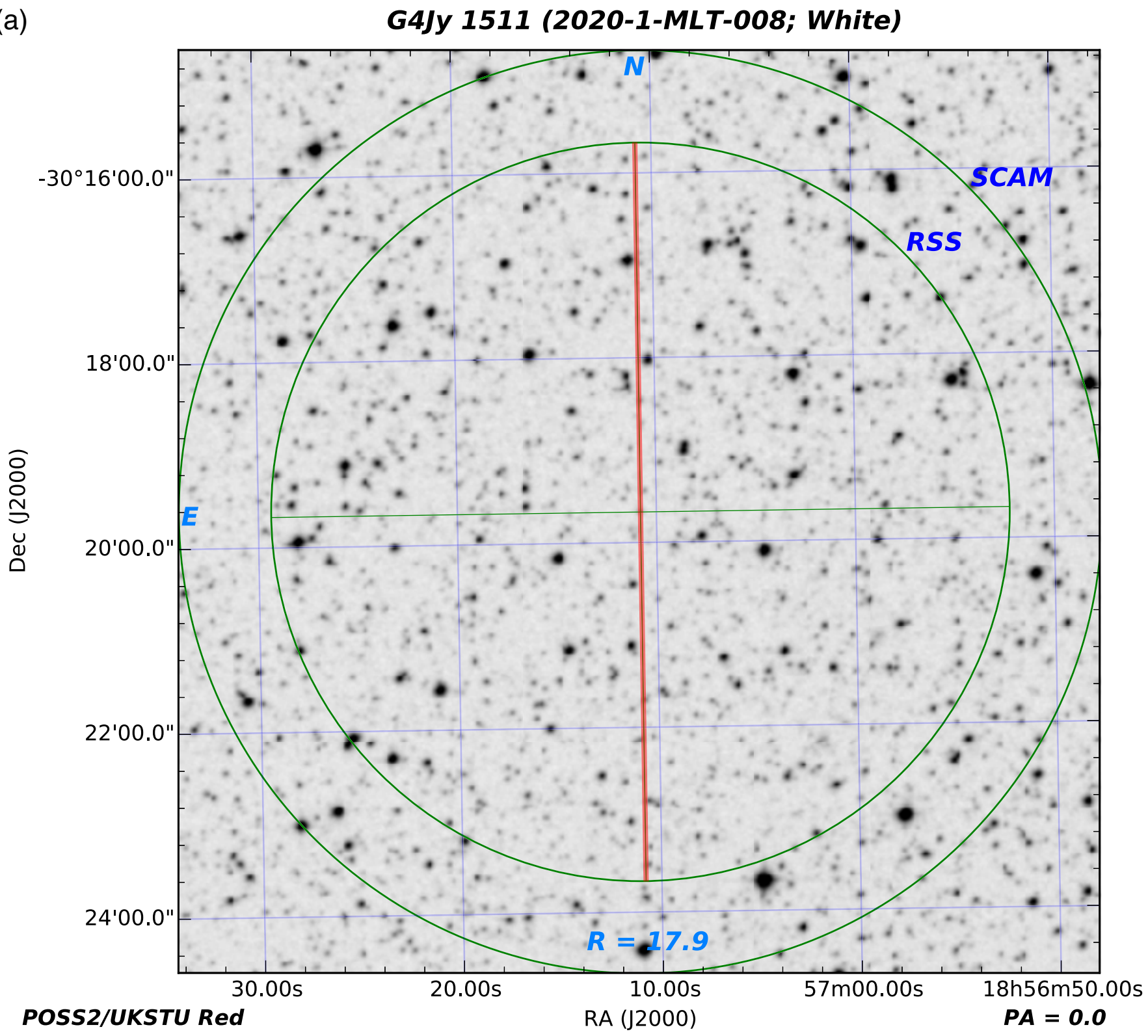

Finder Chart for G4Jy 1511

(b)

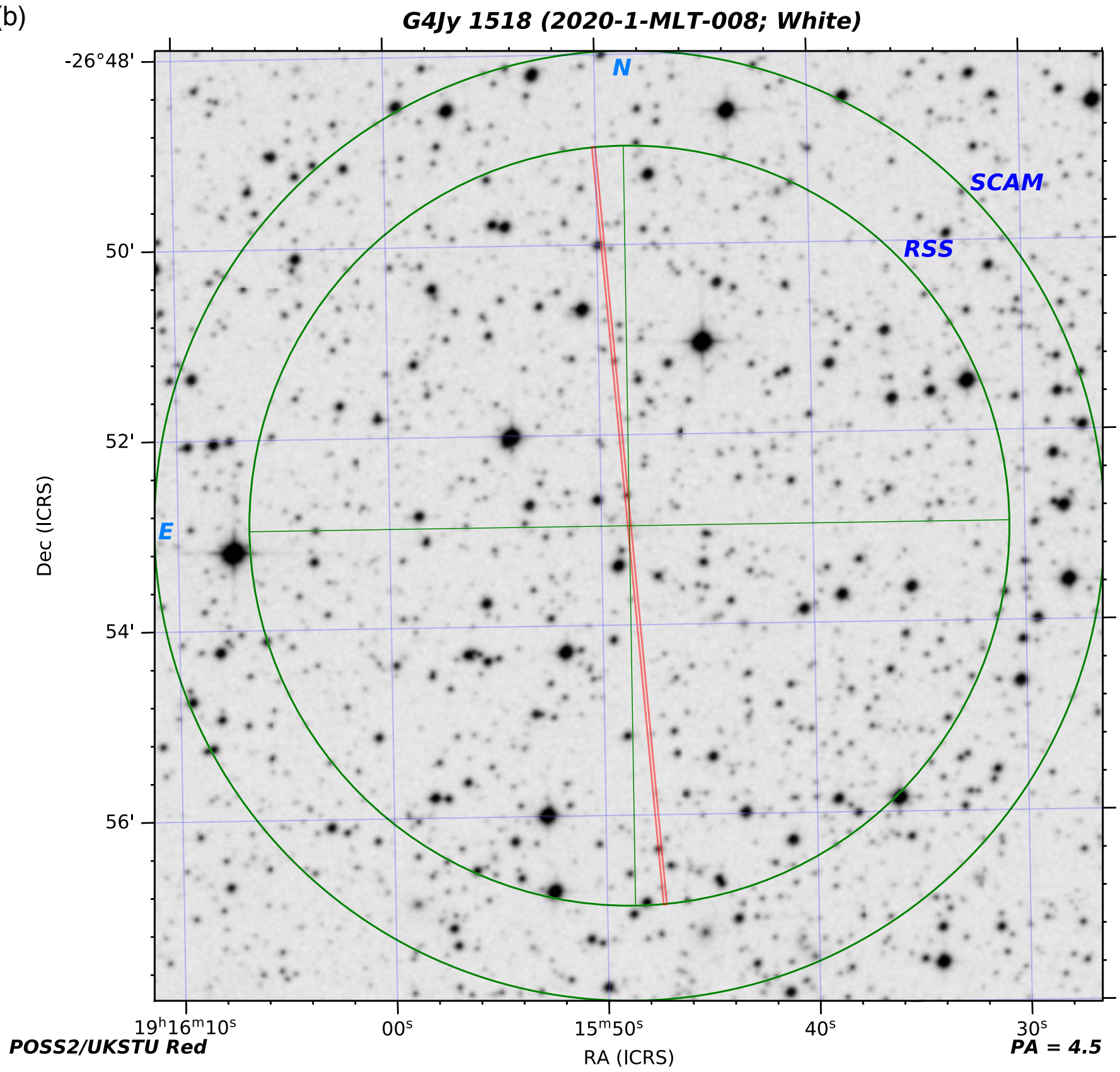

Finder Chart for G4Jy 1518

**Figure B2.** Continued.

(a)

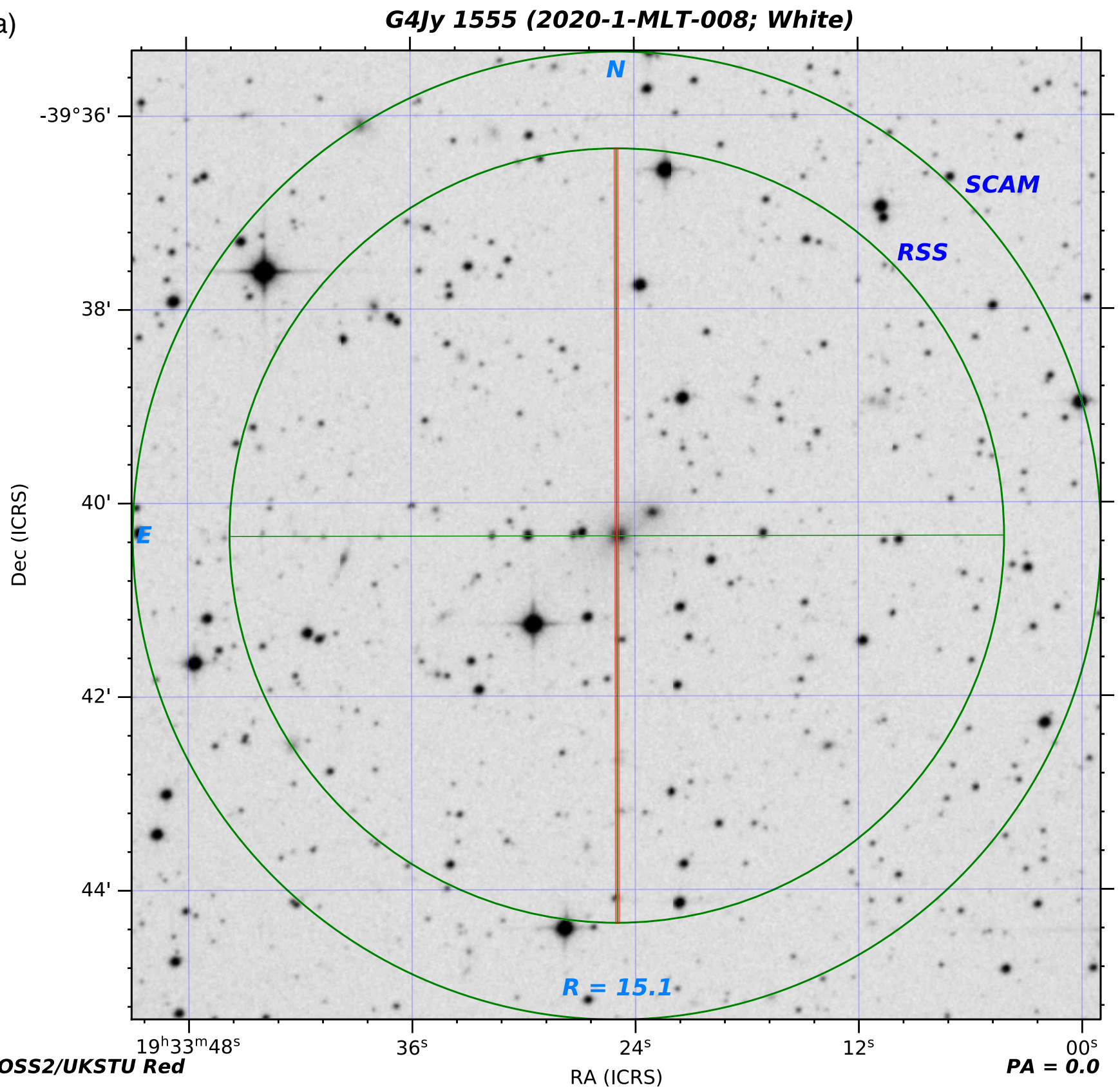

Finder Chart for G4Jy 1555

(b)

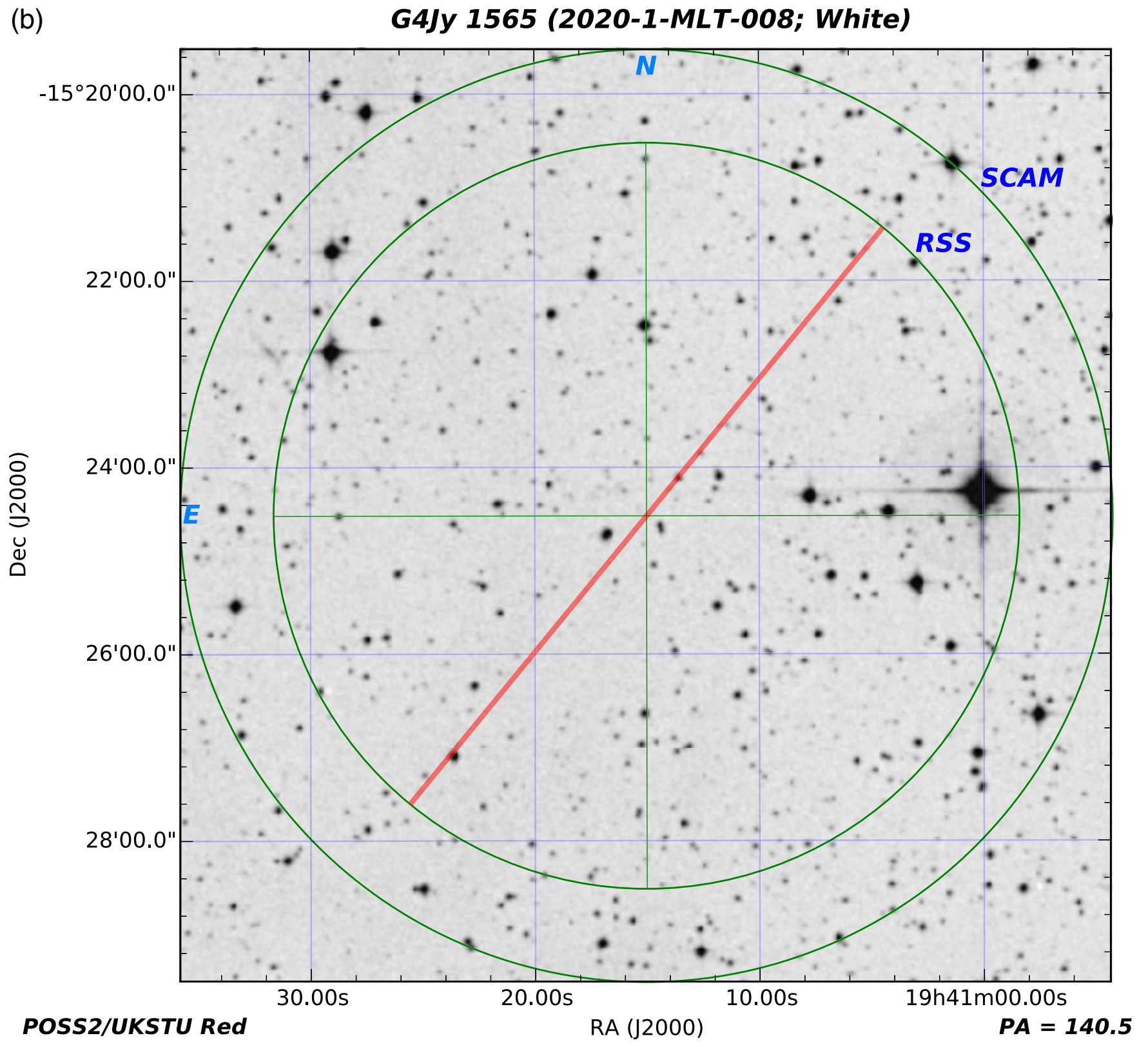

Finder Chart for G4Jy 1565

**Figure B2.** Continued.

(a)

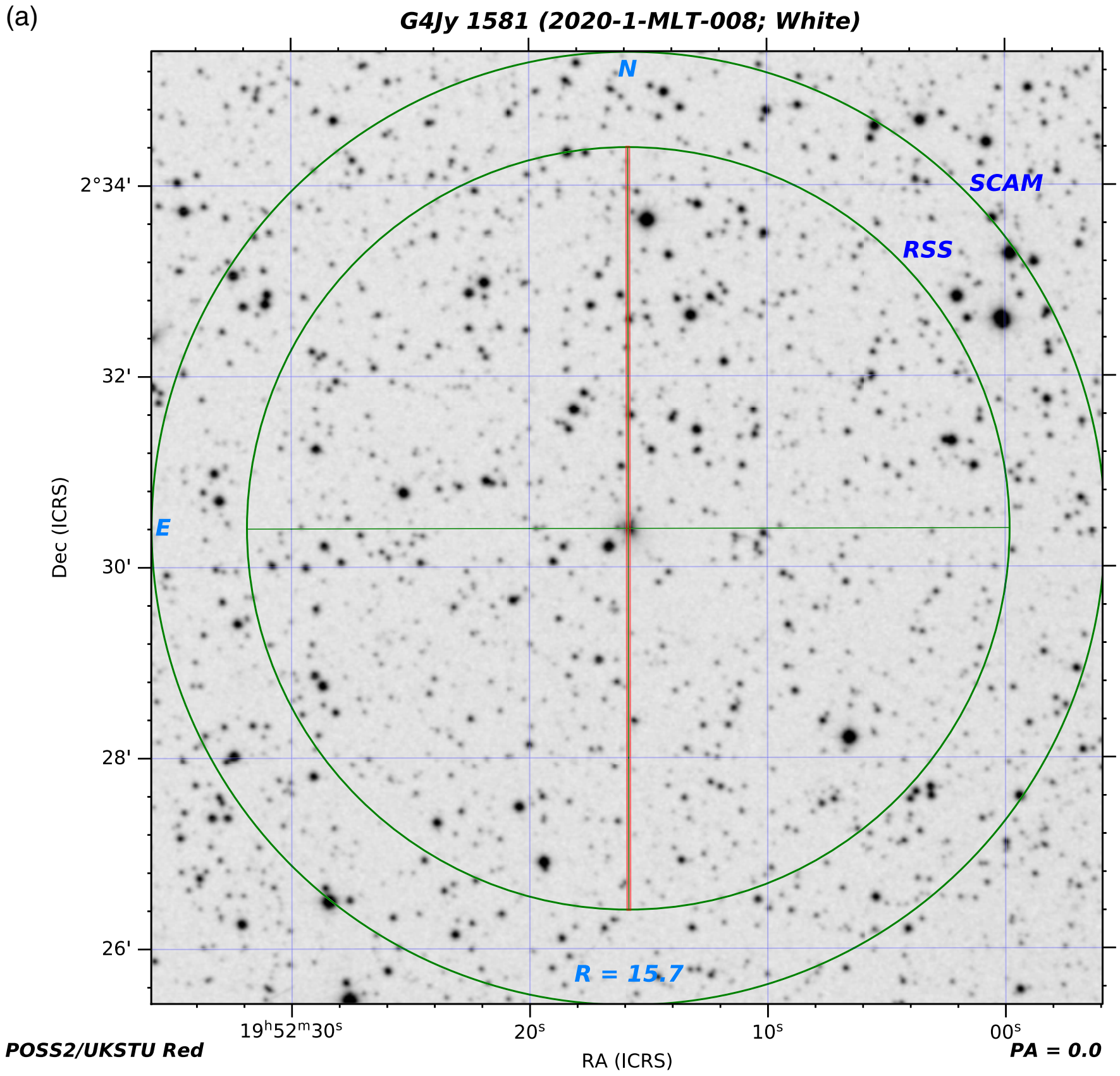

Finder Chart for G4Jy 1581

(b)

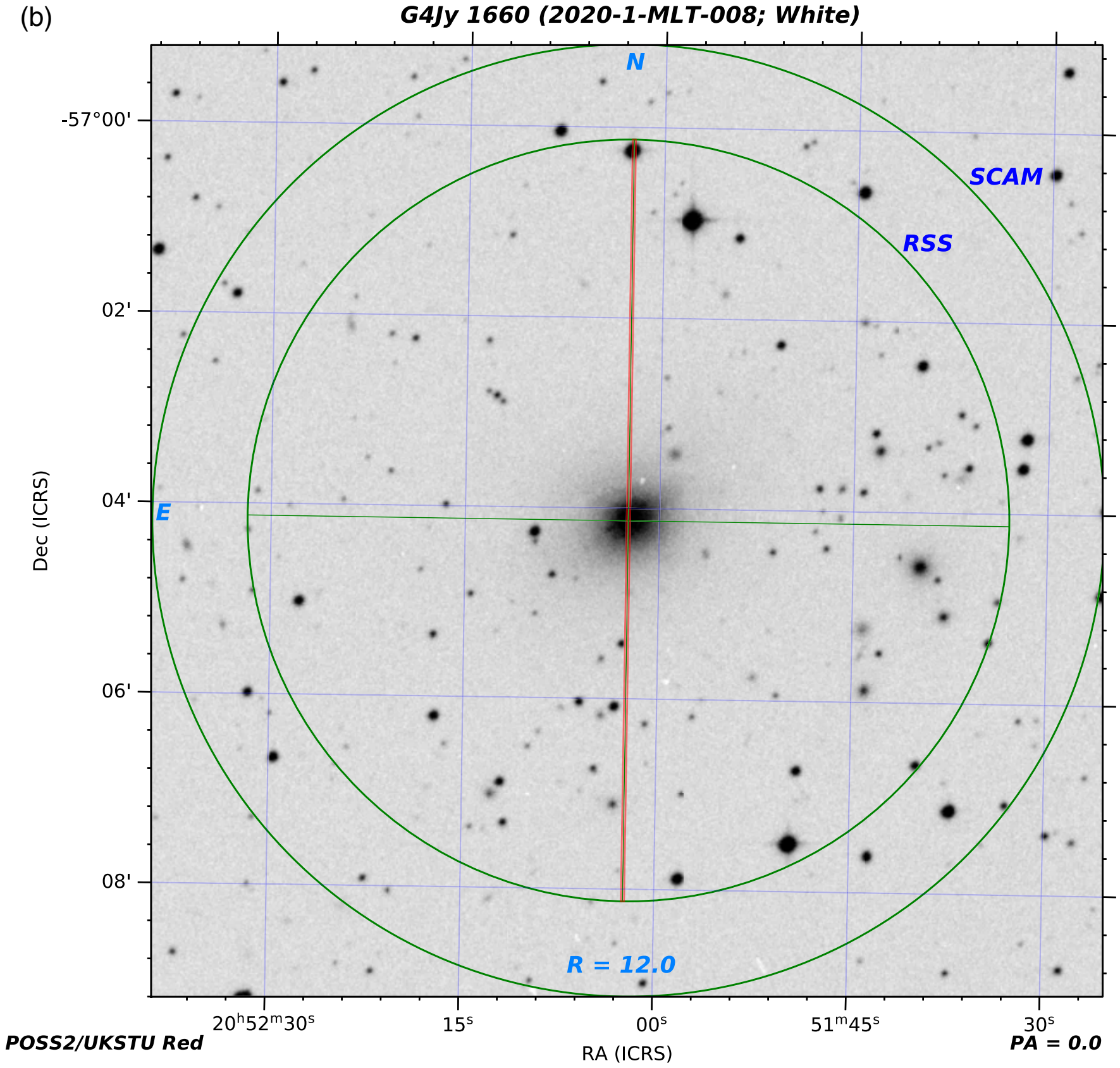

Finder Chart for G4Jy 1660

**Figure B2.** Continued.

(a)

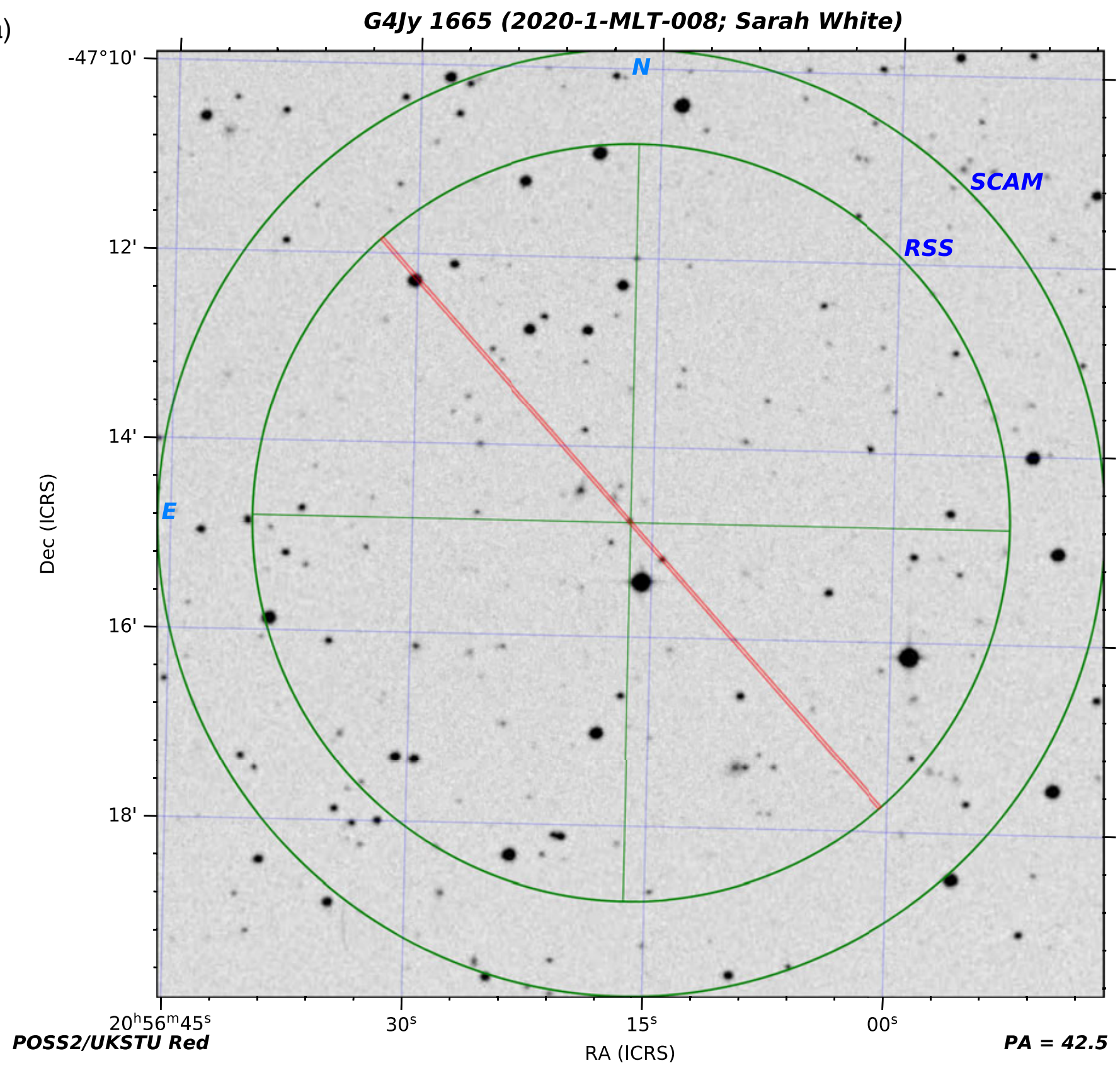

Finder Chart for G4Jy 1665

(b)

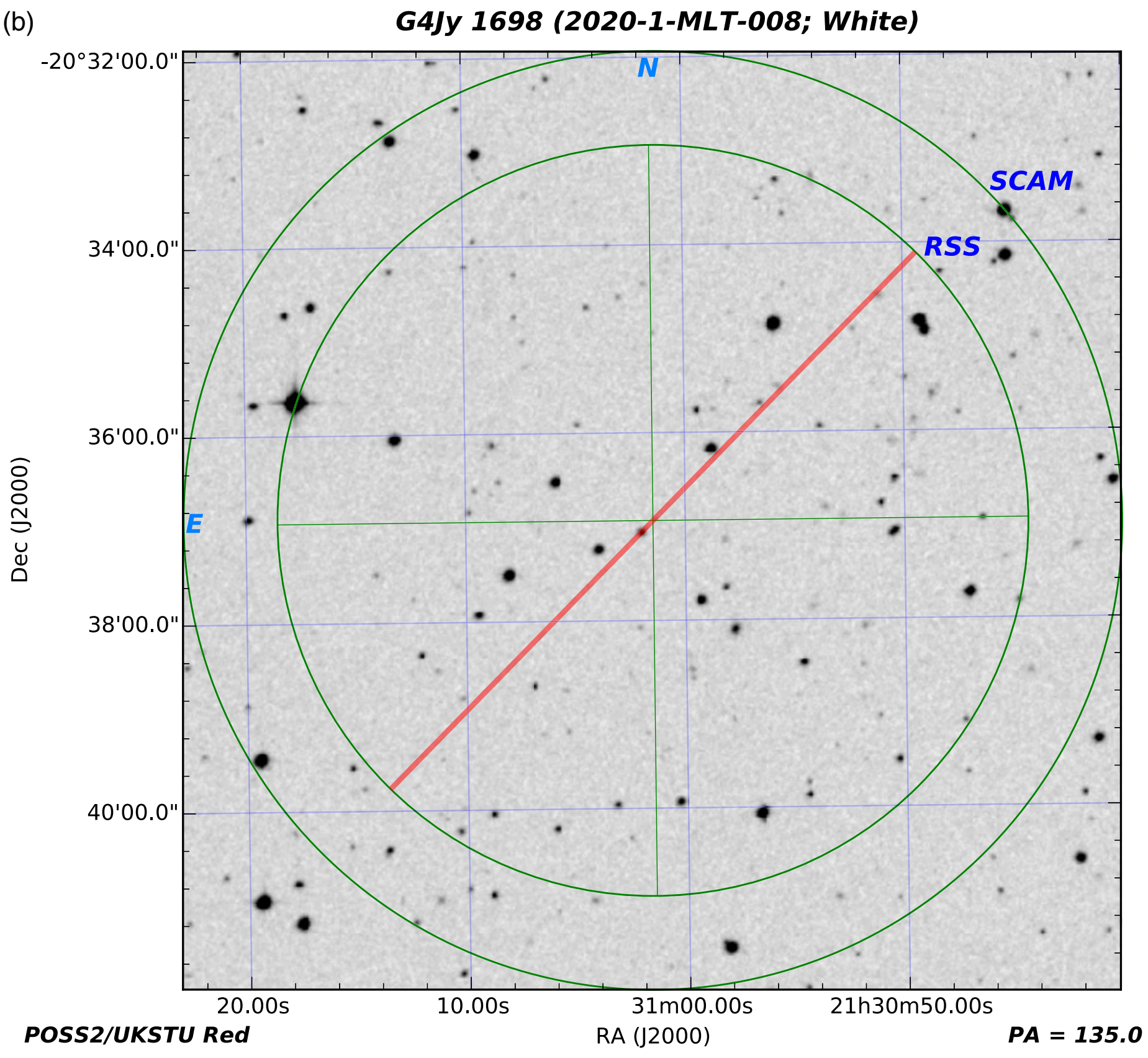

Finder Chart for G4Jy 1698

**Figure B2.** Continued.

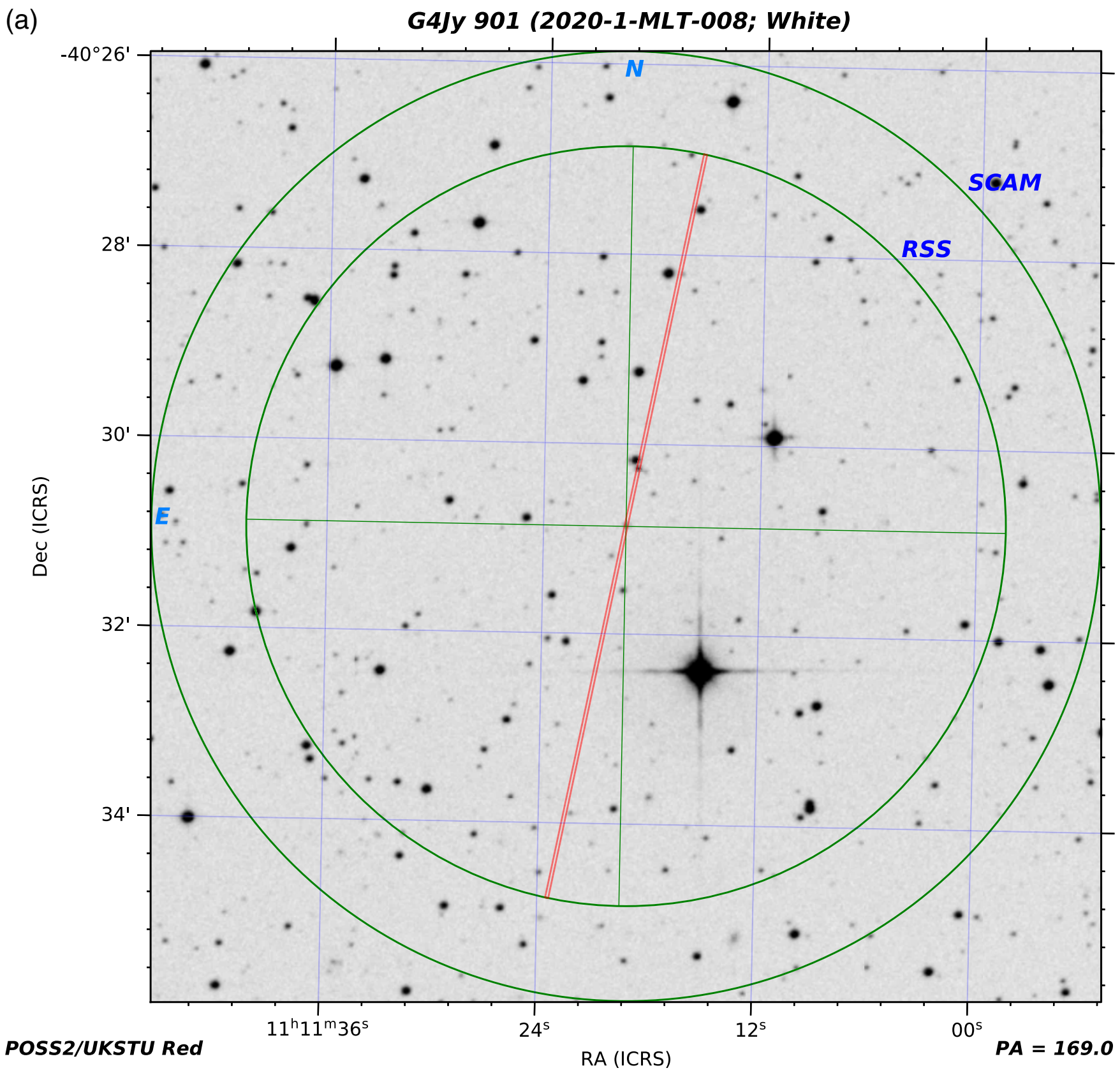

Finder Chart for G4Jy 1704

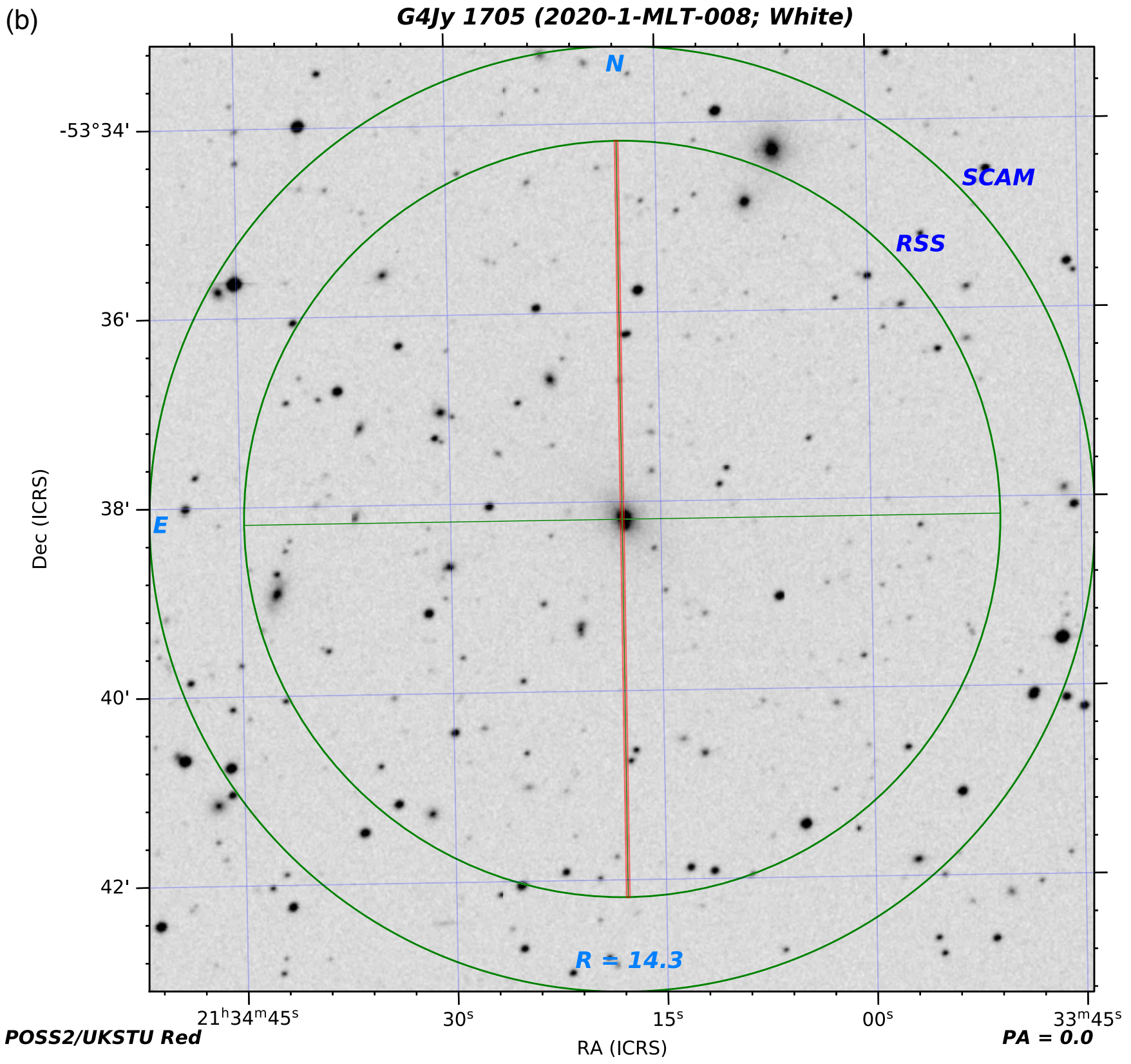

Finder Chart for G4Jy 1705

**Figure B2.** Continued.

(a)

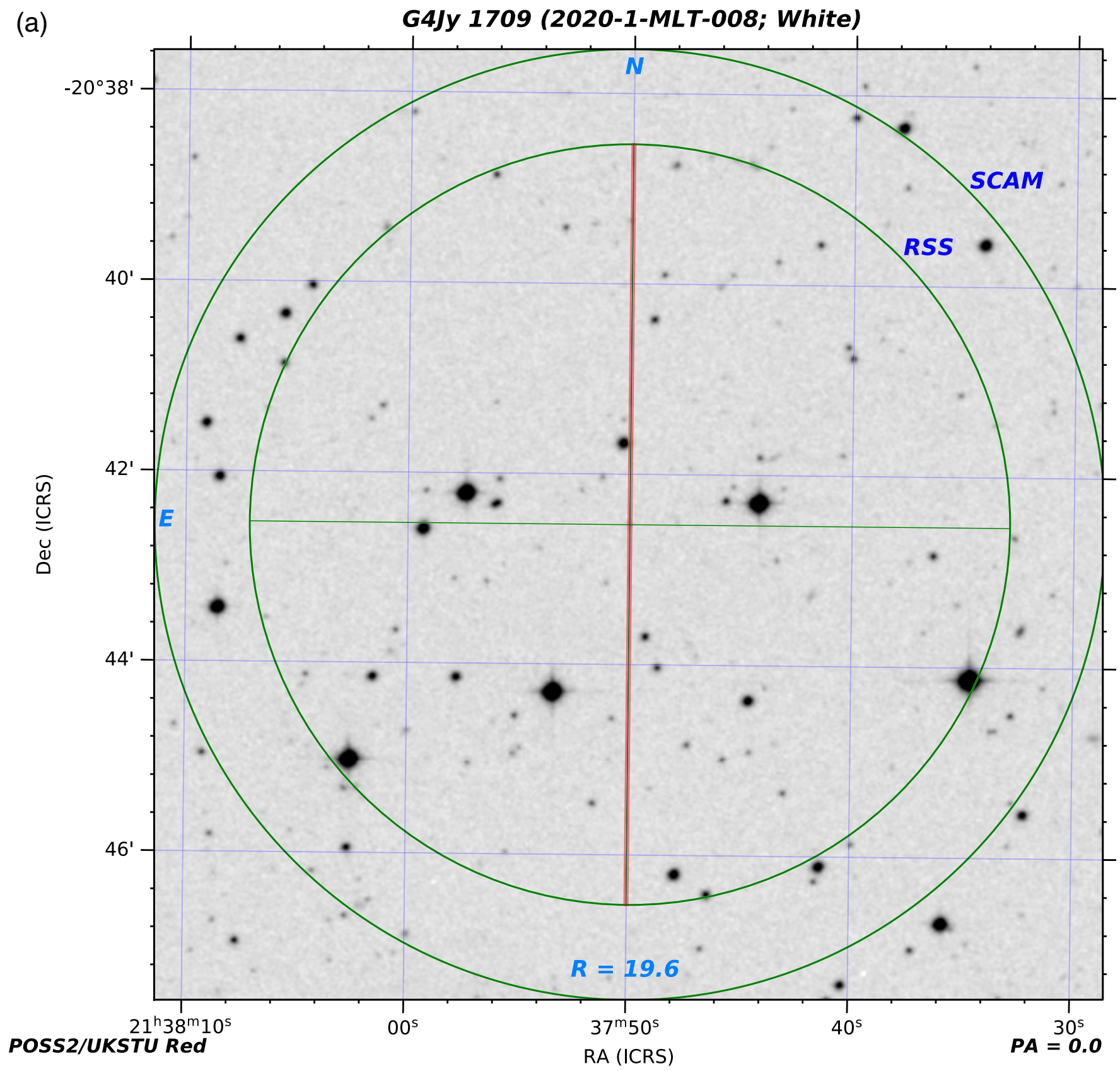

Finder Chart for G4Jy 1709

(b)

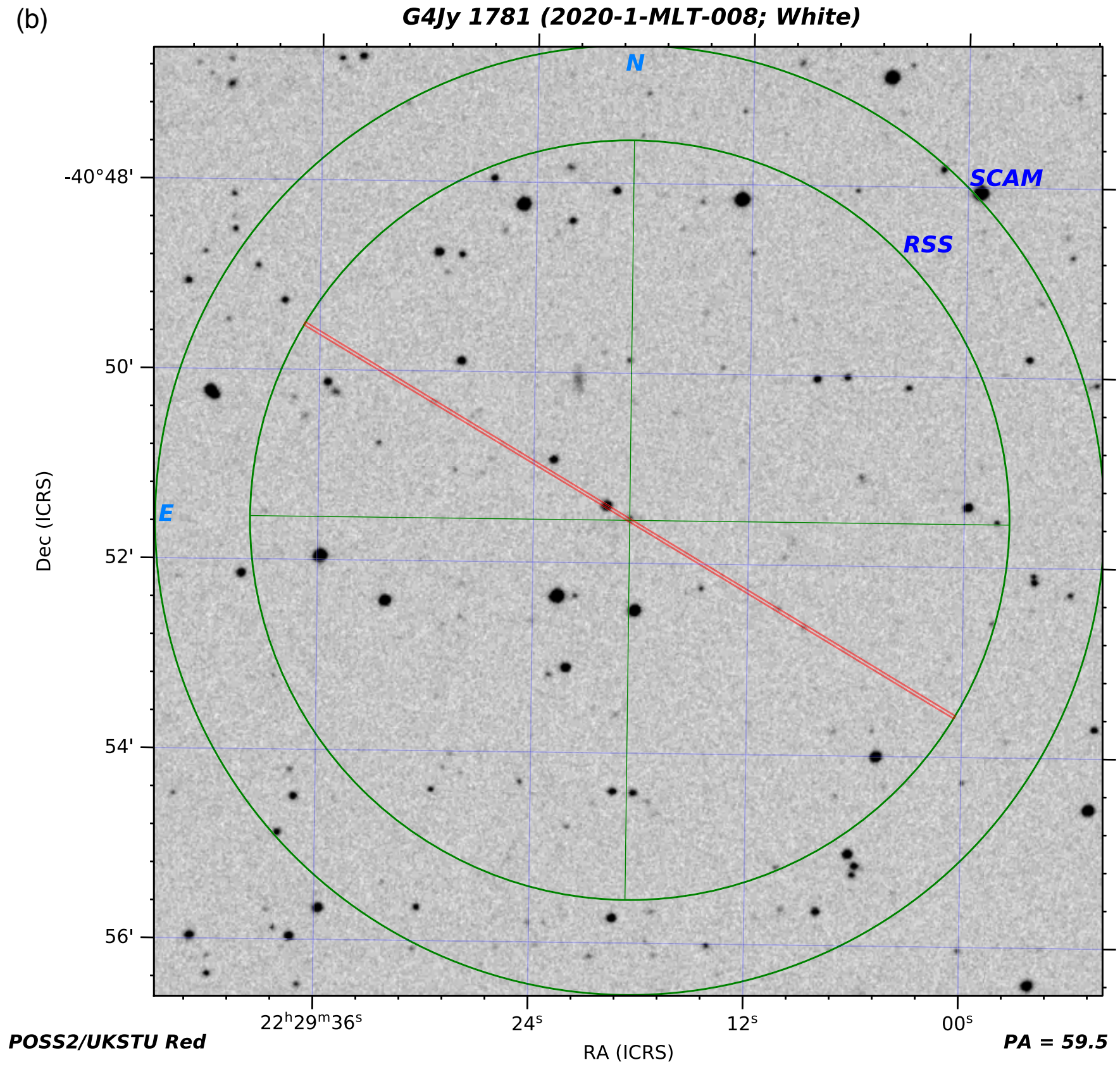

Finder Chart for G4Jy 1781

**Figure B2.** Continued.

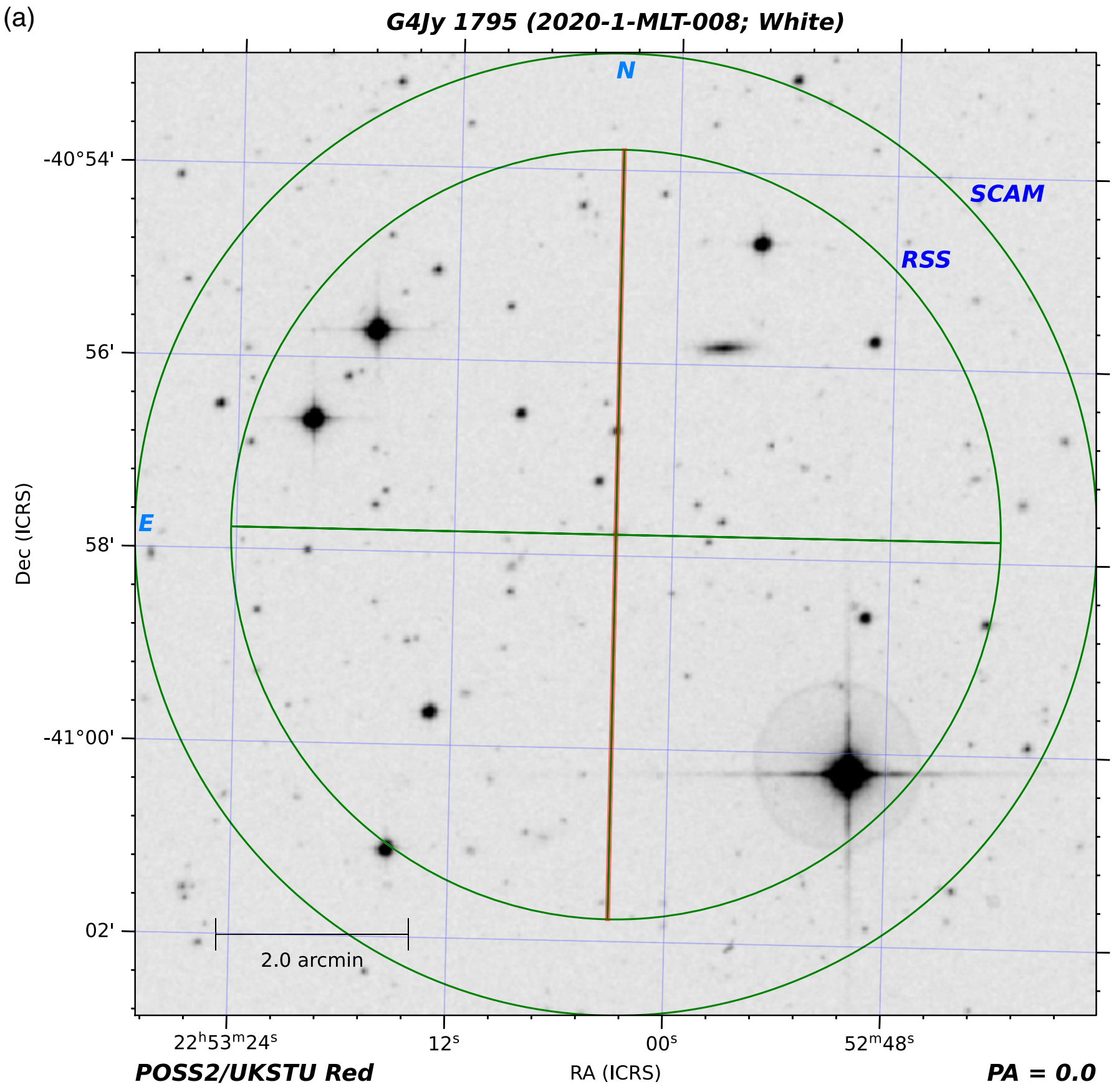

Finder Chart for G4Jy 1795

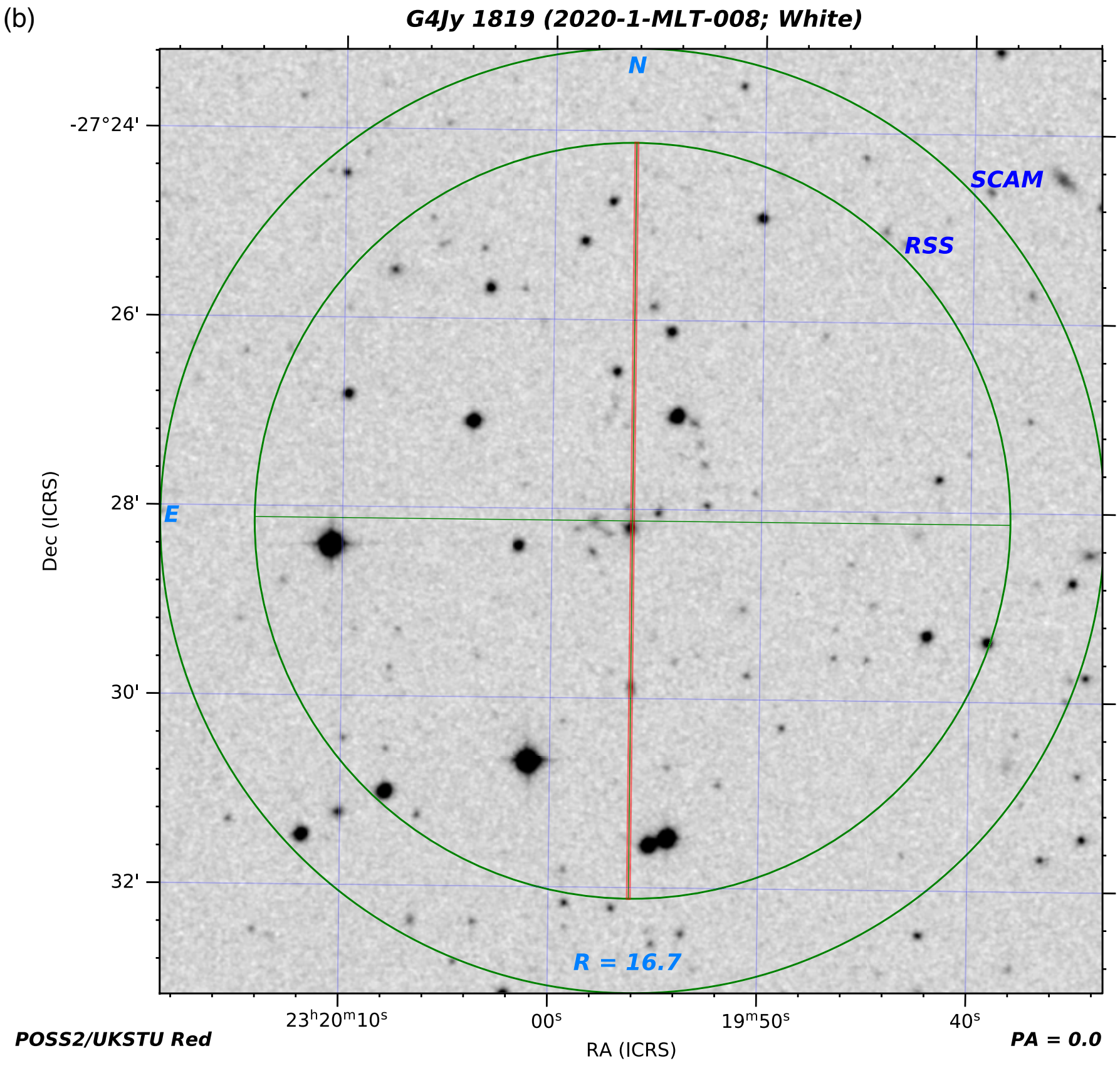

Finder Chart for G4Jy 1819

**Figure B2.** Continued.

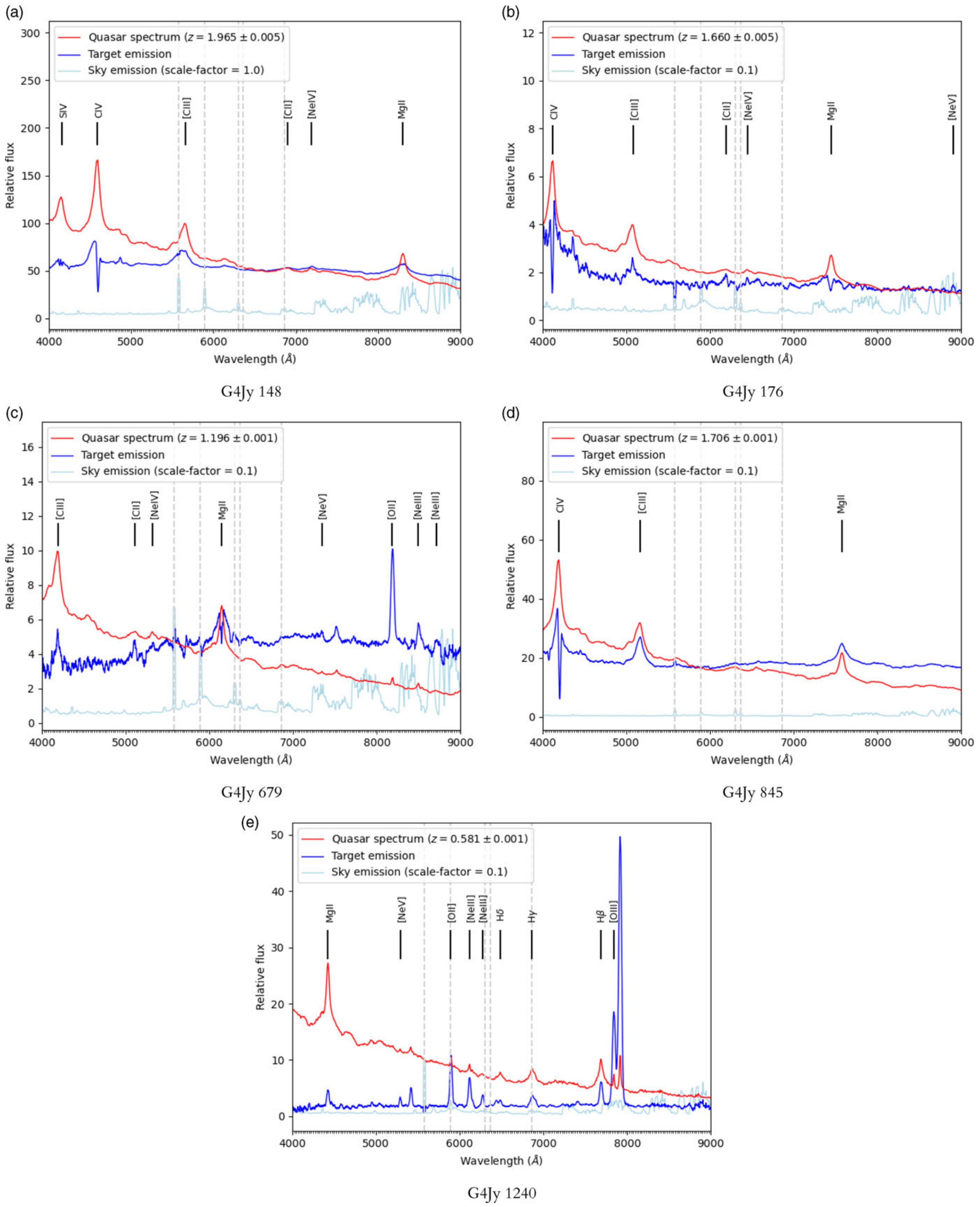

G4Jy 148

G4Jy 176

G4Jy 679

G4Jy 845

G4Jy 1240

**Figure B3.** SDSS spectra (blue lines) of G4Jy sources (Section 3 and 7), with sky-emission spectra represented by lighter-blue lines. (The latter is scaled to aid comparison with the target emission, and the scale factor that has been applied is noted in each legend.) The dashed, grey, vertical lines indicate prominent sky-emission, and the target spectrum is re-fitted with the appropriate template spectrum (red lines).

## Appendix C. Redshifts and radio properties

Table C1 presents observed and *intrinsic* radio properties for 299 G4Jy sources, having collated their redshifts from SALT proposal 2020-1-MLT-008 (PI: White, this work), 6dFGS (Jones et al. 2009), SDSS DR12 (Alam et al. 2015), and SDSS DR16 (Ahumada et al. 2020). We visually inspect all of the spectra and find that five sources have incorrect redshifts in the SDSS database, mainly on account of emission lines being misidentified. The better fittings of these spectra, and the corrected redshifts, are shown in figure B3 and summarised in Table C2.

In the case of G4Jy 148, the peak emission from the presence of C IV was mislabelled as Lyman-$\alpha$, whilst the prominent emission of C III] and Mg II went unlabelled. By re-fitting the spectrum (still with a quasar template) we determine a redshift of $z = 1.965$, and assign an error of 0.005 to accommodate the slight mismatch of the wavelength scaling. For comparison, Osmer, Porter & Green (1994) report a redshift of $z = 1.925$, and Yao et al. (2019) provide a redshift of $z = 1.972$. Meanwhile, Steidel & Sargent (1991) measure $z = 1.9417$ from the C III] line and $z = 1.9635$ from the Mg II line.

For G4Jy 176, the emission lines are correctly labelled by the SDSS data-reduction pipeline but the *wavelength* identifications are slightly offset. This is most likely due to resonance leading to 'negative emission' that affects the 'weighting' of the emission line's central wavelength[o]. Therefore, we re-fit the spectrum with a focus on the (non-resonating) C III] emission-line, and calculate a spectroscopic redshift of $z = 1.660 \pm 0.005$. Again, the relatively large error is to account for the wavelength scaling being slightly different from that of the redshifted template. Whilst the radio emission of this source is well-documented, we cannot find another spectroscopic redshift in published results.

SDSS provided a redshift of $z = 0.24639 \pm 0.00010$ for G4Jy 679 (3C 190), which is primarily based upon strong [O II] emission being misinterpreted as H$\alpha$ emission. With what we believe to be the correct interpretation, supported by the Mg II showing a resonant profile, and multiple emission-line identifications, we find that the spectroscopic redshift is $z = 1.196 \pm 0.001$. This is in good agreement with $z = 1.195649 \pm 0.000368$ from optical spectroscopy (Hewett & Wild 2010), and with $z = 1.1944 \pm 0.0012$ through H I absorption (Grasha et al. 2019).

Like G4Jy 148, G4Jy 845 (3C 243) has its C IV emission mislabelled as Lyman-$\alpha$, and the prominent emission of C III] and Mg II going unidentified. Our re-fitting of the spectrum leads to a redshift ($z = 1.706 \pm 0.001$) that is again in close agreement with that provided by (Hewett & Wild 2010, $z = 1.70756 \pm 0.00049$).

Finally, G4Jy 1240 (3C 316) had originally been identified as the highest-redshift source in the G4Jy Sample, at $z \sim 5.5$(!). However, inspection of the SDSS spectrum showed that the [O III] line at rest-frame $\lambda = 5\,007$ Å had been misidentified as a Lyman-$\alpha$ line. Re-fitting the quasar spectrum gives a redshift of $z = 0.581 \pm 0.001$, which is in close agreement with previous spectroscopic redshifts available via prior work ($z = 0.5795$; Gendre & Wall 2008; Smith et al. 2010). Indeed, via the SDSS helpdesk (with thanks to Joel Brownstein) it was found that, although the lowest-$\chi^2$ fit in SDSS's automated pipeline (Bolton et al. 2012) had $z = 5.49964$, the third next-best fit has $z = 0.580673$. (For interest, we add lower-ranked SDSS redshifts to Table C2.)

[o]https://skyserver.sdss.org/dr16/en/get/SpecById.ashx?id=8654825137051553792.

**Table C1.** Angular sizes, spectral indices, and 151-MHz flux-densities for 299 G4Jy sources, provided in the G4Jy catalogue (White et al. 2020a,b). These G4Jy spectral indices ('G4Jy_alpha') are measured across 20 flux-densities, from 72 MHz to 231 MHz. By combining these data with redshift information from SALT (this work), 6dFGS (Jones et al. 2009), SDSS DR12 (Alam et al. 2015), and SDSS DR16 (Ahumada et al. 2020), we calculate the radio luminosities and linear sizes of these sources. [The '(re-fitted)' label applies to five sources where we corrected the redshift provided in the SDSS database (see Figure B3).] The 'Confusion flag' is also from the G4Jy catalogue, and indicates (via '1') whether the *radio* emission of the G4Jy source may be blended with another (unrelated) radio source. An inequality ('$<$') in the angular-size column indicates that this value should be treated as an upper limit. See (White et al. 2020a) for further details.

| Source name | Confusion flag | Angular size/" | G4Jy spectral-index | Total $S_{151\,\mathrm{MHz}}$ /Jy | Redshift origin | Host-galaxy name | Spectroscopic redshift | $L_{151\,\mathrm{MHz}}$ /W Hz$^{-1}$ | Linear size /kpc |
|---|---|---|---|---|---|---|---|---|---|
| G4Jy 3 | 0 | 115.7 | $-1.13\pm0.01$ | $7.193\pm0.031$ | 6dFGS | g0003130−355614 | 0.04973 | $4.230\times10^{25}\pm1.821\times10^{23}$ | 112.5 |
| G4Jy 28 | 0 | 47.6 | $-1.11\pm0.01$ | $7.489\pm0.025$ | 6dFGS | g0016200−143011 | 0.76776 | $2.170\times10^{28}\pm7.135\times10^{25}$ | 351.9 |
| G4Jy 32 | 0 | 135.4 | $-0.76\pm0.02$ | $4.519\pm0.041$ | SDSS DR12 | J001815.21+214133.4 | $0.30261\pm0.00004$ | $1.248\times10^{27}\pm1.122\times10^{25}$ | 606.9 |
| G4Jy 43 | 1 | 61.0 | $-0.72\pm0.01$ | $9.982\pm0.026$ | SALT | J002308.86−250229.6 | $0.354\pm0.001$ | $3.882\times10^{27}\pm1.004\times10^{25}$ | 303.4 |
| G4Jy 44 | 0 | $<19.1$ | $-0.91\pm0.01$ | $5.553\pm0.020$ | SDSS DR12 | J002319.04−074449.3 | $0.51367\pm0.00005$ | $5.477\times10^{27}\pm1.994\times10^{25}$ | 118.3 |
| G4Jy 45 | 0 | $<42.0$ | $-0.86\pm0.01$ | $16.103\pm0.023$ | SALT | J002430.15−292854.3 | $0.407\pm0.002$ | $9.020\times10^{27}\pm1.311\times10^{25}$ | 228.0 |
| G4Jy 65 | 0 | $<24.2$ | $-0.85\pm0.02$ | $11.459\pm0.034$ | SDSS DR16 | J003606.45+183759.0 | $1.47035\pm0.00024$ | $1.350\times10^{29}\pm4.033\times10^{26}$ | 204.6 |
| G4Jy 67 | 0 | $<26.1$ | $-0.82\pm0.01$ | $19.879\pm0.025$ | 6dFGS | g0037041−010908 | 0.07333 | $2.581\times10^{26}\pm3.191\times10^{23}$ | 36.4 |
| G4Jy 69 | 1 | $<19.9$ | $-0.71\pm0.01$ | $28.249\pm0.033$ | 6dFGS | g0038205−020741 | 0.22040 | $3.812\times10^{27}\pm4.481\times10^{24}$ | 70.8 |
| G4Jy 70 | 0 | $<18.1$ | $-0.90\pm0.01$ | $11.349\pm0.020$ | 6dFGS | g0038269−385948 | 0.59272 | $1.570\times10^{28}\pm2.778\times10^{25}$ | 120.3 |
| G4Jy 74 | 0 | 183.8 | $-0.86\pm0.01$ | $7.188\pm0.033$ | SDSS DR12 | J004052.73+085401.0 | $0.13018\pm0.00002$ | $3.159\times10^{26}\pm1.435\times10^{24}$ | 426.4 |
| G4Jy 84 | 1 | 151.3 | $-0.74\pm0.02$ | $4.267\pm0.025$ | 6dFGS | g0046050−633319 | 0.07446 | $5.685\times10^{25}\pm3.371\times10^{23}$ | 214.0 |
| G4Jy 85 | 0 | 98.0 | $-0.72\pm0.01$ | $30.859\pm0.028$ | 6dFGS | g0046178−420752 | 0.05277 | $2.011\times10^{26}\pm1.802\times10^{23}$ | 100.7 |
| G4Jy 86 | 1 | 996.3 | – | $9.922\pm0.031$ | 6dFGS | g0047331−251719 | 0.00083 | – | 17.2 |
| G4Jy 92 | 1 | $<17.9$ | $-0.76\pm0.01$ | $7.558\pm0.034$ | SDSS DR16 | J005147.16+174710.6 | $1.53626\pm0.00056$ | $9.169\times10^{28}\pm4.165\times10^{26}$ | 151.6 |
| G4Jy 98 | 1 | $<18.8$ | $-0.51\pm0.01$ | $11.394\pm0.027$ | SALT | J005408.43−033355.2 | $0.211\pm0.001$ | $1.344\times10^{27}\pm3.191\times10^{24}$ | 64.7 |
| G4Jy 99 | 1 | $<33.1$ | $-1.08\pm0.02$ | $19.360\pm0.048$ | SDSS DR12 | J005550.61+262437.2 | $0.19545\pm0.00001$ | $2.143\times10^{27}\pm5.294\times10^{24}$ | 107.3 |
| G4Jy 100 | 0 | 225.9 | $-0.73\pm0.02$ | $4.478\pm0.052$ | 6dFGS | g0056027−012004 | 0.04296 | $1.912\times10^{25}\pm2.232\times10^{23}$ | 191.3 |
| G4Jy 101 | 0 | 243.2 | $-0.65\pm0.02$ | $6.019\pm0.068$ | 6dFGS | g0056256−011545 | 0.03819 | $2.014\times10^{25}\pm2.285\times10^{23}$ | 184.2 |
| G4Jy 104 | 0 | 71.4 | $-0.64\pm0.01$ | $18.837\pm0.042$ | 6dFGS | g0057349−012328 | 0.04494 | $8.793\times10^{25}\pm1.967\times10^{23}$ | 63.1 |
| G4Jy 110 | 1 | 71.4 | $-0.77\pm0.01$ | $4.016\pm0.024$ | 6dFGS | g0100467−503208 | 0.06088 | $3.527\times10^{25}\pm2.099\times10^{23}$ | 83.8 |
| G4Jy 128 | 1 | $<57.8$ | $-0.67\pm0.01$ | $4.230\pm0.025$ | 6dFGS | g0111091−135740 | 0.05160 | $2.626\times10^{25}\pm1.567\times10^{23}$ | 58.2 |
| G4Jy 131 | 1 | 472.3 | $-0.76\pm0.02$ | $4.618\pm0.058$ | SDSS DR16 | J011259.57+152928.7 | $0.04385\pm0.00002$ | $2.060\times10^{25}\pm2.605\times10^{23}$ | 407.8 |
| G4Jy 133 | 0 | 543.6 | $-0.84\pm0.01$ | $13.949\pm0.039$ | 6dFGS | g0116251−472241 | 0.14585 | $7.809\times10^{26}\pm2.163\times10^{24}$ | 1388.2 |
| G4Jy 139 | 0 | $<19.3$ | $-0.93\pm0.01$ | $10.849\pm0.028$ | SDSS DR16 | J011818.48+025805.7 | $0.66715\pm0.00004$ | $2.023\times10^{28}\pm5.172\times10^{25}$ | 135.2 |
| G4Jy 140 | 1 | $<23.7$ | $-0.79\pm0.01$ | $6.054\pm0.021$ | 6dFGS | g0118342−184917 | 0.27757 | $1.387\times10^{27}\pm4.746\times10^{24}$ | 100.0 |
| G4Jy 148 | 0 | $<15.6$ | $-0.42\pm0.01$ | $4.749\pm0.023$ | SDSS DR16 (re-fitted) | J012227.89−042127.1 | $1.96500\pm0.00024$ | $6.970\times10^{28}\pm3.427\times10^{26}$ | 130.8 |
| G4Jy 150 | 1 | 190.8 | $-0.92\pm0.02$ | $4.267\pm0.035$ | 6dFGS | g0125443−012246 | 0.01814 | $3.165\times10^{24}\pm2.624\times10^{22}$ | 70.3 |

**Table C1.** Continued.

| Source name | Confusion flag | Angular size/" | G4Jy spectral-index | Total $S_{151\,\mathrm{MHz}}$ /Jy | Redshift origin | Host-galaxy name | Spectroscopic redshift | $L_{151\,\mathrm{MHz}}$ /W Hz$^{-1}$ | Linear size /kpc |
|---|---|---|---|---|---|---|---|---|---|
| G4Jy 151 | 1 | 846.6 | $-0.78 \pm 0.01$ | $14.220 \pm 0.043$ | 6dFGS | g0126006−012043 | 0.01841 | $1.084\times10^{25} \pm 3.291\times10^{22}$ | 316.5 |
| G4Jy 159 | 0 | $< 30.5$ | $-0.76 \pm 0.02$ | $12.821 \pm 0.049$ | SDSS DR12 | J012830.11+290300.8 | $0.39610 \pm 0.00002$ | $6.521\times10^{27} \pm 2.469\times10^{25}$ | 162.9 |
| G4Jy 163 | 0 | $< 18.2$ | $-0.63 \pm 0.01$ | $4.686 \pm 0.029$ | SDSS DR16 | J013107.04+041014.8 | $1.26030 \pm 0.00017$ | $3.221\times10^{28} \pm 2.024\times10^{26}$ | 151.9 |
| G4Jy 167 | 0 | $< 71.7$ | $-0.95 \pm 0.01$ | $10.948 \pm 0.027$ | SDSS DR12 | J013121.65+062343.1 | $0.65932 \pm 0.00003$ | $1.998\times10^{28} \pm 4.952\times10^{25}$ | 499.9 |
| G4Jy 171 | 0 | 689.1 | $-0.71 \pm 0.01$ | $32.108 \pm 0.061$ | 6dFGS | g0133577−362936 | 0.03051 | $6.812\times10^{25} \pm 1.287\times10^{23}$ | 420.7 |
| G4Jy 172 | 1 | $< 45.7$ | $-0.76 \pm 0.01$ | $8.357 \pm 0.018$ | SALT | J013333.18−444417.7 | $0.091 \pm 0.003$ | $1.698\times10^{26} \pm 3.638\times10^{23}$ | 77.5 |
| G4Jy 176 | 0 | 17.6 | $-0.86 \pm 0.02$ | $6.453 \pm 0.035$ | SDSS DR16 (re-fitted) | J013813.78+230115.6 | $1.66000 \pm 0.00086$ | $1.029\times10^{29} \pm 5.639\times10^{26}$ | 149.1 |
| G4Jy 179 | 0 | $< 16.9$ | $-0.31 \pm 0.01$ | $9.129 \pm 0.031$ | SDSS DR16 | J014109.16+135328.3 | $0.23577 \pm 0.00006$ | $1.311\times10^{27} \pm 4.388\times10^{24}$ | 63.3 |
| G4Jy 184 | 0 | $< 26.2$ | $-0.88 \pm 0.01$ | $5.130 \pm 0.028$ | SDSS DR12 | J014316.72−011900.9 | $0.51931 \pm 0.00004$ | $5.133\times10^{27} \pm 2.798\times10^{25}$ | 163.2 |
| G4Jy 192 | 0 | 111.3 | $-0.80 \pm 0.01$ | $16.595 \pm 0.029$ | SALT | J015035.94−293155.3 | $0.413 \pm 0.003$ | $9.436\times10^{27} \pm 1.643\times10^{25}$ | 609.5 |
| G4Jy 203 | 0 | $< 29.3$ | $-0.65 \pm 0.02$ | $4.099 \pm 0.031$ | SDSS DR16 | J015559.18+033158.1 | $1.28009 \pm 0.00024$ | $2.949\times10^{28} \pm 2.237\times10^{26}$ | 245.1 |
| G4Jy 208 | 0 | $< 17.4$ | $-0.69 \pm 0.01$ | $9.483 \pm 0.027$ | 6dFGS | g0157416−104341 | 0.61804 | $1.308\times10^{28} \pm 3.698\times10^{25}$ | 117.9 |
| G4Jy 216 | 0 | $< 19.2$ | $-0.44 \pm 0.01$ | $7.665 \pm 0.026$ | 6dFGS | g0201572−113233 | 0.67006 | $1.121\times10^{28} \pm 3.741\times10^{25}$ | 134.8 |
| G4Jy 229 | 0 | $< 55.3$ | $-0.47 \pm 0.01$ | $6.235 \pm 0.016$ | 6dFGS | g0210462−510102 | 0.03154 | $1.404\times10^{25} \pm 3.552\times10^{22}$ | 34.9 |
| G4Jy 241 | 0 | 316.4 | $-0.78 \pm 0.01$ | $10.815 \pm 0.035$ | 6dFGS | g0216451−474909 | 0.06423 | $1.062\times10^{26} \pm 3.403\times10^{23}$ | 390.7 |
| G4Jy 247 | 0 | 63.8 | $-1.00 \pm 0.01$ | $9.362 \pm 0.024$ | 6dFGS | g0219029−362607 | 0.48848 | $8.492\times10^{27} \pm 2.159\times10^{25}$ | 384.8 |
| G4Jy 260 | 0 | $< 17.1$ | $-0.78 \pm 0.01$ | $10.180 \pm 0.022$ | 6dFGS | g0225028−231248 | 0.23208 | $1.560\times10^{27} \pm 3.417\times10^{24}$ | 63.3 |
| G4Jy 277 | 0 | $< 19.5$ | $-0.75 \pm 0.01$ | $6.811 \pm 0.023$ | SDSS DR16 | J023507.34−040205.3 | $1.44207 \pm 0.00010$ | $7.022\times10^{28} \pm 2.365\times10^{26}$ | 164.7 |
| G4Jy 288 | 0 | $< 16.0$ | $-0.51 \pm 0.01$ | $17.886 \pm 0.031$ | 6dFGS | g0242407−000048 | 0.00379 | $5.661\times10^{23} \pm 9.762\times10^{20}$ | 1.3 |
| G4Jy 292 | 0 | 89.7 | $-0.74 \pm 0.01$ | $5.671 \pm 0.016$ | 6dFGS | g0245541−445939 | 0.28267 | $1.336\times10^{27} \pm 3.786\times10^{24}$ | 383.4 |
| G4Jy 330 | 0 | 89.5 | $-0.80 \pm 0.01$ | $4.071 \pm 0.021$ | 6dFGS | g0310015−301940 | 0.06817 | $4.531\times10^{25} \pm 2.362\times10^{23}$ | 116.7 |
| G4Jy 341 | 0 | 54.1 | $-0.65 \pm 0.01$ | $4.181 \pm 0.022$ | 6dFGS | g0316393−435117 | 0.06274 | $3.879\times10^{25} \pm 2.030\times10^{23}$ | 65.4 |
| G4Jy 344 | 0 | 57.1 | $-1.03 \pm 0.01$ | $7.305 \pm 0.021$ | 6dFGS | g0317577−441417 | 0.07588 | $1.034\times10^{26} \pm 2.951\times10^{23}$ | 82.2 |
| G4Jy 361 | 1 | $< 15.7$ | $-0.79 \pm 0.01$ | $7.137 \pm 0.021$ | 6dFGS | g0328365−284156 | 0.10869 | $2.117\times10^{26} \pm 6.251\times10^{23}$ | 31.2 |
| G4Jy 366 | 0 | 276.9 | $-0.79 \pm 0.01$ | $8.374 \pm 0.046$ | 6dFGS | g0334033−390036 | 0.06227 | $7.716\times10^{25} \pm 4.269\times10^{23}$ | 332.2 |
| G4Jy 373 | 0 | 88.8 | $-0.83 \pm 0.01$ | $10.028 \pm 0.019$ | SALT | J033846.01−352252.0 | $0.113 \pm 0.001$ | $3.242\times10^{26} \pm 6.208\times10^{23}$ | 182.4 |
| G4Jy 377 | 0 | $< 15.2$ | $-0.64 \pm 0.01$ | $7.407 \pm 0.022$ | 6dFGS | g0342054−370322 | 0.28380 | $1.719\times10^{27} \pm 5.067\times10^{24}$ | 65.1 |
| G4Jy 381 | 0 | 283.6 | $-0.75 \pm 0.01$ | $16.692 \pm 0.041$ | 6dFGS | g0346306−342246 | 0.05346 | $1.119\times10^{26} \pm 2.768\times10^{23}$ | 295.3 |
| G4Jy 386 | 0 | 355.0 | $-0.80 \pm 0.01$ | $27.550 \pm 0.052$ | 6dFGS | g0351358−274435 | 0.06565 | $2.835\times10^{26} \pm 5.337\times10^{23}$ | 447.3 |
| G4Jy 392 | 0 | 56.7 | $-0.84 \pm 0.01$ | $20.011 \pm 0.027$ | SDSS DR12 | J035230.55−071102.3 | $0.96354 \pm 0.00011$ | $8.561\times10^{28} \pm 1.147\times10^{26}$ | 449.8 |
| G4Jy 411 | 0 | $< 18.1$ | – | $9.072 \pm 0.025$ | 6dFGS | g0405340−130814 | 0.57428 | – | 118.5 |
| G4Jy 423 | 0 | $< 36.2$ | $-0.71 \pm 0.01$ | $7.596 \pm 0.020$ | 6dFGS | g0411594−643624 | 0.15507 | $4.770\times10^{26} \pm 1.246\times10^{24}$ | 97.3 |

**Table C1.** Continued.

| Source name | Confusion flag | Angular size/" | G4Jy spectral-index | Total $S_{151\,\mathrm{MHz}}$ /Jy | Redshift origin | Host-galaxy name | Spectroscopic redshift | $L_{151\,\mathrm{MHz}}$ /W Hz$^{-1}$ | Linear size /kpc |
|---|---|---|---|---|---|---|---|---|---|
| G4Jy 437 | 0 | $<21.5$ | $-1.10 \pm 0.01$ | $8.600 \pm 0.024$ | 6dFGS | g0417168−055345 | 0.77465 | $2.534\times10^{28} \pm 7.027\times10^{25}$ | 159.6 |
| G4Jy 462 | 0 | $<345.4$ | $-0.67 \pm 0.01$ | $25.569 \pm 0.044$ | 6dFGS | g0429082−534940 | 0.03797 | $8.463\times10^{25} \pm 1.468\times10^{23}$ | 260.1 |
| G4Jy 466 | 1 | 104.7 | $-0.65 \pm 0.01$ | $7.592 \pm 0.031$ | 6dFGS | g0430220−613201 | 0.05546 | $5.460\times10^{25} \pm 2.258\times10^{23}$ | 112.7 |
| G4Jy 475 | 1 | 305.0 | $-0.77 \pm 0.02$ | $4.066 \pm 0.054$ | 6dFGS | g0434104−132212 | 0.03487 | $1.134\times10^{25} \pm 1.500\times10^{23}$ | 211.7 |
| G4Jy 479 | 0 | 115.2 | $-0.67 \pm 0.01$ | $4.390 \pm 0.026$ | 6dFGS | g0436354−222639 | 0.06895 | $4.961\times10^{25} \pm 2.984\times10^{23}$ | 151.9 |
| G4Jy 492 | 0 | 59.1 | $-0.76 \pm 0.01$ | $37.342 \pm 0.034$ | SALT | J044437.70−280954.3 | $0.148 \pm 0.001$ | $2.135\times10^{27} \pm 1.954\times10^{24}$ | 152.8 |
| G4Jy 495 | 1 | $<71.1$ | $-0.96 \pm 0.01$ | $4.284 \pm 0.021$ | 6dFGS | g0445106−383835 | 0.53609 | $4.781\times10^{27} \pm 2.309\times10^{25}$ | 450.0 |
| G4Jy 499 | 1 | 196.8 | $-0.93 \pm 0.01$ | $4.274 \pm 0.029$ | 6dFGS | g0448306−203214 | 0.07393 | $5.687\times10^{25} \pm 3.902\times10^{23}$ | 276.6 |
| G4Jy 510 | 1 | 73.3 | $-0.78 \pm 0.01$ | $10.007 \pm 0.024$ | 6dFGS | g0456089−215909 | 0.27821 | $2.302\times10^{27} \pm 5.557\times10^{24}$ | 309.7 |
| G4Jy 516 | 1 | $<18.8$ | $-0.87 \pm 0.01$ | $9.057 \pm 0.024$ | 6dFGS | g0504531−101453 | 0.04032 | $3.415\times10^{25} \pm 9.069\times10^{22}$ | 15.0 |
| G4Jy 517 | 1 | 2143.5 | – | $9.538 \pm 0.096$ | 6dFGS | g0505492−283519 | 0.03812 | – | 1 620.3 |
| G4Jy 530 | 1 | 109.9 | $-0.86 \pm 0.01$ | $16.768 \pm 0.033$ | SALT | WISEA J051247.41−482416.5 | $0.305 \pm 0.002$ | $4.838\times10^{27} \pm 9.478\times10^{24}$ | 495.2 |
| G4Jy 536 | 0 | 54.3 | $-0.76 \pm 0.01$ | $6.739 \pm 0.024$ | 6dFGS | g0518264−561413 | 0.09480 | $1.491\times10^{26} \pm 5.228\times10^{23}$ | 95.5 |
| G4Jy 540 | 0 | 5.0 | $-0.62 \pm 0.01$ | $55.942 \pm 0.032$ | 6dFGS | g0522580−362731 | 0.05651 | $4.175\times10^{26} \pm 2.362\times10^{23}$ | 5.5 |
| G4Jy 541 | 0 | 65.1 | $-0.81 \pm 0.01$ | $4.793 \pm 0.020$ | SALT | J052320.72−481630.6 | $0.200 \pm 0.007$ | $5.324\times10^{26} \pm 2.229\times10^{24}$ | 214.9 |
| G4Jy 543 | 0 | 283.1 | $-0.77 \pm 0.01$ | $5.806 \pm 0.030$ | 6dFGS | g0525272−324216 | 0.07676 | $8.262\times10^{25} \pm 4.268\times10^{23}$ | 411.7 |
| G4Jy 579 | 0 | 522.5 | $-0.69 \pm 0.02$ | $4.348 \pm 0.036$ | 6dFGS | g0548276−325838 | 0.03716 | $1.378\times10^{25} \pm 1.125\times10^{23}$ | 385.5 |
| G4Jy 590 | 0 | 77.9 | $-0.95 \pm 0.01$ | $9.785 \pm 0.022$ | SALT | J060312.22−342632.6 | $0.529 \pm 0.002$ | $1.054\times10^{28} \pm 2.319\times10^{25}$ | 489.5 |
| G4Jy 592 | 1 | 18.6 | $-0.95 \pm 0.01$ | $5.698 \pm 0.020$ | 6dFGS | g0605540−351808 | 0.14115 | $3.016\times10^{26} \pm 1.038\times10^{24}$ | 46.2 |
| G4Jy 604 | 0 | 881.9 | $-0.76 \pm 0.02$ | $5.962 \pm 0.058$ | SALT | J061813.03−484458.3 | $0.049 \pm 0.003$ | $3.342\times10^{25} \pm 3.235\times10^{23}$ | 845.8 |
| G4Jy 606 | 0 | 47.3 | – | $7.656 \pm 0.019$ | 6dFGS | g0621013−450440 | 0.20818 | – | 161.0 |
| G4Jy 607 | 0 | 178.6 | $-0.83 \pm 0.01$ | $13.938 \pm 0.031$ | 6dFGS | g0621433−524133 | 0.05107 | $8.535\times10^{25} \pm 1.904\times10^{23}$ | 178.1 |
| G4Jy 611 | 0 | 59.7 | $-1.02 \pm 0.01$ | $46.864 \pm 0.037$ | SALT | WISEA J062620.46−534135.1 | $0.055 \pm 0.001$ | $3.379\times10^{26} \pm 2.700\times10^{23}$ | 63.8 |
| G4Jy 613 | 0 | 194.1 | $-0.89 \pm 0.01$ | $19.702 \pm 0.035$ | 6dFGS | g0626496−543234 | 0.05171 | $1.242\times10^{26} \pm 2.195\times10^{23}$ | 195.8 |
| G4Jy 614 | 0 | 94.2 | $-0.73 \pm 0.01$ | $16.577 \pm 0.024$ | 6dFGS | g0627067−352915 | 0.05482 | $1.169\times10^{26} \pm 1.671\times10^{23}$ | 100.4 |
| G4Jy 615 | 0 | 84.6 | $-1.14 \pm 0.01$ | $5.104 \pm 0.020$ | 6dFGS | g0628498−414337 | 0.17826 | $4.649\times10^{26} \pm 1.845\times10^{24}$ | 254.8 |
| G4Jy 619 | 0 | 806.4 | $-0.75 \pm 0.01$ | $41.188 \pm 0.099$ | 6dFGS | g0636323−203453 | 0.05501 | $2.929\times10^{26} \pm 7.034\times10^{23}$ | 862.1 |
| G4Jy 627 | 1 | 204.7 | $-0.78 \pm 0.01$ | $6.732 \pm 0.023$ | 6dFGS | g0644251−434349 | 0.06142 | $6.024\times10^{25} \pm 2.030\times10^{23}$ | 242.5 |
| G4Jy 631 | 1 | 123.4 | $-0.81 \pm 0.01$ | $5.286 \pm 0.029$ | 6dFGS | g0650215−554933 | 0.04969 | $3.056\times10^{25} \pm 1.705\times10^{23}$ | 119.9 |
| G4Jy 633 | 0 | 103.8 | $-0.79 \pm 0.01$ | $5.837 \pm 0.020$ | 6dFGS | g0651549−602217 | 0.13380 | $2.699\times10^{26} \pm 9.333\times10^{23}$ | 246.5 |
| G4Jy 641 | 1 | 255.8 | $-0.94 \pm 0.02$ | $5.504 \pm 0.037$ | SALT | WISEA J070532.94−451308.8 | $0.128 \pm 0.002$ | $2.354\times10^{26} \pm 1.569\times10^{24}$ | 584.9 |
| G4Jy 651 | 0 | 211.8 | $-0.81 \pm 0.01$ | $11.948 \pm 0.035$ | 6dFGS | g0717081−362159 | 0.03134 | $2.684\times10^{25} \pm 7.841\times10^{22}$ | 132.7 |
| G4Jy 672 | 0 | $<25.1$ | $-0.62 \pm 0.01$ | $11.568 \pm 0.027$ | SALT | WISEA J074331.61−672625.5 | $1.510 \pm 0.001$ | $1.182\times10^{29} \pm 2.809\times10^{26}$ | 212.5 |

**Table C1.** Continued.

| Source name | Confusion flag | Angular size/" | G4Jy spectral-index | Total $S_{151\,\mathrm{MHz}}$ /Jy | Redshift origin | Host-galaxy name | Spectroscopic redshift | $L_{151\,\mathrm{MHz}}$ /W Hz$^{-1}$ | Linear size /kpc |
|---|---|---|---|---|---|---|---|---|---|
| G4Jy 679 | 0 | $<16.6$ | $-0.79 \pm 0.01$ | $16.760 \pm 0.032$ | SDSS DR12 (re-fitted) | J080133.55+141442.8 | $1.19600 \pm 0.00010$ | $1.155\times10^{29} \pm 2.218\times10^{26}$ | 137.6 |
| G4Jy 682 | 0 | $<16.1$ | $-0.87 \pm 0.01$ | $14.441 \pm 0.030$ | SDSS DR16 | J080447.96+101523.7 | $1.96818 \pm 0.00016$ | $3.481\times10^{29} \pm 7.222\times10^{26}$ | 135.0 |
| G4Jy 683 | 0 | 97.3 | $-0.74 \pm 0.02$ | $23.065 \pm 0.046$ | SDSS DR12 | J080535.00+240950.3 | $0.05968 \pm 0.00001$ | $1.940\times10^{26} \pm 3.905\times10^{23}$ | 112.3 |
| G4Jy 685 | 1 | 86.2 | $-0.88 \pm 0.01$ | $22.306 \pm 0.034$ | 6dFGS | g0808536−102740 | 0.10891 | $6.705\times10^{26} \pm 1.033\times10^{24}$ | 171.4 |
| G4Jy 692 | 1 | $<18.1$ | $-1.07 \pm 0.01$ | $22.731 \pm 0.033$ | SALT | J081527.81−030826.6 | $0.198 \pm 0.001$ | $2.589\times10^{27} \pm 3.802\times10^{24}$ | 59.3 |
| G4Jy 699 | 0 | $<16.5$ | – | $5.316 \pm 0.032$ | SDSS DR12 | J082144.02+174820.2 | $0.29699 \pm 0.00008$ | – | 73.0 |
| G4Jy 706 | 0 | $<19.3$ | $-0.62 \pm 0.01$ | $17.804 \pm 0.022$ | SALT | J082717.41−202624.8 | $0.828 \pm 0.004$ | $4.645\times10^{28} \pm 5.828\times10^{25}$ | 146.6 |
| G4Jy 722 | 0 | $<19.5$ | $-0.73 \pm 0.01$ | $11.758 \pm 0.026$ | SDSS DR12 | J084047.58+131223.5 | $0.68038 \pm 0.00007$ | $2.074\times10^{28} \pm 4.603\times10^{25}$ | 137.7 |
| G4Jy 725 | 0 | $<63.9$ | $-0.90 \pm 0.02$ | $6.178 \pm 0.048$ | SDSS DR12 | J084309.85+294404.6 | $0.39797 \pm 0.00002$ | $3.332\times10^{27} \pm 2.601\times10^{25}$ | 342.1 |
| G4Jy 728 | 1 | $<79.3$ | – | $5.119 \pm 0.027$ | SDSS DR16 | J084605.09+145928.3 | $0.67560 \pm 0.00004$ | – | 558.5 |
| G4Jy 741 | 0 | $<16.5$ | – | $12.711 \pm 0.042$ | SDSS DR16 | J085810.02+275053.8 | $0.23017 \pm 0.00008$ | – | 60.7 |
| G4Jy 745 | 0 | 57.7 | $-0.77 \pm 0.02$ | $6.473 \pm 0.032$ | SDSS DR16 | J090048.22+183226.0 | $0.60078 \pm 0.00005$ | $8.704\times10^{27} \pm 4.266\times10^{25}$ | 385.9 |
| G4Jy 750 | 1 | $<44.2$ | $-0.78 \pm 0.02$ | $4.263 \pm 0.023$ | SDSS DR12 | J090604.06+110328.7 | $0.41658 \pm 0.00005$ | $2.453\times10^{27} \pm 1.332\times10^{25}$ | 243.4 |
| G4Jy 751 | 0 | $<39.6$ | – | $11.496 \pm 0.027$ | SDSS DR12 | J090631.87+164611.8 | $0.41106 \pm 0.00004$ | – | 216.3 |
| G4Jy 755 | 0 | $<19.2$ | – | $4.596 \pm 0.029$ | SDSS DR16 | J091022.55+241919.5 | $0.90674 \pm 0.00011$ | – | 149.9 |
| G4Jy 760 | 0 | $<48.0$ | – | $7.789 \pm 0.030$ | SDSS DR12 | J091405.21+171554.3 | $0.51998 \pm 0.00004$ | – | 299.1 |
| G4Jy 770 | 1 | $<51.6$ | – | $5.637 \pm 0.024$ | SDSS DR12 | J092507.27+144425.6 | $0.89615 \pm 0.00007$ | – | 401.5 |
| G4Jy 786 | 0 | 56.0 | $-0.67 \pm 0.01$ | $4.791 \pm 0.028$ | SDSS DR12 | J094319.23−000424.8 | $0.46432 \pm 0.00005$ | $3.392\times10^{27} \pm 1.978\times10^{25}$ | 328.5 |
| G4Jy 794 | 0 | 188.3 | – | $37.273 \pm 0.042$ | SDSS DR16 | J094745.14+072520.5 | $0.08556 \pm 0.00004$ | – | 302.2 |
| G4Jy 797 | 0 | $<53.7$ | – | $16.766 \pm 0.028$ | SDSS DR16 | J095010.79+142000.6 | $0.55183 \pm 0.00002$ | – | 344.9 |
| G4Jy 809 | 0 | $<34.4$ | $-0.60 \pm 0.02$ | $4.597 \pm 0.026$ | SDSS DR12 | J100021.80+223318.6 | $0.41893 \pm 0.00002$ | $2.519\times10^{27} \pm 1.423\times10^{25}$ | 190.0 |
| G4Jy 816 | 1 | 111.6 | $-0.97 \pm 0.01$ | $5.030 \pm 0.022$ | 6dFGS | g1004400−321642 | 0.08842 | $9.792\times10^{25} \pm 4.192\times10^{23}$ | 184.4 |
| G4Jy 817 | 0 | 58.3 | $-0.83 \pm 0.02$ | $4.403 \pm 0.030$ | SDSS DR16 | J100445.74+222519.2 | $0.98061 \pm 0.00005$ | $1.954\times10^{28} \pm 1.326\times10^{26}$ | 464.5 |
| G4Jy 820 | 1 | 86.5 | – | $5.595 \pm 0.026$ | SDSS DR16 | J100726.10+124856.2 | $0.24074 \pm 0.00008$ | – | 328.8 |
| G4Jy 829 | 0 | 49.6 | $-0.98 \pm 0.01$ | $5.735 \pm 0.018$ | 6dFGS | g1013297−283126 | 0.25482 | $1.133\times10^{27} \pm 3.605\times10^{24}$ | 196.8 |
| G4Jy 832 | 0 | $<17.3$ | – | $4.129 \pm 0.044$ | SDSS DR16 | J101418.79+290400.3 | $0.37828 \pm 0.00002$ | – | 89.8 |
| G4Jy 834 | 1 | $<16.7$ | $-0.74 \pm 0.02$ | $5.687 \pm 0.037$ | SDSS DR16 | J101749.37+273204.1 | $0.46792 \pm 0.00002$ | $4.220\times10^{27} \pm 2.721\times10^{25}$ | 98.3 |
| G4Jy 842 | 0 | $<19.5$ | $-0.89 \pm 0.02$ | $4.199 \pm 0.023$ | SDSS DR16 | J102459.80+062453.7 | $1.73621 \pm 0.00080$ | $7.679\times10^{28} \pm 4.173\times10^{26}$ | 165.0 |
| G4Jy 843 | 0 | $<75.1$ | $-0.87 \pm 0.02$ | $7.397 \pm 0.032$ | SDSS DR16 | J102520.80+201020.0 | $0.43687 \pm 0.00004$ | $4.911\times10^{27} \pm 2.140\times10^{25}$ | 425.2 |
| G4Jy 845 | 0 | $<16.8$ | $-1.08 \pm 0.01$ | $9.493 \pm 0.024$ | SDSS DR12 (re-fitted) | J102631.95+062733.0 | $1.70600 \pm 0.00015$ | $2.012\times10^{29} \pm 5.145\times10^{26}$ | 142.2 |
| G4Jy 852 | 0 | 53.8 | – | $5.543 \pm 0.036$ | SDSS DR16 | J103244.38+250223.0 | $0.52416 \pm 0.00003$ | – | 336.8 |
| G4Jy 864 | 0 | $<17.0$ | $-0.56 \pm 0.01$ | $9.295 \pm 0.027$ | SDSS DR16 | J104138.98+024231.4 | $0.53395 \pm 0.00008$ | $8.662\times10^{27} \pm 2.486\times10^{25}$ | 107.4 |
| G4Jy 865 | 0 | $<17.9$ | $-0.89 \pm 0.01$ | $13.049 \pm 0.027$ | SDSS DR12 | J104244.60+120331.2 | $1.02860 \pm 0.00011$ | $6.749\times10^{28} \pm 1.414\times10^{26}$ | 144.3 |
| G4Jy 876 | 1 | 76.7 | $-0.82 \pm 0.01$ | $10.110 \pm 0.024$ | 6dFGS | g1051299−091810 | 0.34506 | $3.830\times10^{27} \pm 9.147\times10^{24}$ | 375.2 |

**Table C1.** Continued.

| Source name | Confusion flag | Angular size/″ | G4Jy spectral-index | Total $S_{151\,\mathrm{MHz}}$ /Jy | Redshift origin | Host-galaxy name | Spectroscopic redshift | $L_{151\,\mathrm{MHz}}$ /W Hz$^{-1}$ | Linear size /kpc |
|---|---|---|---|---|---|---|---|---|---|
| G4Jy 881 | 0 | $<35.0$ | $-1.09 \pm 0.01$ | $4.950 \pm 0.027$ | SDSS DR12 | J105517.27+020544.9 | $0.87529 \pm 0.00027$ | $1.974\times10^{28} \pm 1.096\times10^{26}$ | 270.5 |
| G4Jy 882 | 0 | $<25.5$ | $-0.64 \pm 0.01$ | $5.964 \pm 0.019$ | 6dFGS | g1055334−283134 | 0.06103 | $5.225\times10^{25} \pm 1.630\times10^{23}$ | 30.0 |
| G4Jy 884 | 0 | 30.2 | – | $8.476 \pm 0.033$ | SDSS DR12 | J105817.90+195150.9 | $1.11592 \pm 0.00007$ | – | 247.2 |
| G4Jy 885 | 0 | $<19.6$ | $-0.31 \pm 0.01$ | $4.833 \pm 0.025$ | SDSS DR16 | J105829.60+013358.8 | $0.89193 \pm 0.00062$ | $1.223\times10^{28} \pm 6.436\times10^{25}$ | 152.3 |
| G4Jy 886 | 1 | 237.4 | $-0.79 \pm 0.01$ | $7.410 \pm 0.029$ | 6dFGS | g1058548−361921 | 0.07047 | $8.834\times10^{25} \pm 3.514\times10^{23}$ | 319.3 |
| G4Jy 891 | 0 | 56.9 | $-0.89 \pm 0.01$ | $4.382 \pm 0.016$ | 6dFGS | g1103316−325117 | 0.35526 | $1.811\times10^{27} \pm 6.648\times10^{24}$ | 283.8 |
| G4Jy 893 | 1 | 61.7 | $-1.02 \pm 0.01$ | $8.861 \pm 0.019$ | 6dFGS | g1106121−244444 | 0.04997 | $5.236\times10^{25} \pm 1.111\times10^{23}$ | 60.3 |
| G4Jy 895 | 0 | $<19.3$ | $-0.84 \pm 0.01$ | $4.848 \pm 0.024$ | SDSS DR16 | J110631.77−005252.3 | $0.42370 \pm 0.00002$ | $2.966\times10^{27} \pm 1.498\times10^{25}$ | 107.3 |
| G4Jy 901 | 0 | 20.8 | $-0.72 \pm 0.01$ | $5.714 \pm 0.020$ | SALT | WISEA J111119.43−403051.9 | $0.725 \pm 0.002$ | $1.158\times10^{28} \pm 3.996\times10^{25}$ | 150.7 |
| G4Jy 903 | 0 | 56.1 | $-0.82 \pm 0.01$ | $8.932 \pm 0.026$ | SDSS DR16 | J111332.49−021253.3 | $0.47543 \pm 0.00003$ | $7.099\times10^{27} \pm 2.074\times10^{25}$ | 333.5 |
| G4Jy 906 | 1 | $<67.9$ | $-0.73 \pm 0.03$ | $7.323 \pm 0.057$ | SDSS DR12 | J111634.61+291517.1 | $0.04868 \pm 0.00001$ | $4.044\times10^{25} \pm 3.148\times10^{23}$ | 64.7 |
| G4Jy 909 | 0 | 62.2 | $-0.74 \pm 0.01$ | $8.948 \pm 0.024$ | SALT | J111826.95−463414.9 | $0.714 \pm 0.001$ | $1.770\times10^{28} \pm 4.720\times10^{25}$ | 447.9 |
| G4Jy 910 | 0 | $<18.3$ | – | $4.228 \pm 0.030$ | SDSS DR16 | J111857.29+123441.7 | $2.12528 \pm 0.00027$ | – | 152.0 |
| G4Jy 916 | 1 | $<16.5$ | $-0.81 \pm 0.01$ | $6.335 \pm 0.037$ | SDSS DR16 | J112437.40+045618.9 | $0.28280 \pm 0.00002$ | $1.521\times10^{27} \pm 8.977\times10^{24}$ | 70.5 |
| G4Jy 917 | 0 | 103.6 | $-0.73 \pm 0.01$ | $11.328 \pm 0.022$ | 6dFGS | g1125529−352340 | 0.03262 | $2.755\times10^{25} \pm 5.316\times10^{22}$ | 67.4 |
| G4Jy 919 | 1 | 15.8 | $-0.94 \pm 0.01$ | $6.408 \pm 0.028$ | SDSS DR16 | J112627.12+122034.8 | $1.09920 \pm 0.00013$ | $4.053\times10^{28} \pm 1.798\times10^{26}$ | 129.1 |
| G4Jy 924 | 0 | 16.6 | $-0.82 \pm 0.02$ | $4.340 \pm 0.028$ | SDSS DR16 | J113259.49+102342.2 | $0.53963 \pm 0.00004$ | $4.640\times10^{27} \pm 2.989\times10^{25}$ | 105.4 |
| G4Jy 933 | 1 | $<15.9$ | $-0.78 \pm 0.01$ | $19.350 \pm 0.025$ | 6dFGS | g1139107−135043 | 0.55607 | $2.181\times10^{28} \pm 2.799\times10^{25}$ | 102.5 |
| G4Jy 936 | 0 | $<78.4$ | – | $5.772 \pm 0.034$ | SDSS DR12 | J114027.73+120308.2 | $0.08115 \pm 0.00001$ | – | 119.9 |
| G4Jy 944 | 0 | $<26.5$ | $-1.00 \pm 0.02$ | $7.935 \pm 0.053$ | SDSS DR12 | J114257.21+212911.2 | $1.37303 \pm 0.00010$ | $9.096\times10^{28} \pm 6.070\times10^{26}$ | 223.1 |
| G4Jy 949 | 0 | 219.9 | – | $22.096 \pm 0.077$ | SDSS DR12 | J114505.01+193622.8 | $0.02160 \pm 0.00001$ | – | 96.1 |
| G4Jy 957 | 0 | 227.6 | $-0.86 \pm 0.01$ | $9.692 \pm 0.042$ | SALT | J114906.68−120433.4 | $0.119 \pm 0.002$ | $3.513\times10^{26} \pm 1.510\times10^{24}$ | 488.8 |
| G4Jy 965 | 1 | $<17.5$ | – | $8.868 \pm 0.018$ | SALT | J115421.78−350529.0 | $0.258 \pm 0.001$ | – | 70.0 |
| G4Jy 966 | 0 | 176.9 | $-0.79 \pm 0.03$ | $6.082 \pm 0.081$ | SDSS DR12 | J115820.13+262112.0 | $0.11202 \pm 0.00002$ | $1.924\times10^{26} \pm 2.556\times10^{24}$ | 360.6 |
| G4Jy 968 | 0 | $<14.0$ | $-0.78 \pm 0.04$ | $4.077 \pm 0.071$ | SDSS DR12 | J115931.83+291443.8 | $0.72475 \pm 0.00010$ | $8.535\times10^{27} \pm 1.477\times10^{26}$ | 101.4 |
| G4Jy 975 | 0 | $<17.8$ | $-0.80 \pm 0.01$ | $8.563 \pm 0.041$ | SDSS DR12 | J120620.12+040610.7 | $0.53604 \pm 0.00005$ | $8.913\times10^{27} \pm 4.300\times10^{25}$ | 112.7 |
| G4Jy 979 | 1 | 64.8 | $-0.78 \pm 0.01$ | $5.509 \pm 0.029$ | 6dFGS | g1214188−415954 | 0.06918 | $6.312\times10^{25} \pm 3.278\times10^{23}$ | 85.7 |
| G4Jy 984 | 0 | $<24.8$ | $-0.95 \pm 0.01$ | $7.872 \pm 0.047$ | SDSS DR12 | J121738.77+033948.2 | $0.07784 \pm 0.00002$ | $1.169\times10^{26} \pm 7.046\times10^{23}$ | 36.5 |
| G4Jy 985 | 0 | $<14.7$ | $-0.81 \pm 0.03$ | $4.781 \pm 0.054$ | SDSS DR12 | J121756.88+252927.5 | $0.67890 \pm 0.00007$ | $8.727\times10^{27} \pm 9.935\times10^{25}$ | 103.7 |
| G4Jy 987 | 0 | 275.1 | $-0.84 \pm 0.01$ | $14.127 \pm 0.051$ | 6dFGS | g1218380−102035 | 0.08734 | $2.650\times10^{26} \pm 9.621\times10^{23}$ | 449.6 |
| G4Jy 988 | 0 | 77.4 | $-0.72 \pm 0.03$ | $4.648 \pm 0.054$ | SDSS DR12 | J121859.14+195528.0 | $0.42442 \pm 0.00002$ | $2.732\times10^{27} \pm 3.145\times10^{25}$ | 430.9 |

**Table C1.** Continued.

| Source name | Confusion flag | Angular size/″ | G4Jy spectral-index | Total $S_{151\,\mathrm{MHz}}$ /Jy | Redshift origin | Host-galaxy name | Spectroscopic redshift | $L_{151\,\mathrm{MHz}}$ /W Hz$^{-1}$ | Linear size /kpc |
|---|---|---|---|---|---|---|---|---|---|
| G4Jy 991 | 1 | $<17.3$ | $-0.99 \pm 0.02$ | $8.746 \pm 0.045$ | SDSS DR12 | J122027.98+092827.3 | $1.08103 \pm 0.00023$ | $5.519{\times}10^{28} \pm 2.827{\times}10^{26}$ | 140.9 |
| G4Jy 993 | 0 | $<18.7$ | $-0.77 \pm 0.03$ | $4.167 \pm 0.053$ | SDSS DR12 | J122044.89+223404.7 | $1.87646 \pm 0.00011$ | $8.098{\times}10^{28} \pm 1.027{\times}10^{27}$ | 157.5 |
| G4Jy 994 | 1 | 57.1 | – | $7.344 \pm 0.029$ | 6dFGS | g1223434−423532 | 0.02658 | – | 30.5 |
| G4Jy 995 | 1 | $<59.0$ | $-0.90 \pm 0.03$ | $4.292 \pm 0.066$ | SDSS DR16 | J122433.28+261315.0 | $0.58266 \pm 0.00009$ | $5.684{\times}10^{27} \pm 8.798{\times}10^{25}$ | 389.1 |
| G4Jy 997 | 0 | $<14.7$ | $-0.69 \pm 0.02$ | $6.103 \pm 0.049$ | SDSS DR12 | J122454.46+212246.4 | $0.43383 \pm 0.00006$ | $3.740{\times}10^{27} \pm 2.974{\times}10^{25}$ | 82.9 |
| G4Jy 1001 | 0 | 73.6 | $-0.61 \pm 0.03$ | $4.830 \pm 0.058$ | SDSS DR16 | J122811.73+202352.4 | $0.70004 \pm 0.00006$ | $8.540{\times}10^{27} \pm 1.028{\times}10^{26}$ | 525.8 |
| G4Jy 1005 | 0 | $<14.1$ | $-1.00 \pm 0.01$ | $7.876 \pm 0.030$ | SDSS DR12 | J123200.01−022404.7 | $1.04313 \pm 0.00011$ | $4.570{\times}10^{28} \pm 1.762{\times}10^{26}$ | 114.0 |
| G4Jy 1007 | 0 | 16.4 | $-0.87 \pm 0.01$ | $4.028 \pm 0.020$ | 6dFGS | g1232431−244552 | 0.25682 | $7.896{\times}10^{26} \pm 3.975{\times}10^{24}$ | 65.4 |
| G4Jy 1008 | 0 | 118.1 | – | $18.191 \pm 0.068$ | SDSS DR16 | J123526.66+212034.7 | $0.42266 \pm 0.00004$ | – | 656.0 |
| G4Jy 1009 | 1 | 100.4 | $-0.86 \pm 0.01$ | $12.563 \pm 0.028$ | 6dFGS | g1235378−251217 | 0.35496 | $5.138{\times}10^{27} \pm 1.138{\times}10^{25}$ | 500.3 |
| G4Jy 1011 | 0 | 191.0 | $-0.92 \pm 0.02$ | $7.802 \pm 0.083$ | SDSS DR12 | J123625.77+163218.3 | $0.06846 \pm 0.00001$ | $8.832{\times}10^{25} \pm 9.395{\times}10^{23}$ | 250.2 |
| G4Jy 1012 | 0 | $<24.8$ | – | $4.709 \pm 0.066$ | SDSS DR16 | J123631.31+263508.6 | $1.39871 \pm 0.00035$ | – | 209.1 |
| G4Jy 1019 | 0 | $<15.8$ | – | $16.680 \pm 0.061$ | SDSS DR16 | J124357.65+162253.3 | $0.55523 \pm 0.00004$ | – | 101.8 |
| G4Jy 1029 | 0 | $<48.2$ | $-0.79 \pm 0.02$ | $4.006 \pm 0.039$ | SDSS DR12 | J125222.61+031554.0 | $0.09885 \pm 0.00002$ | $9.712{\times}10^{25} \pm 9.420{\times}10^{23}$ | 88.0 |
| G4Jy 1032 | 0 | $<31.9$ | – | $9.493 \pm 0.075$ | SDSS DR12 | J125412.00+273733.9 | $0.08582 \pm 0.00001$ | – | 51.3 |
| G4Jy 1034 | 0 | $<172.8$ | $-0.73 \pm 0.01$ | $30.746 \pm 0.053$ | 6dFGS | g1254357−123407 | 0.01317 | $1.190{\times}10^{25} \pm 2.032{\times}10^{22}$ | 46.5 |
| G4Jy 1035 | 1 | $<24.9$ | $-1.19 \pm 0.01$ | $13.365 \pm 0.034$ | 6dFGS | g1254410−291340 | 0.05736 | $1.062{\times}10^{26} \pm 2.731{\times}10^{23}$ | 27.7 |
| G4Jy 1039 | 0 | 23.2 | $-0.82 \pm 0.01$ | $6.164 \pm 0.025$ | 6dFGS | g1257206−333444 | 0.18987 | $6.115{\times}10^{26} \pm 2.445{\times}10^{24}$ | 73.5 |
| G4Jy 1040 | 1 | 40.9 | $-1.07 \pm 0.01$ | $6.550 \pm 0.027$ | 6dFGS | g1257219−302149 | 0.05422 | $4.598{\times}10^{25} \pm 1.906{\times}10^{23}$ | 43.1 |
| G4Jy 1044 | 0 | 69.7 | $-0.73 \pm 0.02$ | $4.713 \pm 0.026$ | 6dFGS | g1301008−322629 | 0.01701 | $3.059{\times}10^{24} \pm 1.676{\times}10^{22}$ | 24.1 |
| G4Jy 1051 | 0 | $<46.1$ | – | $6.588 \pm 0.039$ | SDSS DR12 | J130753.92+064213.8 | $0.60139 \pm 0.00007$ | – | 308.5 |
| G4Jy 1060 | 0 | 190.3 | – | $6.211 \pm 0.044$ | SDSS DR12 | J131617.01+070246.6 | $0.05022 \pm 0.00001$ | – | 186.8 |
| G4Jy 1062 | 0 | $<15.9$ | $-0.63 \pm 0.01$ | $5.995 \pm 0.028$ | SDSS DR12 | J131938.76−004940.0 | $0.89114 \pm 0.00006$ | $1.857{\times}10^{28} \pm 8.814{\times}10^{25}$ | 123.5 |
| G4Jy 1065 | 1 | $<15.2$ | $-0.69 \pm 0.01$ | $8.900 \pm 0.038$ | SDSS DR12 | J132118.83+110649.9 | $2.17827 \pm 0.00015$ | $2.210{\times}10^{29} \pm 9.377{\times}10^{26}$ | 125.8 |
| G4Jy 1066 | 0 | 74.5 | $-0.89 \pm 0.01$ | $7.607 \pm 0.037$ | SDSS DR16 | J132318.81+030807.1 | $0.26893 \pm 0.00004$ | $1.662{\times}10^{27} \pm 8.165{\times}10^{24}$ | 307.2 |
| G4Jy 1068 | 1 | 70.6 | $-0.89 \pm 0.01$ | $6.611 \pm 0.027$ | SDSS DR12 | J132738.22−020309.9 | $0.18269 \pm 0.00002$ | $6.096{\times}10^{26} \pm 2.520{\times}10^{24}$ | 216.7 |
| G4Jy 1075 | 0 | 109.3 | $-0.73 \pm 0.01$ | $10.569 \pm 0.044$ | SDSS DR16 | J133253.26+020045.6 | $0.21582 \pm 0.00003$ | $1.368{\times}10^{27} \pm 5.743{\times}10^{24}$ | 382.5 |
| G4Jy 1079 | 1 | 807.8 | $-0.95 \pm 0.02$ | $13.707 \pm 0.173$ | 6dFGS | g1334186−100929 | 0.08372 | $2.372{\times}10^{26} \pm 2.996{\times}10^{24}$ | 1 270.8 |
| G4Jy 1081 | 0 | 197.6 | – | $8.366 \pm 0.070$ | 6dFGS | g1337005−295160 | 0.00170 | – | 7.0 |
| G4Jy 1084 | 0 | $<18.5$ | $-0.71 \pm 0.02$ | $4.535 \pm 0.033$ | SDSS DR16 | J133932.44+014522.5 | $0.56669 \pm 0.00004$ | $5.164{\times}10^{27} \pm 3.806{\times}10^{25}$ | 120.4 |
| G4Jy 1086 | 0 | $<19.5$ | $-0.73 \pm 0.01$ | $6.266 \pm 0.035$ | SDSS DR12 | J134243.62+050432.2 | $0.13648 \pm 0.00001$ | $2.997{\times}10^{26} \pm 1.654{\times}10^{24}$ | 47.1 |
| G4Jy 1094 | 0 | $<77.5$ | $-1.17 \pm 0.01$ | $4.116 \pm 0.027$ | 6dFGS | g1348542−252724 | 0.12612 | $1.754{\times}10^{26} \pm 1.149{\times}10^{24}$ | 175.0 |
| G4Jy 1108 | 0 | $<26.4$ | – | $7.931 \pm 0.042$ | SDSS DR12 | J135704.43+191907.3 | $0.71967 \pm 0.00001$ | – | 190.7 |
| G4Jy 1110 | 0 | 368.3 | $-0.77 \pm 0.01$ | $8.850 \pm 0.047$ | 6dFGS | g1401419−113625 | 0.03797 | $2.940{\times}10^{25} \pm 1.568{\times}10^{23}$ | 277.4 |

**Table C1.** Continued.

| Source name | Confusion flag | Angular size/” | G4Jy spectral-index | Total $S_{151\,\mathrm{MHz}}$ /Jy | Redshift origin | Host-galaxy name | Spectroscopic redshift | $L_{151\,\mathrm{MHz}}$ /W Hz$^{-1}$ | Linear size /kpc |
|---|---|---|---|---|---|---|---|---|---|
| G4Jy 1112 | 0 | $<34.4$ | $-0.84\pm0.01$ | $5.030\pm0.032$ | SALT | J140231.57+021546.5 | $0.180\pm0.003$ | $4.456\times10^{26}\pm2.855\times10^{24}$ | 104.4 |
| G4Jy 1122 | 0 | $<47.6$ | $-0.86\pm0.01$ | $7.044\pm0.027$ | SDSS DR16 | J140648.61−015416.4 | $0.63993\pm0.00012$ | $1.148\times10^{28}\pm4.430\times10^{25}$ | 327.6 |
| G4Jy 1127 | 0 | $<43.8$ | $-1.06\pm0.02$ | $4.339\pm0.072$ | 6dFGS | g1413319−300244 | 0.06486 | $4.427\times10^{25}\pm7.356\times10^{23}$ | 54.6 |
| G4Jy 1135 | 0 | 186.5 | $-0.89\pm0.01$ | $11.445\pm0.079$ | SALT | J141633.15−364053.7 | $0.075\pm0.001$ | $1.564\times10^{26}\pm1.081\times10^{24}$ | 265.6 |
| G4Jy 1137 | 0 | 317.0 | $-0.55\pm0.01$ | $11.805\pm0.053$ | SDSS DR12 | J141652.94+104826.7 | $0.02471\pm0.00001$ | $1.624\times10^{25}\pm7.275\times10^{22}$ | 157.8 |
| G4Jy 1145 | 0 | 61.6 | $-0.87\pm0.01$ | $9.797\pm0.023$ | 6dFGS | g1419497−192826 | 0.11990 | $3.614\times10^{26}\pm8.323\times10^{23}$ | 133.2 |
| G4Jy 1157 | 1 | 59.8 | $-1.06\pm0.01$ | $15.118\pm0.066$ | SALT | J142416.47−382647.6 | $0.406\pm0.001$ | $9.022\times10^{27}\pm3.967\times10^{25}$ | 324.5 |
| G4Jy 1163 | 0 | $<21.3$ | – | $7.566\pm0.047$ | SDSS DR12 | J142550.71+240403.3 | $0.65300\pm0.00004$ | – | 147.9 |
| G4Jy 1173 | 0 | 683.4 | – | $10.567\pm0.093$ | SDSS DR12 | J142955.37+071512.8 | $0.05509\pm0.00001$ | – | 731.6 |
| G4Jy 1177 | 0 | $<19.5$ | – | $4.225\pm0.040$ | SDSS DR16 | J143556.61+172934.6 | $1.20212\pm0.00015$ | – | 161.8 |
| G4Jy 1182 | 0 | $<46.4$ | – | $5.590\pm0.053$ | SDSS DR16 | J143821.86+034013.2 | $0.22468\pm0.00004$ | – | 167.5 |
| G4Jy 1184 | 0 | $<18.6$ | $-0.55\pm0.01$ | $8.604\pm0.023$ | 6dFGS | g1439287−165905 | 0.14570 | $4.621\times10^{26}\pm1.232\times10^{24}$ | 47.5 |
| G4Jy 1189 | 0 | $<36.5$ | – | $4.066\pm0.037$ | SDSS DR16 | J144631.52+072900.0 | $0.74211\pm0.00005$ | – | 266.8 |
| G4Jy 1199 | 0 | 70.8 | – | $4.015\pm0.035$ | SDSS DR12 | J145353.90+093423.2 | $0.62711\pm0.00002$ | – | 483.1 |
| G4Jy 1200 | 1 | 294.2 | – | $7.184\pm0.065$ | SDSS DR16 | J145423.45+162119.0 | $0.04528\pm0.00001$ | – | 261.9 |
| G4Jy 1202 | 0 | 42.9 | $-0.78\pm0.02$ | $4.840\pm0.050$ | 6dFGS | g1454274−374733 | 0.31384 | $1.458\times10^{27}\pm1.517\times10^{25}$ | 197.1 |
| G4Jy 1203 | 1 | 124.3 | $-1.03\pm0.01$ | $19.718\pm0.060$ | SALT | J145428.22−364004.7 | $0.420\pm0.001$ | $1.260\times10^{28}\pm3.831\times10^{25}$ | 687.7 |
| G4Jy 1205 | 0 | 90.6 | $-0.95\pm0.02$ | $6.970\pm0.064$ | 6dFGS | g1455096−365508 | 0.09459 | $1.562\times10^{26}\pm1.436\times10^{24}$ | 159.1 |
| G4Jy 1210 | 1 | 89.4 | – | $7.392\pm0.044$ | SDSS DR12 | J145605.67+162654.8 | $0.28713\pm0.00006$ | – | 386.2 |
| G4Jy 1212 | 0 | 158.5 | $-0.86\pm0.04$ | $5.077\pm0.092$ | SDSS DR12 | J145753.81+283218.7 | $0.14404\pm0.00001$ | $2.775\times10^{26}\pm5.056\times10^{24}$ | 400.6 |
| G4Jy 1216 | 0 | 188.2 | – | $68.229\pm0.102$ | SDSS DR12 | J150457.12+260058.4 | $0.05398\pm0.00001$ | – | 197.7 |
| G4Jy 1227 | 0 | 115.5 | – | $23.746\pm0.049$ | SDSS DR12 | J151100.02+075150.2 | $0.45938\pm0.00004$ | – | 672.9 |
| G4Jy 1231 | 0 | 55.6 | $-0.87\pm0.02$ | $4.638\pm0.040$ | SDSS DR16 | J151215.74+020316.9 | $0.21985\pm0.00002$ | $6.417\times10^{26}\pm5.468\times10^{24}$ | 197.3 |
| G4Jy 1239 | 0 | $<25.2$ | $-0.93\pm0.01$ | $55.856\pm0.052$ | SDSS DR12 | J151644.48+070117.8 | $0.03453\pm0.00001$ | $1.536\times10^{26}\pm1.418\times10^{23}$ | 17.3 |
| G4Jy 1240 | 1 | $<18.8$ | – | $6.454\pm0.044$ | SDSS DR16 (re-fitted) | J151656.59+183021.5 | $0.58100\pm0.00032$ | – | 123.8 |
| G4Jy 1244 | 0 | $<18.0$ | – | $14.881\pm0.047$ | SDSS DR12 | J152005.47+201605.4 | $1.57250\pm0.00021$ | – | 152.5 |
| G4Jy 1245 | 0 | $<80.5$ | $-0.80\pm0.01$ | $6.504\pm0.031$ | 6dFGS | g1520060−283420 | 0.12264 | $2.498\times10^{26}\pm1.192\times10^{24}$ | 177.4 |
| G4Jy 1255 | 0 | $<16.4$ | – | $4.203\pm0.034$ | SDSS DR16 | J152356.96+105543.4 | $0.41392\pm0.00007$ | – | 90.0 |
| G4Jy 1267 | 0 | 19.4 | – | $5.469\pm0.039$ | SDSS DR12 | J153315.07+133225.1 | $0.77477\pm0.00004$ | – | 144.0 |
| G4Jy 1270 | 0 | $<14.2$ | – | $6.875\pm0.043$ | SDSS DR12 | J153732.43+134448.1 | $0.67176\pm0.00012$ | – | 99.8 |
| G4Jy 1272 | 1 | 47.7 | $-0.92\pm0.02$ | $4.816\pm0.044$ | SDSS DR12 | J154112.87+005032.0 | $1.13738\pm0.00064$ | $3.244\times10^{28}\pm2.948\times10^{26}$ | 392.2 |
| G4Jy 1278 | 0 | 58.9 | $-0.82\pm0.02$ | $11.052\pm0.053$ | SDSS DR16 | J154743.53+205216.6 | $0.26428\pm0.00002$ | $2.285\times10^{27}\pm1.101\times10^{25}$ | 240.0 |
| G4Jy 1286 | 0 | 54.0 | $0.04\pm0.02$ | $7.146\pm0.043$ | 6dFGS | g1556589−791404 | 0.05896 | $5.607\times10^{25}\pm3.383\times10^{23}$ | 61.6 |

**Table C1.** Continued.

| Source name | Confusion flag | Angular size/″ | G4Jy spectral-index | Total $S_{151\,\mathrm{MHz}}$ /Jy | Redshift origin | Host-galaxy name | Spectroscopic redshift | $L_{151\,\mathrm{MHz}}$ /W Hz$^{-1}$ | Linear size /kpc |
|---|---|---|---|---|---|---|---|---|---|
| G4Jy 1290 | 0 | $< 76.6$ | $-0.95 \pm 0.02$ | $4.207 \pm 0.045$ | SDSS DR16 | J155907.05+121030.4 | $0.31736 \pm 0.00004$ | $1.364\times10^{27} \pm 1.472\times10^{25}$ | 354.6 |
| G4Jy 1291 | 0 | 102.1 | $-0.76 \pm 0.02$ | $4.052 \pm 0.051$ | SDSS DR12 | J160027.78+083743.0 | $0.22687 \pm 0.00003$ | $5.883\times10^{26} \pm 7.462\times10^{24}$ | 371.3 |
| G4Jy 1297 | 0 | 89.5 | $-0.87 \pm 0.02$ | $4.351 \pm 0.058$ | SDSS DR12 | J160240.36+154521.5 | $0.03738 \pm 0.00001$ | $1.405\times10^{25} \pm 1.869\times10^{23}$ | 66.4 |
| G4Jy 1306 | 0 | 2.5 | $-0.66 \pm 0.01$ | $8.798 \pm 0.049$ | SALT | J160612.69+000027.2 | $0.058 \pm 0.001$ | $6.947\times10^{25} \pm 3.897\times10^{23}$ | 2.8 |
| G4Jy 1332 | 0 | 23.6 | $-0.66 \pm 0.02$ | $13.679 \pm 0.068$ | SDSS DR12 | J162439.08+234512.1 | $0.92715 \pm 0.00011$ | $4.731\times10^{28} \pm 2.354\times10^{26}$ | 185.4 |
| G4Jy 1339 | 0 | $< 63.9$ | $-0.77 \pm 0.02$ | $13.033 \pm 0.090$ | SDSS DR12 | J162803.97+274139.3 | $0.44881 \pm 0.00003$ | $8.868\times10^{27} \pm 6.123\times10^{25}$ | 367.5 |
| G4Jy 1343 | 0 | $< 58.4$ | $-0.88 \pm 0.01$ | $13.024 \pm 0.046$ | SALT | J163141.60−265651.3 | $0.168 \pm 0.001$ | $9.984\times10^{26} \pm 3.539\times10^{24}$ | 167.7 |
| G4Jy 1348 | 0 | 56.7 | $-0.76 \pm 0.03$ | $7.952 \pm 0.082$ | SDSS DR12 | J163636.50+264809.2 | $0.56148 \pm 0.00004$ | $9.092\times10^{27} \pm 9.331\times10^{25}$ | 367.1 |
| G4Jy 1358 | 0 | 19.5 | $-0.56 \pm 0.01$ | $12.424 \pm 0.057$ | SDSS DR12 | J164348.60+171549.4 | $0.16300 \pm 0.00002$ | $8.494\times10^{26} \pm 3.872\times10^{24}$ | 54.6 |
| G4Jy 1360 | 0 | 207.0 | $-0.48 \pm 0.02$ | $25.322 \pm 0.060$ | 6dFGS | g1644161−771549 | 0.04297 | $1.071\times10^{26} \pm 2.522\times10^{23}$ | 175.3 |
| G4Jy 1364 | 0 | $< 18.1$ | $-0.74 \pm 0.01$ | $11.270 \pm 0.069$ | SALT | J164542.39+021145.0 | $0.094 \pm 0.001$ | $2.446\times10^{26} \pm 1.502\times10^{24}$ | 31.6 |
| G4Jy 1383 | 0 | $< 87.9$ | $-0.43 \pm 0.02$ | $7.644 \pm 0.040$ | 6dFGS | g1702410−774157 | 0.09461 | $1.635\times10^{26} \pm 8.527\times10^{23}$ | 154.3 |
| G4Jy 1402 | 0 | 216.6 | $-0.64 \pm 0.01$ | $232.344 \pm 0.181$ | 6dFGS | g1720282−005847 | 0.03038 | $4.875\times10^{26} \pm 3.794\times10^{23}$ | 131.7 |
| G4Jy 1423 | 0 | 308.8 | $-0.84 \pm 0.01$ | $39.071 \pm 0.062$ | 6dFGS | g1737358−563403 | 0.09839 | $9.425\times10^{26} \pm 1.505\times10^{24}$ | 561.5 |
| G4Jy 1487 | 0 | 17.7 | – | $30.436 \pm 0.053$ | SALT | J183058.92−360230.7 | $0.078 \pm 0.001$ | – | 26.1 |
| G4Jy 1504 | 0 | 88.7 | $-0.99 \pm 0.01$ | $19.111 \pm 0.037$ | 6dFGS | g1843146−483623 | 0.11072 | $6.022\times10^{26} \pm 1.151\times10^{24}$ | 179.0 |
| G4Jy 1511 | 0 | $< 20.1$ | $-0.91 \pm 0.01$ | $10.105 \pm 0.037$ | SALT | J185710.80−301940.1 | $1.554 \pm 0.003$ | $1.439\times10^{29} \pm 5.204\times10^{26}$ | 170.3 |
| G4Jy 1518 | 1 | 43.5 | $-1.00 \pm 0.01$ | $15.518 \pm 0.048$ | SALT | WISEA J191548.68−265257.4 | $0.231 \pm 0.003$ | $2.463\times10^{27} \pm 7.547\times10^{24}$ | 160.3 |
| G4Jy 1526 | 0 | 82.7 | $-0.87 \pm 0.01$ | $8.137 \pm 0.041$ | 6dFGS | g1919280−295808 | 0.16670 | $6.116\times10^{26} \pm 3.081\times10^{24}$ | 235.9 |
| G4Jy 1533 | 0 | $< 16.5$ | $0.14 \pm 0.02$ | $5.299 \pm 0.038$ | 6dFGS | g1924510−291430 | 0.35238 | $1.572\times10^{27} \pm 1.122\times10^{25}$ | 81.9 |
| G4Jy 1537 | 1 | 141.1 | $-1.00 \pm 0.02$ | $5.964 \pm 0.057$ | 6dFGS | g1926057−574017 | 0.06092 | $5.317\times10^{25} \pm 5.076\times10^{23}$ | 165.9 |
| G4Jy 1544 | 1 | 198.7 | $-0.72 \pm 0.02$ | $7.580 \pm 0.053$ | 6dFGS | g1928170−293144 | 0.02440 | $1.021\times10^{25} \pm 7.174\times10^{22}$ | 97.7 |
| G4Jy 1549 | 1 | $< 55.3$ | $-1.06 \pm 0.01$ | $8.636 \pm 0.035$ | 6dFGS | g1931382−335442 | 0.09781 | $2.101\times10^{26} \pm 8.508\times10^{23}$ | 100.0 |
| G4Jy 1555 | 0 | 100.2 | $-0.81 \pm 0.01$ | $10.753 \pm 0.041$ | SALT | J193325.00−394020.7 | $0.074 \pm 0.001$ | $1.421\times10^{26} \pm 5.410\times10^{23}$ | 140.9 |
| G4Jy 1565 | 1 | $< 18.5$ | $-0.63 \pm 0.01$ | $25.361 \pm 0.043$ | SALT | J194115.02−152431.2 | $0.454 \pm 0.001$ | $1.683\times10^{28} \pm 2.877\times10^{25}$ | 107.1 |
| G4Jy 1581 | 0 | 148.5 | $-0.88 \pm 0.01$ | $25.613 \pm 0.065$ | SALT | J195215.79+023024.1 | $0.059 \pm 0.001$ | $2.121\times10^{26} \pm 5.384\times10^{23}$ | 169.5 |
| G4Jy 1592 | 1 | 16.8 | $-0.74 \pm 0.01$ | $9.100 \pm 0.038$ | 6dFGS | g1959038−353431 | 0.35292 | $3.542\times10^{27} \pm 1.471\times10^{25}$ | 83.4 |
| G4Jy 1606 | 0 | 48.5 | $-1.11 \pm 0.02$ | $5.048 \pm 0.072$ | 6dFGS | g2011275−564407 | 0.05514 | $3.679\times10^{25} \pm 5.260\times10^{23}$ | 52.0 |
| G4Jy 1613 | 0 | 494.2 | $-0.98 \pm 0.02$ | $10.348 \pm 0.125$ | 6dFGS | g2018013−553932 | 0.06059 | $9.109\times10^{25} \pm 1.101\times10^{24}$ | 578.1 |
| G4Jy 1619 | 0 | $< 20.1$ | $-0.94 \pm 0.02$ | $6.130 \pm 0.030$ | 6dFGS | g2024353−051641 | 0.08218 | $1.020\times10^{26} \pm 4.935\times10^{23}$ | 31.1 |
| G4Jy 1635 | 1 | $< 61.0$ | $-0.79 \pm 0.01$ | $11.789 \pm 0.029$ | 6dFGS | g2033166−225317 | 0.13140 | $5.241\times10^{26} \pm 1.286\times10^{24}$ | 142.6 |

**Table C1.** Continued.

| Source name | Confusion flag | Angular size/″ | G4Jy spectral-index | Total $S_{151\,\mathrm{MHz}}$ /Jy | Redshift origin | Host-galaxy name | Spectroscopic redshift | $L_{151\,\mathrm{MHz}}$ /W Hz$^{-1}$ | Linear size /kpc |
|---|---|---|---|---|---|---|---|---|---|
| G4Jy 1638 | 0 | $<24.1$ | $-0.86 \pm 0.01$ | $8.809 \pm 0.037$ | 6dFGS | g2034447−354902 | 0.08873 | $1.711\times10^{26} \pm 7.140\times10^{23}$ | 40.0 |
| G4Jy 1643 | 0 | 220.8 | $-0.65 \pm 0.01$ | $9.344 \pm 0.047$ | 6dFGS | g2043457−263301 | 0.04074 | $3.568\times10^{25} \pm 1.805\times10^{23}$ | 177.8 |
| G4Jy 1658 | 0 | $<27.6$ | $-0.92 \pm 0.03$ | $4.066 \pm 0.086$ | SDSS DR12 | J205125.70+165253.1 | $0.52857 \pm 0.00002$ | $4.305\times10^{27} \pm 9.156\times10^{25}$ | 173.4 |
| G4Jy 1660 | 0 | 40.5 | $-0.54 \pm 0.01$ | $5.406 \pm 0.036$ | SALT | J205202.35−570407.5 | $0.012 \pm 0.001$ | $1.731\times10^{24} \pm 1.146\times10^{22}$ | 9.9 |
| G4Jy 1664 | 0 | $<37.5$ | $-0.71 \pm 0.01$ | $12.294 \pm 0.027$ | 6dFGS | g2056043−195635 | 0.15651 | $7.871\times10^{26} \pm 1.715\times10^{24}$ | 101.6 |
| G4Jy 1665 | 0 | 84.9 | $-0.73 \pm 0.01$ | $5.727 \pm 0.023$ | SALT | J205616.36−471447.6 | $1.488 \pm 0.003$ | $6.245\times10^{28} \pm 2.548\times10^{26}$ | 718.3 |
| G4Jy 1671 | 1 | 440.4 | $-0.73 \pm 0.01$ | $24.589 \pm 0.063$ | 6dFGS | g2101377−280154 | 0.03942 | $8.805\times10^{25} \pm 2.247\times10^{23}$ | 343.6 |
| G4Jy 1677 | 0 | 263.4 | $-0.82 \pm 0.01$ | $18.556 \pm 0.051$ | 6dFGS | g2107163−252808 | 0.03680 | $5.792\times10^{25} \pm 1.607\times10^{23}$ | 192.5 |
| G4Jy 1678 | 0 | 76.6 | – | $23.187 \pm 0.051$ | 6dFGS | g2107257−252543 | 0.03878 | – | 58.8 |
| G4Jy 1686 | 0 | $<14.1$ | $-0.75 \pm 0.01$ | $12.094 \pm 0.015$ | 6dFGS | g2118106−301912 | 0.97711 | $5.031\times10^{28} \pm 6.079\times10^{25}$ | 112.3 |
| G4Jy 1698 | 0 | $<18.2$ | $-0.75 \pm 0.01$ | $11.697 \pm 0.018$ | SALT | J213101.47−203656.1 | $1.630 \pm 0.002$ | $1.599\times10^{29} \pm 2.473\times10^{26}$ | 154.2 |
| G4Jy 1704 | 0 | 195.7 | $-0.83 \pm 0.02$ | $4.443 \pm 0.047$ | SALT | J213406.70−533418.7 | $0.078 \pm 0.001$ | $6.569\times10^{25} \pm 6.899\times10^{23}$ | 288.8 |
| G4Jy 1705 | 0 | 72.0 | $-0.90 \pm 0.02$ | $4.187 \pm 0.047$ | SALT | J213417.69−533811.1 | $0.076 \pm 0.002$ | $5.888\times10^{25} \pm 6.564\times10^{23}$ | 103.7 |
| G4Jy 1708 | 0 | 110.8 | $-0.78 \pm 0.01$ | $20.556 \pm 0.032$ | 6dFGS | g2137452−143256 | 0.19979 | $2.265\times10^{27} \pm 3.495\times10^{24}$ | 365.4 |
| G4Jy 1709 | 0 | $<17.0$ | – | $11.341 \pm 0.019$ | SALT | J213750.00−204231.7 | $0.638 \pm 0.004$ | – | 116.8 |
| G4Jy 1714 | 0 | 64.8 | $-1.11 \pm 0.01$ | $4.550 \pm 0.017$ | 6dFGS | g2140141−441228 | 0.07124 | $5.671\times10^{25} \pm 2.170\times10^{23}$ | 88.0 |
| G4Jy 1718 | 0 | 171.5 | $-1.55 \pm 0.01$ | $6.282 \pm 0.037$ | 6dFGS | g2143592−563721 | 0.08189 | $1.088\times10^{26} \pm 6.458\times10^{23}$ | 264.5 |
| G4Jy 1719 | 0 | 51.6 | – | $20.703 \pm 0.142$ | SDSS DR12 | J214411.71+281018.8 | $0.21461 \pm 0.00002$ | – | 179.8 |
| G4Jy 1732 | 0 | 366.1 | $-0.74 \pm 0.02$ | $7.227 \pm 0.054$ | 6dFGS | g2151299−552013 | 0.03876 | $2.501\times10^{25} \pm 1.884\times10^{23}$ | 281.2 |
| G4Jy 1741 | 0 | 346.9 | $-0.85 \pm 0.02$ | $4.234 \pm 0.031$ | 6dFGS | g2154230−455232 | 0.14519 | $2.352\times10^{26} \pm 1.734\times10^{24}$ | 882.7 |
| G4Jy 1744 | 1 | 86.0 | $-1.06 \pm 0.02$ | $5.085 \pm 0.084$ | SDSS DR16 | J215541.97+123128.6 | $0.19305 \pm 0.00004$ | $5.462\times10^{26} \pm 9.044\times10^{24}$ | 276.0 |
| G4Jy 1748 | 0 | 65.9 | $-0.50 \pm 0.01$ | $100.007 \pm 0.077$ | 6dFGS | g2157060−694124 | 0.02848 | $1.833\times10^{26} \pm 1.415\times10^{23}$ | 37.7 |
| G4Jy 1751 | 0 | 96.0 | $-0.86 \pm 0.01$ | $7.887 \pm 0.019$ | 6dFGS | g2201171−374624 | 0.03341 | $2.023\times10^{25} \pm 4.949\times10^{22}$ | 63.9 |
| G4Jy 1753 | 0 | $<26.7$ | $-0.63 \pm 0.01$ | $4.568 \pm 0.014$ | 6dFGS | g2201564−332103 | 0.15359 | $2.777\times10^{26} \pm 8.782\times10^{23}$ | 71.2 |
| G4Jy 1760 | 0 | 15.5 | $-0.31 \pm 0.01$ | $4.840 \pm 0.021$ | 6dFGS | g2209202−233153 | 0.08644 | $8.504\times10^{25} \pm 3.611\times10^{23}$ | 25.1 |
| G4Jy 1763 | 0 | $<19.0$ | $-0.74 \pm 0.01$ | $4.959 \pm 0.023$ | 6dFGS | g2211241−132810 | 0.39178 | $2.449\times10^{27} \pm 1.135\times10^{25}$ | 100.8 |
| G4Jy 1775 | 1 | 538.6 | $-0.86 \pm 0.01$ | $30.199 \pm 0.112$ | 6dFGS | g2223495−020613 | 0.05584 | $2.229\times10^{26} \pm 8.250\times10^{23}$ | 583.9 |
| G4Jy 1781 | 0 | 7.0 | $-0.50 \pm 0.01$ | $10.262 \pm 0.017$ | SALT | J222918.63−405132.8 | $0.448 \pm 0.002$ | $6.290\times10^{27} \pm 1.070\times10^{25}$ | 40.3 |
| G4Jy 1786 | 0 | 90.1 | $-0.72 \pm 0.01$ | $9.259 \pm 0.024$ | 6dFGS | g2239114−172028 | 0.07411 | $1.220\times10^{26} \pm 3.128\times10^{23}$ | 126.9 |
| G4Jy 1788 | 1 | 62.4 | $-0.76 \pm 0.01$ | $4.285 \pm 0.022$ | 6dFGS | g2242237−604430 | 0.28157 | $1.005\times10^{27} \pm 5.108\times10^{24}$ | 265.9 |
| G4Jy 1795 | 0 | 32.7 | $-0.78 \pm 0.01$ | $26.175 \pm 0.024$ | SALT | J225303.11−405746.7 | $0.307 \pm 0.003$ | $7.502\times10^{27} \pm 7.016\times10^{24}$ | 148.0 |
| G4Jy 1796 | 0 | $<14.6$ | $-0.78 \pm 0.01$ | $6.194 \pm 0.018$ | 6dFGS | g2254599−391304 | 0.26132 | $1.236\times10^{27} \pm 3.624\times10^{24}$ | 59.0 |
| G4Jy 1802 | 0 | 120.4 | $-0.79 \pm 0.01$ | $5.147 \pm 0.020$ | 6dFGS | g2303030−184126 | 0.12884 | $2.193\times10^{26} \pm 8.514\times10^{23}$ | 276.8 |
| G4Jy 1812 | 0 | 70.0 | $-1.17 \pm 0.01$ | $4.020 \pm 0.017$ | 6dFGS | g2313586−424339 | 0.05641 | $3.081\times10^{25} \pm 1.291\times10^{23}$ | 76.6 |

**Table C1.** Continued.

| Source name | Confusion flag | Angular size/" | G4Jy spectral-index | Total $S_{151\,\mathrm{MHz}}$ /Jy | Redshift origin | Host-galaxy name | Spectroscopic redshift | $L_{151\,\mathrm{MHz}}$ /W Hz$^{-1}$ | Linear size /kpc |
|---|---|---|---|---|---|---|---|---|---|
| G4Jy 1817 | 0 | 90.5 | $-0.71 \pm 0.02$ | $4.280 \pm 0.031$ | 6dFGS | g2319059−420649 | 0.05428 | $2.954\times10^{25} \pm 2.120\times10^{23}$ | 95.5 |
| G4Jy 1819 | 0 | 167.8 | $-0.88 \pm 0.01$ | $13.648 \pm 0.031$ | SALT | J231956.26−272807.4 | $0.174 \pm 0.001$ | $1.129\times10^{27} \pm 2.587\times10^{24}$ | 495.5 |
| G4Jy 1829 | 0 | $< 16.9$ | $-0.84 \pm 0.01$ | $14.654 \pm 0.024$ | 6dFGS | g2325197−120727 | 0.08212 | $2.415\times10^{26} \pm 3.995\times10^{23}$ | 26.1 |
| G4Jy 1831 | 0 | $< 92.5$ | $-0.79 \pm 0.01$ | $10.289 \pm 0.041$ | SDSS DR16 | J232653.77−020213.7 | $0.18874 \pm 0.00002$ | $1.002\times10^{27} \pm 3.955\times10^{24}$ | 291.6 |
| G4Jy 1838 | 0 | $< 18.4$ | $-0.77 \pm 0.01$ | $7.340 \pm 0.026$ | SDSS DR16 | J233225.59−095756.2 | $1.67157 \pm 0.00064$ | $1.084\times10^{29} \pm 3.904\times10^{26}$ | 155.8 |
| G4Jy 1842 | 0 | 60.6 | $-0.84 \pm 0.01$ | $5.008 \pm 0.018$ | 6dFGS | g2334449−525119 | 1.03189 | $2.517\times10^{28} \pm 9.075\times10^{25}$ | 488.8 |
| G4Jy 1856 | 0 | 18.1 | $-0.84 \pm 0.01$ | $7.566 \pm 0.040$ | 6dFGS | g2351561−010913 | 0.17362 | $6.186\times10^{26} \pm 3.248\times10^{24}$ | 53.4 |
| G4Jy 1858 | 0 | $< 62.4$ | $-1.12 \pm 0.01$ | $22.438 \pm 0.024$ | 6dFGS | g2357007−344533 | 0.04941 | $1.302\times10^{26} \pm 1.403\times10^{23}$ | 60.3 |
| G4Jy 1863 | 0 | 372.9 | $-0.65 \pm 0.01$ | $125.123 \pm 0.087$ | 6dFGS | g2359043−605460 | 0.09573 | $2.799\times10^{27} \pm 1.940\times10^{24}$ | 661.6 |

**Table C2.** SDSS spectroscopic redshifts for five G4Jy sources, with new redshifts presented as a result of re-fitting the target spectrum (Appendix C and figure B3). The 'SDSS best-fit' values are where the reduced-$\chi^2$ metric is the global minimum (per source). Each of the sources listed are quasars.

| Source name | Redshift origin | SDSS objID | SDSS best-fit redshift | Alternative SDSS redshift | Reduced-$\chi^2$ ranking | Re-fitted redshift (this work) |
|---|---|---|---|---|---|---|
| G4Jy 148 | SDSS DR16 | 1237679323928789061 | $2.78413 \pm 0.00024$ | $1.97376 \pm 0.00034$ | 2nd | $1.965 \pm 0.005$ |
| G4Jy 176 | SDSS DR16 | 1237666274203271475 | $1.68729 \pm 0.00086$ | $1.63661 \pm 0.00034$ | 5th | $1.660 \pm 0.005$ |
| G4Jy 679 | SDSS DR12 | 1237667254525952687 | $0.24639 \pm 0.00010$ | $1.19644 \pm 0.00060$ | 6th | $1.196 \pm 0.001$ |
| G4Jy 845 | SDSS DR12 | 1237658423552376915 | $2.46126 \pm 0.00015$ | $1.71835 \pm 0.00016$ | 4th | $1.706 \pm 0.001$ |
| G4Jy 1240 | SDSS DR12 | 1237667967497077036 | $5.49964 \pm 0.00032$ | $0.58067 \pm 0.00005$ | 3rd | $0.581 \pm 0.001$ |